\PassOptionsToPackage{usenames,dvipsnames}{xcolor}
\documentclass[a4paper,11pt,1p]{elsarticle}

\usepackage[T1]{fontenc}
\usepackage[utf8]{inputenc}
\usepackage[british]{babel}
\addto\captionsbritish{\renewcommand{\appendixname}{Appendix }}
\usepackage{slashbox} 
\usepackage{geometry}
\usepackage{graphicx} 
\usepackage{amssymb} 
\usepackage{amsmath}
\usepackage{fancyvrb}
\usepackage{multicol}
\usepackage{multirow}
\usepackage{mathtools}
\usepackage{bbm}
\usepackage{tabularx}
\usepackage{hyperref}
\usepackage{adjustbox}  
\usepackage{url}
\usepackage{booktabs}
\usepackage{mathrsfs}
\usepackage{textcomp}
\usepackage{pgfplotstable}
\pgfplotsset{compat=1.6}
\usepackage[version=3]{mhchem}
\usepackage{framed}
\usepackage{pgfplots}
\usepackage{youngtab}
\usepackage{tikz}
\usepackage{tikz-3dplot}
\usetikzlibrary{trees, arrows} 
\usetikzlibrary{calc}
\usetikzlibrary{plotmarks}
\usepackage[all]{xy}
\usepackage{amsfonts}
\usetikzlibrary{arrows}
\usepackage{anysize}
\marginsize{2.4cm}{2.4cm}{1.2cm}{2.6cm}

\DeclarePairedDelimiter\floor{\lfloor}{\rfloor}
\newcommand{\norm}[1]{\left\lVert#1\right\rVert}
\newcommand{\sign}{\text{sign}}

\makeatletter
\def\ps@pprintTitle{
   \let\@oddhead\@empty
   \let\@evenhead\@empty
   \def\@oddfoot{\reset@font\hfil\thepage\hfil}
   \let\@evenfoot\@oddfoot
}
\makeatother

\newsavebox{\leftbox}
\newsavebox{\rightbox}

\definecolor{DarkMidnightBlue}{rgb}{0.0, 0.04, 0.14}

\hypersetup{hidelinks,
backref=true,
pagebackref=true,
hyperindex=true,
breaklinks=true,
colorlinks=true,
urlcolor=DarkMidnightBlue,
bookmarks=true,
bookmarksopen=false,
pdftitle={Title},
pdfauthor={Author}}

\VerbatimFootnotes

\title{\textbf{\textsf{Breaking and restoration of rotational symmetry in the low-energy spectrum of light alpha-conjugate nuclei on the lattice~I: \\ \ce{^8Be} and \ce{^{12}C}}}}

\author[hiskp,bctp]{Gianluca~Stellin}
\author[hiskp,bctp,kmu]{Serdar~Elhatisari}
\author[hiskp,bctp,fzj,tsu]{Ulf-G.~Mei\ss{}ner}
\address[hiskp]{Helmholtz Institut für Strahlen- und Kernphysik, Universität Bonn, Nu\ss{}allee 14-16, 53115 Bonn, Germany}
\address[bctp]{Bethe Center for Theoretical Physics, Universität Bonn, Nu\ss{}allee 12, 53115 Bonn, Germany}
\address[kmu]{Karamano\u{g}lu Mehmetbey Üniversitesi, 70200 Karaman, Turkey}
\address[fzj]{Insitute for Advanced Simulation, Institut f\"ur Kernphysik and J\"ulich Center for Hadron Physics, \\
Forschungszentrum J\"ulich, 52425 J\"ulich, Germany}
\address[tsu]{Ivane Javakhishvili Tbilisi State University,  0186 Tbilisi, Georgia}

\date{\today}

\begin{document}



\begin{abstract}
\begin{small}
The breaking of rotational symmetry on the lattice for bound eigenstates of the two lightest alpha conjugate nuclei is explored. Moreover, a macroscopic alpha-cluster model is used for investigating the general problems associated with the representation of a physical many-body problem on a cubic lattice. In view of the descent from the 3D rotation group to the cubic group symmetry, the role of the squared total angular momentum operator in the classification of the lattice eigenstates in terms of SO(3) irreps is discussed. In particular, the behaviour of the average values of the latter operator, the Hamiltonian and the inter-particle distance as a function of lattice spacing and size is studied by considering the $0^+$, $2^+$, $4^+$ and $6^+$ (artificial) bound states of \ce{^8Be} and the lowest $0^+$, $2^+$ and $3^-$ multiplets of \ce{^{12}C}.
\end{small}
\end{abstract}

\maketitle

\clearpage

\tableofcontents

\clearpage

\section{\textsf{Preamble}}\label{S-1.0}

The wealth of available literature on lattice calculations is, perhaps, self-explanatory on the role that the latter play in the investigation of relativistic field theories and quantum few-body and many-body systems. After the first study of nuclear matter on the lattice in Ref.~\cite{BrF92} in the framework of quantum hadrodynamics~\cite{Wal74}, lattice simulations have begun to be employed for several other systems involving nuclear matter, fostered by the development of effective field theories \cite{MKS00,DLe09} such as Chiral Effective Field Theories (ChEFT) \cite{LBS04,EHM09,DLe09}.\\
In the lattice framework, the continuous space-time is discretized and compactified on a hypercubic box so that differential operators become matrices and the relevant path-integrals are evaluated numerically. When periodic boundary conditions are imposed in all the space directions, the whole configuration space is reduced to a three-dimensional torus and translational invariance is preserved. Nevertheless, the average values of physical observables on the lattice eigenstates will, in general, depend on the features of the box employed for the description of the physical system rather than obey to their continuum and infinite-volume counterparts.\\
Starting from L\"uscher's early works \cite{Lue86-01,Lue86-02,Lue91}, in the last three decades much effort has been devoted to investigate the finite-volume dependence of physical observables on the lattice, with a special attention for the energy of bound states. \\
The original formula connecting the leading-order finite-volume correction for the energy eigenvalues to the asymptotic properties of the two-particle bound wavefunctions in the infinite volume in Ref.~\cite{Lue86-01} has been extended in several directions including non-zero angular momenta \cite{LuS11,KLH11,KLH12,DMO12}, moving frames \cite{DMO12,KSS05,RuG95,BKL11,DaS11,GHL12,REK14}, generalized boundary conditions \cite{SaV05,DMO11,BDL14,KoL16,SNR16,CSW17}, particles with intrinsic spin \cite{BLM08,BrH15} and perturbative Coulomb corrections \cite{BeS14}. In addition, considerable advances have been made in the derivation of analogous formulas for the energy corrections of bound states of three-body \cite{PoR12,MRR15} and N-body systems \cite{KoL18}.\\
While closed expressions for leading-order finite-volume corrections to certain physical observables already exist, artifacts due to the finite lattice spacing remain more difficult to keep under control. \\
Nevertheless, systematic schemes for the improvement of discretized expressions of quantities of physical interest have been developed. In these approaches, correction terms are identified using continuum language and are added with suitable coefficients, so that corrections up to the desired order in the lattice spacing vanish. \\  In the context of field theories, namely Yang-Mills theories, discretization effetcs can be reduced via the Symanzik improvement program \cite{Sym80,Wei83,Sym83,Par85,GaL10}. The latter is based on the systematic inclusion of higher-dimensional operators into the lattice action, whose coefficients are determined through a perturbative or nonperturbative matching procedure \cite{GaL10}. A similar approach, reviewed in the appendix, can be implemented for differential operators applied to wavefunctions, in which the derivation of the coefficients in front of the corrective terms stems only from algebraic considerations \cite{GaL10}, differently from the previous case. \\
Another consequence of transposing a physical system into a cubic lattice is given by the reduction of the rotational symmetry group to the finite group of the rotations of a cube. If the former is ruled by central forces, the rotation group on three dimensions, SO(3), shrinks into the rotation subgroup, $\mathcal{O}$, of the octahedral group $\mathcal{O}_h$. Therefore, lattice eigenstates of a few-body Hamiltonian cannot be unambiguously classified in terms of irreducible representations of  SO(3) or SU(2) \cite{Joh82}. In the transition between infinitesimal and finite spacing, the $2\ell+1$-fold degeneracy in the energies of the members of a multiplet of states transforming according to the same irreducible representation $\ell$ of SO(3) reduces to 1-,2- or 3-fold degeneracy, depending on the cubic-group irreps that appear in the decomposition of the original representation of the rotation group (cf. Tab.~\ref{T-4.0-01} in Sec.~\ref{S-4.0}). In particular, the energy separation between the ensuing $\mathcal{O}$ multiplets grows smoothly with increasing lattice spacings. \\ 
This descent in symmetry has been recapitulated in Ref.~\cite{Joh82}, where the the problem of the identification of the cubic lattice eigenstates in terms of SO(3) irreps has been first outilined. The increasing importance of the discretization of the euclidean spacetime in the context of gauge theories \cite{Wil74,OsS78,DrZ83} led soon to an extension of Johnson's work to the case of an hypercubic lattice \cite{MZG83}. In the meantime, investigations explicitly devoted to rotational symmetry breaking appeared in the context of scalar $\lambda\varphi^4$ \cite{Lan90} and gauge field theories \cite{LaR82,BeB83} on the lattice. More recently, 
quantitative estimations of rotational symmetry breaking have been performed in both the frameworks in Ref~\cite{DaS12} and in Lattice QCD for exotic mesons in Ref.~\cite{DEP09}, via the construction of operators with sharply defined angular momentum. \\
Nevertheless, the restoration of the full rotational invariance on the lattice can be achieved by projecting the lattice wavefunctions onto angular momentum quantum numbers via the construction of projectors on SO(3) irreps. The use of such a technique has been firstly reported  in Ref.~\cite{BaH84}, in the context of cranked Hartree-Fock self-consistent calculations for \ce{^{24}Mg}.\\ 
However, in the present paper we aim at investigating rotational symmetry breaking in bound states of \ce{^8Be} and \ce{^{12}C} nuclei on the lattice rather than at removing these effects. At the same time, the analysis of the low-energy spectra of the two light $\alpha$-conjugate nuclei provides us an occasion to highlight the general issues associated to finite volume and discretization in energies, angular momenta and average interparticle distances.\\
Since the framework allows for a robust analysis over a wide range of lattice spacings and cubic box sizes, for the purpose we adopt a simplified description in terms of $\alpha$ particles instead of individual nucleons, following on the recent literature on the same subject, cf. Refs.~\cite{BNL14,BNL15}.\\
Even if they can explain only a part of the spectra of $4N$ self-conjugate nuclei, $\alpha$-cluster models have strong foundations \cite{Whe37} and influence even in the recent literature \cite{SFV16,FSV17} and succeeded in describing certain ground-state properties of this class of nuclei (cf. the linear behaviour of the binding energy as a function of the number of the bounds between the alpha particles \cite{HaT38}) as well as the occurrence of decay thresholds into lighter $\alpha$-conjugate nuclei  (cf. the \emph{Ikeda diagram} \cite{ITH68,IHS80}). For a recent review, see Ref.~\cite{FHK18}.\\ 
The interaction between $\alpha$ particles can be realistically described by microscopically based potentials within the method of generator coordinates \cite{GrW57}, the resonating group model \cite{Tan69,BFW77}, the orthogonality condition model \cite{Sai69}, the WKB model of Ref.~\cite{Hor80}, the energy-density or the folding model \cite{SPH93}. Alternatively, phenomenological potentials constructed from $\alpha-\alpha$ scattering data, like the Woods-Saxon ones of Ref.~\cite{NKK71} and Ref.~\cite{KNS75}, or the Gaussian ones of Ref.~\cite{ABo66}, can be considered.\\
Furthermore, our two-body interaction presented in Sec.~\ref{S-2.0}, builds on the work of Ref.~\cite{BNL14} and consists of an isotropic Ali-Bodmer type potential, \textit{i.e.} a superposition of a positive and a negative-amplitude Gaussian. The other part of the Hamiltonian operator including the kinetic term is presented in Secs.~\ref{S-2.0} and \ref{S-3.1}. Moreover, for the implementation of the second-order derivative operators of the latter on the lattice, the improvement scheme summarized in \ref{S-10.1} and \ref{S-10.2} has been adopted.\\
The sought extension of the finite-volume and discretization analysis in Sec.~III.~A and  B of Ref.~\cite{BNL14} to higher angular momentum multiplets has been here achieved through the introduction of an additional tool, the discretized version of the squared total angular momentum operator. If the lattice spacing is not too large (\textit{e.g.} $a \lesssim 1.5\hspace{1mm}\mathrm{fm}$ and $a \lesssim 0.65\hspace{1mm}\mathrm{fm}$ in the two \ce{^8Be} configurations considered in Sec.~\ref{S-7.0}) and the lattice volume is large enough (\textit{e.g.} $L \equiv Na \gtrsim 18\hspace{1mm}\mathrm{fm}$ and $Na \gtrsim 12\hspace{1mm}\mathrm{fm}$ respectively), the average values of the squared total angular momentum operator on the states turn out to provide precise information on the SO(3) multiplets to which the eigenstates belong in the continuum and infinite-volume limit. The capability of the latter operator of drawing this information also from the lowest energy bound states of \ce{^{12}C} is tested and discussed in Sec.~\ref{S-8.0}. A similar analysis on the bound eigenstates of the \ce{^{16}O} nucleus in the same $\alpha$-cluster model is the subject of a forthcoming paper.\\

\section{\textsf{Theoretical framework}}\label{S-2.0}

\subsection{\textsf{The Hamiltonian}}\label{S-2.1}

In the phenomenological picture considered here, individual nucleons are grouped into \ce{^4He} clusters, that are treated as spinless spherically-charged particles of mass $m \equiv m_{\ce{^4He}}$ subject to both two-body $V^{\rm II}$ and three-body potentials $V^{\rm III}$. Therefore, the Hamiltonian of the system reads 
\begin{equation}
H = -\frac{\hbar^2}{2m}\sum_{i=1}^M \nabla_i^2 + \sum_{i<j} V^{\rm II}(\textbf{r}_i,\textbf{r}_j)+\sum_{i<j<k} V^{\rm III}(\textbf{r}_i,\textbf{r}_j,\textbf{r}_k)~.\label{2.1-01}
\end{equation}
The global effects of the strong force between two $\alpha$ particles at a distance $\mathbf{r}$ are described by the phenomenological Ali-Bodmer potential, 
\begin{equation}
V_{AB}(\mathbf{r}) = V_0 e^{-\eta_0^2 r^2}+V_1 e^{-\eta_1^2r^2}~,\label{2.1-02}
\end{equation}
consisting of a superposition of a long range attractive Gaussian and a short range repulsive one with the parameters
\begin{eqnarray}
\eta_0^{-1} = 2.29~\mbox{rm}~,~~~V_0 = -216.3~\mbox{MeV}~,\nonumber\\
\eta_1^{-1} = 1.89~\mbox{rm}~,~~~V_1 = -353.5~\mbox{MeV}~.\label{2.1-03}
\end{eqnarray}
Moreover, the range parameter of the attractive part of this isotropic Ali-Bodmer potential agrees with the ones fitting the $\alpha-\alpha$ scattering lengths with $\ell = 0$, $2$ and $4$ to their experimental values \cite{ABo66}, whereas the compatibility of $V_0$ with the best fits of the latter (cf. $d_0'$, $d_2$ and $d_4$ in Ref.~\cite{ABo66}) is poorer ($\approx 30~\%$). As the repulsive part of this potential is strongly angular momentum dependent, its parameters reproduce within $10\%$ likelihood only the ones for $D$-wave scattering lengths, $d_2$ \cite{ABo66}.
Assuming that the charge distribution of the $\alpha$-particles is spherical and obeys a Gaussian law with an rms radius $R_{\alpha}=1.44\hspace{1mm}\mathrm{fm}$ \cite{ABo66}, the Coulomb interaction between the \ce{^4He} nuclei takes the form 
\begin{equation}
V_C(\mathbf{r})=\frac{4e^2}{4\pi \varepsilon_0} \frac{1}{r}\mathrm{erf} \left(\frac{\sqrt{3}r}{2R_{\alpha}}\right)~.\label{2.1-04}
\end{equation}
in terms of the error function, $\mathrm{erf}(x)= (1/\sqrt{\pi})\cdot $ $\cdot \int_x^x e^{-t^2} dt$. The three-body term of the Hamiltonian, $V^{\rm III}$, consists of a Gaussian attractive potential, 
\begin{equation}
V_T(\mathbf{r}_{ij},\mathbf{r}_{jk},\mathbf{r}_{ik})=V_0 e^{-\lambda(r_{ij}^2+r_{jk}^2+r_{ik}^2)}~,\label{2.1-04}
\end{equation}
whose range $\lambda = 0.005~\mathrm{fm}^{-2}$ and amplitude parameters $V_0 = -4.41~\textrm{MeV}$ were originarily fitted to reproduce, respectively, the binding energy of the \ce{^{12}C} and the spacing between the Hoyle state, \textit{i.e.} the $0_2^+$ at $7.65~\mathrm{MeV}$ and $2_1^+$  one at $4.44~\mathrm{MeV}$ \cite{PoC79} of the same nuclide in the case the original angular momentum dependent Ali-Bodmer potential, \textit{i.e.} a superposition of three pairs of Gaussians of the form \eqref{2.1-02} with parameters $d_0'$, $d_2$ and $d_4$ \cite{ABo66}, was adopted. However, in the present case, the three pairs of quadratic exponentials, corresponding to the best fitting potentials for the $S$, $D$ and $G$-wave $\alpha-\alpha$ scattering amplitudes \cite{ABo66}, have been resummed into a single pair of Gaussians that adjusts the zero of the energy on the Hoyle state rather than on the $3\alpha$ decay threshold. Since the spacing between the latter two is experimentally well-established, the possibility of reproducing the binding energy of the nucleus still remains.\\

\begin{figure}[t!]
\centering
\includegraphics[width=0.85\columnwidth]{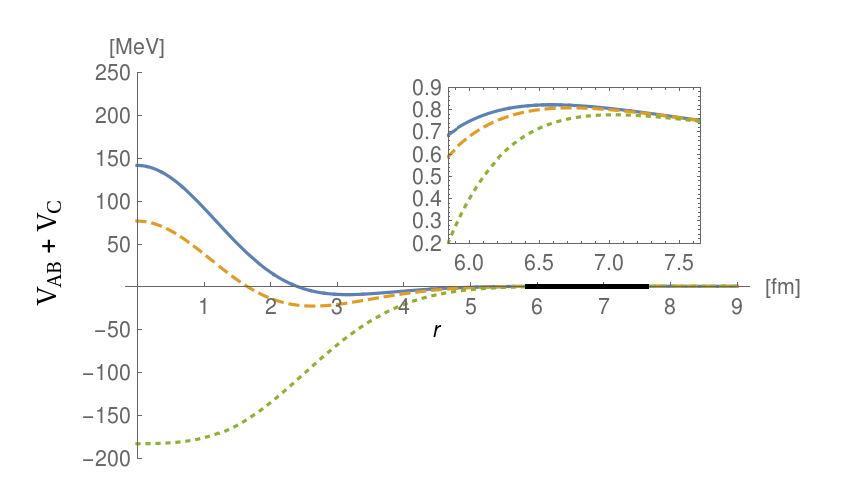}
\begin{small}
\caption{Behaviour of the two-body potentials for a system of two particles in presence of Coulomb (cf. Eq.~\ref{2.1-04}) and Ali-Bodmer (cf. Eq.~\ref{2.1-02}) interactions with $V_0$ equal to $100\%$ (solid line), $130\%$ (dashed line) and $250\%$ (dotted line) of its value presented in Eq.~\ref{2.1-03}. The latter two potentials with artificially enhanced strength parameter have been introduced in order to generate a set of low-lying bound states with different angular momenta, at the root of the analysis presented in Sec.~\ref{S-7.0}. The increase of $V_0$ leads to the disappearance of the absolute maximum at zero interaction distance simulating the short-range Pauli repulsion between the $\alpha$-particles. In particular, the shape of the dotted curve resembles the one of a Woods-Saxon potential except for the remaining shallow maximum at $7.0$~fm, highlighted in the magnification.}
\end{small}
\label{F-2.1-01}
\end{figure}

\section{\textsf{Operators on the lattice}}\label{S-3.0}

Now, let us construct the operators of physical interest acting on a discretized and finite configuration space, \textit{i.e.} a lattice with $N$ points per dimension and spacing $a$.

\subsection{\textsf{Kinetic energy}}\label{S-3.1}

\begin{figure}[h]
\begin{minipage}[c]{0.61\columnwidth}
\begin{center}
\begin{tikzpicture}
\begin{axis}[xlabel={\small $p_x$}, ylabel={\small $2m\mathcal{T}(p_x)/\hbar^2$},xmin = -3, xmax=3,ymin=0,ymax=5, samples=50, legend style ={at={(0.32,0.93)}, anchor=north west, draw=black,fill=white,align=left},
	legend entries ={\small continuum, \small $K=4$, \small $K=3$,\small $K=2$, \small $K=1$}]
 \addlegendimage{black, solid, very thick};
 \addlegendimage{magenta, solid, thick};
 \addlegendimage{red, densely dotted, thick};
 \addlegendimage{green, dotted, thick};
 \addlegendimage{blue, loosely dotted, thick};
  \addplot [black, solid, very thick] {x^2};
  \addplot [blue, loosely dotted, thick] {2*(1-cos(deg(x)))};
  \addplot [green, dotted, thick] {5/2-(8/3)*cos(deg(x))+(1/6)*cos(deg(2*x))};
  \addplot [red, densely dotted, thick] {49/18-(3)*cos(deg(x))+(3/10)*cos(deg(2*x))-(1/45)*cos(deg(3*x))};
  \addplot [magenta, solid, thick] {205/72-(16/5)*cos(deg(x))+(2/5)*cos(deg(2*x))-(16/315)*cos(deg(3*x))+(1/280)*cos(deg(4*x))};
\end{axis}
\end{tikzpicture}
\end{center}
\end{minipage}
\begin{minipage}[c]{0.29\columnwidth}
\caption{Behaviour of the eingenvaules of a free particle in one dimension, $x$, as function of the lattice momentum $p_x$ for four different values of the second derivative improvement index and unit spacing. For increasing values of $K$ the eigenvalues of $\mathcal{T}(p_x)$ approach the continuum ones with increasing likelihood.}
\end{minipage}
\label{F-3.1-01}
\end{figure}
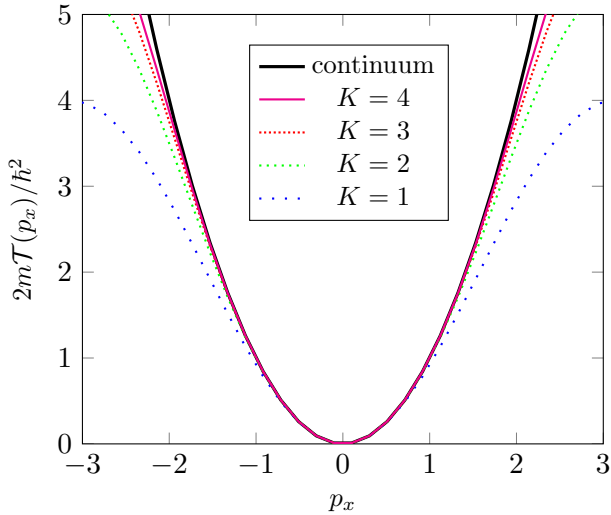

Applying the many-body kinetic energy operator
\begin{equation}
T = -\frac{\hbar^2}{2m}\sum_{i=1}^M \nabla_i^2  \label{3.1-01}
\end{equation}
on the most general many-body wavefunction
\begin{equation}
\Psi(\mathbf{r}_1,\mathbf{r}_2,...\mathbf{r}_M) = \langle \Psi| \mathbf{r}_1,\mathbf{r}_2,...\mathbf{r}_M \rangle \label{3.1-02}
\end{equation}
and replacing the the exact derivatives with their discretized version (cf. Eq.~\eqref{10.1-11}), the explicit form of lattice counterpart of $T$ can be derived. To this aim, it is customary to introduce ladder operators, $a_i^{\dagger}(\mathbf{r}_i)$ and $a_i(\mathbf{r}_i)$, acting on the discretized version of the kets of Eq.~\eqref{3.1-02}, whose meaning is respectively the creation and the destruction of the particle $i$ at the position $\mathbf{r}_i$. Therefore, applying the discretization scheme outlined in \ref{S-10.1} \cite{GaL10} with improvement index $K$, the kinetic energy operator on the cubic lattice $\mathscr{N}$ becomes
\begin{equation}
\hat{\mathcal{T}} =  -\frac{\hbar^2}{2m}\sum_{\substack{ \alpha \in \\ x,y,z}}\sum_{i=1}^M \sum_{\mathbf{r}_i \in \mathscr{N}}\sum_{k=1}^KC_k^{(2P,K)}\left[-2 a_i^{\dagger}(\mathbf{r}_i) a_i(\mathbf{r}_i) + a_i^{\dagger}(\mathbf{r}_i) a_i(\mathbf{r}_i + k a \mathbf{e}_{\alpha}) +  a_i^{\dagger}(\mathbf{r}_i) a_i(\mathbf{r}_i - k a \mathbf{e}_{\alpha})\right]
\label{3.1-03}
\end{equation}
where $e_{\alpha}$ are unit-vectors parallel to the axes of the lattice. The latter equation can be more succinctly rewritten as
\begin{equation}
\hat{\mathcal{T}} =  -\frac{\hbar^2}{2m}\sum_{\substack{ \alpha \in \\ x,y,z}}\sum_{i=1}^M \sum_{\mathbf{r}_i \in \mathscr{N}}\sum_{l=-K}^KC_{|l|}^{(2P,K)} a_i^{\dagger}(\mathbf{r}_i) a_i(\mathbf{r}_i + l a \mathbf{e}_{\alpha})~.
\label{3.1-04}
\end{equation}
After defining dimensionless lattice momenta as 
\begin{equation}
 \textbf{p}_i = \frac{2\pi \textbf{n}_i}{N} \hspace{1.0cm} \textbf{n}_i \in \mathscr{N} \subset \mathbb{Z}^3~,\label{3.1-07}
\end{equation}
by imposing periodic boundary conditions, we can switch to the momentum space via the discrete Fourier transform of the lattice ladder operators, 
\begin{equation}
\hat{\mathcal{T}} = \sum_{i=1}^M \sum_{\mathbf{p_i} \in \mathscr{N}} a_i^{\dagger}(\mathbf{p}_i) \mathcal{T}_{p_i} a_i(\mathbf{p}_i)~.\label{3.1-05}
\end{equation}
Therefore, we can extract the analytical expression of the eigenvalues of a system of free particles from the original expression of $\hat{\mathcal{T}}$  in configuration space (cf. Eq.~\eqref{3.1-04}),
\begin{equation}
\begin{split}
\mathcal{T}_{p_i} = \frac{\hbar^2}{2m}\sum_{\substack{ \alpha \in \\ x,y,z}} \sum_{k=1}^K C_k^{(2P,K)}\left[2 - \cosh\left(k~p_{i,\alpha}\right) \right]\label{3.1-06}
\end{split}
\end{equation}
 (cf. Fig.~\ref{F-3.1-01}). From the final form of lattice dispersion relation in Eq.~\eqref{3.1-06}, we can conclude that Galilean invariance is broken on the lattice, since the dependence of the former on the $\mathbf{p}_i$'s is not quadratic \cite{LeT07}. 

The extent of the configuration space and the dimension of the corresponding kinetic energy matrix, whose elements are 
\begin{equation}
 \mathcal{T}^{(a)}_{\mathbf{r},\mathbf{r'}} \equiv \langle \mathbf{r}_1, \mathbf{r}_2, ... \mathbf{r}_M|\hat{\mathcal{T}}|\mathbf{r'}_1, \mathbf{r'}_2, ... \mathbf{r'}_M\rangle\label{3.1-08}
\end{equation}
in the absolute basis of states \footnote{Notice that dimensionless position vectors $\mathbf{n}_i$, such that $\mathbf{r}_i = a\mathbf{n}_i$, have been introduced.},
\begin{equation}
|\mathbf{n}_1, \mathbf{n}_2, ...\mathbf{n}_M\rangle = \prod_{i=1}^M \left(\sum_{p_i \in \mathscr{N}} e^{-i\mathbf{n}_i\cdot\mathbf{p}_i} \right) |\mathbf{p}_1, \mathbf{p}_2, ...\mathbf{p}_M\rangle~,\label{3.1-09}
\end{equation}
can be reduced from $N^{3M}$ to $N^{3M-3}$ by singling out the center of mass motion of the $M$ alpha particles. Accordingly, we introduce the following non-orthogonal transformation into relative coordinates
\begin{equation}
\mathbf{r}_{jM} \equiv \mathbf{r}_j - \mathbf{r}_M\hspace{5mm}\mathbf{r}_{CM}= \sum_{i=1}^M\frac{\mathbf{r}_i}{M}\hspace{0.9cm}j=1,2,...M-1\label{3.1-10}
\end{equation}
together with the associated basis of Fock states, 
\begin{equation}
|\mathbf{n}_{1M}, \mathbf{n}_{2M},...\mathbf{n}_{M-1M},\mathbf{n}_{CM}\rangle  = \prod_{i=1}^{M-1} \left( \sum_{\substack{p_{iM} \in \mathscr{N}}} e^{-i\mathbf{n}_ {iM}\cdot\mathbf{p}_{iM}} \right) \cdot e^{-i\mathbf{n}_ {CM}\cdot\mathbf{p}_{CM}} |\mathbf{p}_{1M}, \mathbf{p}_{2M},...,\mathbf{p}_{M-1M},\mathbf{p}_{CM}\rangle~.
\label{3.1-11}
\end{equation}
Therefore, the matrix elements of the kinetic energy operator in the relative basis just introduced become 
\begin{equation}
\begin{gathered}
\mathcal{T}_{\mathbf{n},\mathbf{n'}}^{(r)} \equiv \langle \mathbf{n}_{1M}, \mathbf{n}_{2M}, ...\mathbf{n}_{CM} |\hat{\mathcal{T}}| \mathbf{n'}_{1M}, \mathbf{n'}_{2M}, ...,\mathbf{n'}_{CM} \rangle = -\frac{\hbar^2}{2m a^2} \sum_{\alpha}\sum_{\substack{l=-K \\ l \neq 0}}^K C_{|l|}^{(2P,K)}\times \\ \left[ - \text{\footnotesize $\langle \mathbf{n}_{1M},\mathbf{n}_{2M}, ...\mathbf{n}_{CM} | \mathbf{n}'_{1M}, \mathbf{n}'_{2M}, ...\mathbf{n}'_{CM}\rangle + \langle \mathbf{n}_{1M}, ...\mathbf{n}_{CM} | \mathbf{n}'_{1M} + l\mathbf{e}_{\alpha}, \mathbf{n}'_{2M}, ...\mathbf{n}'_{CM} + l\mathbf{e}_{\alpha}/M\rangle$ } \right. \\  \left. \text{\footnotesize $+ \langle \mathbf{n}_{1M}, ... \mathbf{n}_{CM} | \mathbf{n}'_{1M}, \mathbf{n}'_{2M} + l\mathbf{e}_{\alpha}, ... \mathbf{n}'_{CM} + l\mathbf{e}_{\alpha}/M\rangle + \dots + \langle \mathbf{n}_{1M}, ... \mathbf{n}_{CM} | \mathbf{n}'_{1M}, ...\mathbf{n}'_{M-1M} + l\mathbf{e}_{\alpha},\mathbf{n}'_{CM} + l\mathbf{e}_{\alpha}/M\rangle $ } \right. \\ \left. \text{\footnotesize $+\langle \mathbf{n}_{1M}, ...\mathbf{n}_{CM} | \mathbf{n}'_{1M}  + l\mathbf{e}_{\alpha}, \mathbf{n}'_{2M} + l\mathbf{e}_{\alpha}, ... \mathbf{n}'_{CM} - l\mathbf{e}_{\alpha}(M-1)/M\rangle$} \right]~.
\end{gathered}
\label{3.1-12}
\end{equation}
Replacing the brakets with the pertinent Kronecker deltas, we finally obtain
\begin{equation}
\begin{gathered}
\mathcal{T}_{\mathbf{n},\mathbf{n'}}^{(r)} = -\frac{\hbar^2}{2m a^2} \sum_{\alpha}\sum_{\substack{l=-K \\ l \neq 0}}^K C_{|l|}^{(2P,K)} \left[ \delta_{\mathbf{n}_{CM},\mathbf{n'}_{CM}-l\mathbf{e}_{\alpha}\frac{M-1}{M}}\left(\prod_{i=1}^{M-1}\delta_{\mathbf{n}_{iM},\mathbf{n'}_{iM}+l\mathbf{e}_{\alpha}}\right) \right. \\  \left. -  \delta_{\mathbf{n}_{CM},\mathbf{n'}_{CM}}\prod_{i=1}^{M-1} \delta_{\mathbf{n}_{iM}, \mathbf{n'}_{iM}} + \delta_{\mathbf{n}_{CM},\mathbf{n'}_{CM}+l\mathbf{e}_{\alpha}\frac{1}{M}}\sum_{i=1}^{M-1} \delta_{\mathbf{n}_{iM},\mathbf{n'}_{iM}+l\mathbf{e}_{\alpha}} \prod_{\substack{j=1\\ j\neq i}}^{M-2} \delta_{\mathbf{n}_{jM},\mathbf{n'}_{jM}}\right]~. 
\end{gathered}
\label{3.1-13}
\end{equation}
Choosing a reference frame in which the center of mass is at rest (\textit{i.e.} $\mathbf{p}_{CM}$ = 0), the matrix elements of $\hat{\mathcal{T}}$ become independent on the position of the center of the nucleus and the relevant deltas can be dropped from the last formula, thus
\begin{equation}
\begin{gathered}
\mathcal{T}_{\mathbf{n},\mathbf{n'}}^{(r,0)} \equiv \langle \mathbf{n}_{1M}, \mathbf{n}_{2M}, ...\mathbf{n}_{CM} |\hat{\mathcal{T}}| \mathbf{n'}_{1M}, \mathbf{n'}_{2M}, ...,\mathbf{n'}_{CM} \rangle_{\mathbf{p}_{_{CM}}=0} \\ = -\frac{\hbar^2}{2m a^2} \sum_{\alpha}\sum_{\substack{l=-K \\ l \neq 0}}^K C_{|l|}^{(2P,K)} \text{\footnotesize$\left[ \prod_{i=1}^{M-1}\delta_{\mathbf{n}_{iM},\mathbf{n'}_{iM}+l\mathbf{e}_{\alpha}} - \prod_{i=1}^{M-1} \delta_{\mathbf{n}_{iM}, \mathbf{n'}_{iM}} +  \sum_{i=1}^{M-1} \delta_{\mathbf{n}_{iM},\mathbf{n'}_{iM}+l\mathbf{e}_{\alpha}} \prod_{\substack{j=1\\ j\neq i}}^{M-2} \delta_{\mathbf{n}_{jM},\mathbf{n'}_{jM}}\right]$}~. 
\end{gathered}
\label{3.1-14}
\end{equation}   
After the reduction of the system to $N^{3M-3}$ degrees of freedom, one may wonder whether the matrix elements of $\mathcal{T}^{(r)}$ are invariant when the coordinate change (cf. Eq.~\eqref{3.1-10}) is performed \textit{before} the discretization of $T$ (cf. Eq.~\eqref{3.1-03}). The answer to this point is negative and the reason can be traced back to the non-orthogonality of the transformation into relative coordinates (cf. Eq.~\eqref{3.1-10}). Denoting the latter as $\mathbf{r'}_i \equiv \mathbf{r}_{iM}$ for $i <  M$ and $\mathbf{r'}_M \equiv \mathbf{r}_{CM}$ and computing the Jacobian matrix of the transformation, $\mathbb{J}$, 
\begin{equation}
\mathbb{J} \equiv
\begin{pmatrix}
1 & 0 & \dots & \dots & 0 & -1\\
0 & 1 & 0 & \dots & 0 & -1\\
\vdots & \ddots & \ddots &  \ddots & \vdots & \vdots \\
0 & \dots & 0 & 1 & 0 & -1\\
0 & \dots & \dots & 0 & 1 & -1\\
1/M & 1/M & \dots & \dots & \dots & 1/M\\
\end{pmatrix}~,
\label{3.1-16}
\end{equation}
 the resulting kinetic energy operator, in fact, is non-diagonal in the particle space,
\begin{equation}
T = -\frac{\hbar^2}{2m}\sum_{i,j,k=1}^M \mathbb{J}_{ji}^{-1}\mathbb{J}_{ki}^{-1}\nabla'_j\cdot\nabla'_k ~.
\label{3.1-17}
\end{equation}
It is exactly the presence of different kinds of differential operators, namely pure and mixed second derivatives, that prevents the final rewriting of the matrix elements of Eq.~\eqref{3.1-16}, after the cancellation of the center of mass momentum, to be consistent with Eq.~\eqref{3.1-14}. Nevertheless, the equivalence between the latter two can be approached in the large volume and small lattice spacing limit ($L \equiv Na \geq 18$~fm). \\ Eventually, if Jacobi coordinates instead of the relative ones in Eq.~\ref{3.1-10} were adopted, the coordinate transformation should have been effected \textit{before} the discretization of $T$ (cf. Eq.~\eqref{3.1-03}). The application of $\mathcal{T}$ in absolute coordinates on the transformed basis of states, in fact, would have generated fractional displacements on both the CM coordinates and in all the other relative ones, thus implying the existence of nonzero matrix elements between non-existing lattice sites.\\

\subsection{\textsf{Potentials}}\label{S-3.2}

Unlike the kinetic term, the definition of the lattice counterpart of the potentials \eqref{2.1-03} and \eqref{2.1-04} is straightforward, due to their locality and independence on spatial derivatives. 

\subsection{\textsf{Angular momentum}}\label{S-3.3}

An crucial role in the analysis that follows is played by the square of the collective angular momentum operator, $L_{\rm tot}^2$, whose importance resides in the identification of the multiplets of eigenstates of the lattice Hamiltonian that share the same orbital quantum number and the same energy in the continuum limit.\\
Differently from the previous case, the functional form of this operator is left invariant by linear transformations of the coordinates $\mathbb{J}$,
\begin{equation}
\text{\small$L_{\rm tot,\alpha}  = \sum_{i=1}^M L_{i,\alpha} =-i\hbar \epsilon_{\alpha\beta\gamma}\sum_{i=1}^M \beta_i \frac{\partial}{\partial \gamma_i} = -i\hbar \epsilon_{\alpha\beta\gamma}\sum_{i,j,k=1}^M \mathbb{J}_{ij}^{-1}\mathbb{J}_{ki}\beta'_j \frac{\partial}{\partial \gamma'_i} =  -i\hbar \epsilon_{\alpha\beta\gamma}\sum_{j,k=1}^M\delta_{kj}\beta'_j \frac{\partial}{\partial \gamma'_i} = \sum_{i=1}^M L'_{i,\alpha}$}~,
\label{3.3-01}
\end{equation}
where $\alpha,\beta,\gamma \in x,y,z$, $\epsilon_{\alpha\beta\gamma}$ is the Levi-Civita tensor with $\epsilon_{xyz}=1$ and summations over repeated greek indexes are understood. Accordingly, the square of the collective angular momentum operator can be written irrespectively of the coordinate system as
\begin{equation}
\begin{split}
 L_{\rm tot}^2  = 2\sum_{i<j} \mathbf{L}_i\cdot\mathbf{L}_j + \sum_iL_i^2   =-\hbar^2 \sum_{\beta,\gamma}\sum_{i<j}\left(2\beta_i\beta_j\right.&\left.\frac{\partial^2}{\partial\gamma_i\partial\gamma_j} - 2\beta_i\gamma_j\frac{\partial^2}{\partial\beta_j\partial\gamma_i}\right) \\ & -\hbar^2\sum_{\beta,\gamma}\sum_i\left(\beta_i^2\frac{\partial^2}{\partial \gamma_i^2} - \gamma_i\frac{2}{3}\frac{\partial}{\partial \gamma_i}-\gamma_i\beta_i\frac{\partial^2}{\partial \beta_i \partial \gamma_i}\right)~.
\end{split}
\label{3.3-02}
\end{equation} 
Since all the contributions from the second-derivative terms with $\beta=\gamma$ on the right hand side of Eq.~\eqref{3.3-02} vanish, each of the first three terms on the same side of the formula is hermitian. On the other hand, this property is not fulfilled by the remaining two terms unless they are summed together. \\ Applying the improvement scheme outlined in \ref{S-10.1} with index $K$, the subsequent discretization of the $\gamma_i \partial / \partial\gamma_i$ term of one-body part of Eq.~\eqref{3.3-02} gives 
\begin{equation}
\mathcal{L}_i^2 \Big\lvert_1 \equiv 2\hbar^2\sum_{\mathbf{n}_i\in \mathscr{N}}\sum_{\gamma}\sum_{k=1}^{K} C_k^{(1,K)} (\mathbf{n}_i)_{\gamma}\left[a_i^{\dagger}(\mathbf{n}_i+k\mathbf{e}_{\gamma})a_i(\mathbf{n}_i) - a_i^{\dagger}(\mathbf{n}_i-k\mathbf{e}_{\gamma})a_i(\mathbf{n}_i)\right]~,
\label{3.3-04}
\end{equation}
whereas the one of the remaining one-body part of the same operator gives 
\begin{equation}
\begin{gathered}
\mathcal{L}_i^2 \Big\lvert_2 \equiv \text{\small$-\hbar^2\sum_{\mathbf{n}_i\in \mathscr{N}}\sum_{\beta\neq\gamma}\sum_{k=1}^{K} C_k^{(2M,K)}$}\Big\{\text{\footnotesize $4(\mathbf{n}_i)_{\beta}^2$}\left[\text{\footnotesize $- 2a_i^{\dagger}(\mathbf{n}_i)a_i(\mathbf{n}_i)+a_i^{\dagger}(\mathbf{n}_i+k\mathbf{e}_{\gamma})a_i(\mathbf{n}_i) + a_i^{\dagger}(\mathbf{n}_i-k\mathbf{e}_{\gamma})a_i(\mathbf{n}_i)$}\right] \text{\footnotesize $- (\mathbf{n}_i)_{\beta}(\mathbf{n}_i)_{\gamma} $ } \\ \times \left[ \text{\footnotesize $a_i^{\dagger}(\mathbf{n}_i+k\mathbf{e}_{\beta}+k\mathbf{e}_{\gamma})a_i(\mathbf{n}_i) + a_i^{\dagger}(\mathbf{n}_i-k\mathbf{e}_{\beta}-k\mathbf{e}_{\gamma})a_i(\mathbf{n}_i) - a_i^{\dagger}(\mathbf{n}_i+k\mathbf{e}_{\beta}-k\mathbf{e}_{\gamma})a_i(\mathbf{n}_i)- a_i^{\dagger}(\mathbf{n}_i-k\mathbf{e}_{\beta}+k\mathbf{e}_{\gamma})a_i(\mathbf{n}_i) $ } \right] \Big\}~.
\end{gathered}
\label{3.3-05}
\end{equation}
Before introducing the ladder operators, all the diagonal terms in the greek indexes of this part of $L_{\rm tot}^2$ have been ruled out: the presence of two different kinds of differential operators prevents, in fact, the cancellation of one half of the hopping terms coming from the second pure and mixed derivatives. For what concerns the two-body part of Eq.~\eqref{3.3-02}, the discretization process gives 
\begin{equation}
\begin{gathered}
\mathbf{\mathcal{L}}_i\cdot\mathbf{\mathcal{L}}_j \Big\lvert_1 = -\hbar^2\sum_{\mathbf{n}_i,\mathbf{n}_j \in \mathscr{N}}\sum_{\beta,\gamma}\sum_{k=1}^{K} C_k^{(2M,K)}(\mathbf{n}_i)_{\beta}(\mathbf{n}_j)_{\beta} \\ \cdot \left[ a_i^{\dagger}(\mathbf{n}_i+k\mathbf{e}_{\gamma})a_j^{\dagger}(\mathbf{n}_j+k\mathbf{e}_{\gamma}) a_j(\mathbf{n}_j)a_i(\mathbf{n}_i) + a_i^{\dagger}(\mathbf{n}_i-k\mathbf{e}_{\gamma})a_j^{\dagger}(\mathbf{n}_j-k\mathbf{e}_{\gamma}) a_j(\mathbf{n}_j)a_i(\mathbf{n}_i) \right. \\ \left. - a_i^{\dagger}(\mathbf{n}_i+k\mathbf{e}_{\gamma})a_j^{\dagger}(\mathbf{n}_j-k\mathbf{e}_{\gamma}) a_j(\mathbf{n}_j)a_i(\mathbf{n}_i) - a_i^{\dagger}(\mathbf{n}_i-k\mathbf{e}_{\gamma})a_j^{\dagger}(\mathbf{n}_j+k\mathbf{e}_{\gamma}) a_j(\mathbf{n}_j)a_i(\mathbf{n}_i) \right] 
\end{gathered}
\label{3.3-03}
\end{equation}
and
\begin{equation}
\begin{gathered}
\mathbf{\mathcal{L}}_i\cdot\mathbf{\mathcal{L}}_j \Big\lvert_2 = \hbar^2\sum_{\mathbf{n}_i,\mathbf{n}_j \in \mathscr{N}}\sum_{\beta,\gamma}\sum_{k=1}^{K} C_k^{(2M,K)} (\mathbf{n}_i)_{\beta}(\mathbf{n}_j)_{\gamma} \\ \cdot \left[ a_i^{\dagger}(\mathbf{n}_i+k\mathbf{e}_{\gamma})a_j^{\dagger}(\mathbf{n}_j+k\mathbf{e}_{\beta}) a_j(\mathbf{n}_j)a_i(\mathbf{n}_i) + a_i^{\dagger}(\mathbf{n}_i-k\mathbf{e}_{\gamma})a_j^{\dagger}(\mathbf{n}_j-k\mathbf{e}_{\beta}) a_j(\mathbf{n}_j)a_i(\mathbf{n}_i) \right. \\  \left. - a_i^{\dagger}(\mathbf{n}_i+k\mathbf{e}_{\gamma})a_j^{\dagger}(\mathbf{n}_j-k\mathbf{e}_{\beta}) a_j(\mathbf{n}_j)a_i(\mathbf{n}_i) - a_i^{\dagger}(\mathbf{n}_i-k\mathbf{e}_{\gamma})a_j^{\dagger}(\mathbf{n}_j+k\mathbf{e}_{\beta}) a_j(\mathbf{n}_j)a_i(\mathbf{n}_i) \right]~.
\end{gathered}
\label{3.3-03bis}
\end{equation}
Due to the invariance of $L^2_{\rm tot}$, we are allowed to apply the square of the collective angular momentum operator in relative coordinates to the relative basis of states (cf. Eq.~\eqref{3.1-10}), exploiting the results already presented (cf. Eqs.~\eqref{3.3-03}-\eqref{3.3-05}). The subsequent cancelation of center of mass momentum, $\mathbf{p}_{_{CM}}=0$, yields finally the expression of the matrix element of the operator in the $N^{3M-3}\times N^{3M-3}$ lattice,
\begin{equation}
\begin{split}
\mathcal{L}_{\mathbf{n},\mathbf{n'}}^{'2\hspace{0.5mm}(r,0)} = \sum_{i}\langle \mathbf{n}_{1M}\dots\mathbf{n}_{CM}| \mathcal{L}_i^{'2} | \mathbf{n'}_{1M}\dots\mathbf{n'}_{CM} &\rangle_ {\mathbf{p}_{_{CM}}=0} \\ & +  2\sum_{i<j}\langle \mathbf{n}_{1M} \dots\mathbf{n}_{CM}| \mathbf{\mathcal{L}'}_i\cdot\mathbf{\mathcal{L}'}_j | \mathbf{n'}_{1M} \dots\mathbf{n'}_{CM}\rangle_ {\mathbf{p}_{_{CM}}=0}~,
\end{split}
\label{3.3-06}
\end{equation}
where the one-body contribution is given by
\begin{equation}
\begin{gathered}
\langle \mathbf{n}_{1M} \mathbf{n}_{2M}\dots\mathbf{n}_{CM}| \mathcal{L}_i^{'2} | \mathbf{n'}_{1M} \mathbf{n'}_{2M}\dots\mathbf{n'}_{CM}\rangle_ {\mathbf{p}_{_{CM}}=0} = - \hbar^2 \sum_{\beta \neq \gamma}\sum_{k=1}^K C_k^{(2M,K)}\text{\footnotesize $\left(\prod_{\substack{l=1 \\ l\neq i}}^{M-1} \delta_{\mathbf{n'}_{lM},\mathbf{n}_{lM}} \right)$ } \\\times\left[4(\mathbf{n}_{iM})_{\beta}^2 \times \left(-2\delta_{\mathbf{n}_{iM},\mathbf{n'}_{iM}} + \delta_{\mathbf{n}_{iM},\mathbf{n'}_{iM}+k\mathbf{e}_{\gamma}}+\delta_{\mathbf{n}_{iM},\mathbf{n'}_{iM}-k\mathbf{e}_{\gamma}}\right) - \right. \\ \left. \frac{4a}{3}(\mathbf{n}_{iM})_{\gamma}\left(\delta_{\mathbf{n}_{iM},\mathbf{n'}_{iM}+k\mathbf{e}_{\gamma}} - \delta_{\mathbf{n}_{iM},\mathbf{n'}_{iM}-k\mathbf{e}_{\gamma}}\right) \right. \\ \left. - (\mathbf{n}_{iM})_{\beta}(\mathbf{n}_{iM})_{\gamma} \left(\delta_{\mathbf{n}_{iM}, \mathbf{n'}_{iM}+k\mathbf{e}_{\beta}+k\mathbf{e}_{\gamma}} +\delta_{\mathbf{n}_{iM}, \mathbf{n'}_{iM}-k\mathbf{e}_{\beta}-k\mathbf{e}_{\gamma}} -\delta_{\mathbf{n}_{iM}, \mathbf{n'}_{iM}-k\mathbf{e}_{\beta}+k\mathbf{e}_{\gamma}}-\delta_{\mathbf{n}_{iM}, \mathbf{n'}_{iM}+k\mathbf{e}_{\beta}-k\mathbf{e}_{\gamma}}\right)\right]~.
\end{gathered}
\label{3.3-07}
\end{equation}
and the two-body one coincides with
\begin{equation}
\begin{gathered}
\langle \mathbf{n}_{1M} \mathbf{n}_{2M}\dots\mathbf{n}_{CM}| \mathbf{\mathcal{L}'}_i\cdot\mathbf{\mathcal{L}'}_j | \mathbf{n'}_{1M} \mathbf{n'}_{2M}\dots\mathbf{n'}_{CM}\rangle_ {\mathbf{p}_{_{CM}}=0} = -\hbar^2\sum_{\beta,\gamma}\sum_{k=1}^K C_k^{(2M,K)} \text{\footnotesize $\left(\prod_{\substack{l=1 \\ l\neq i \neq j}}^{M-1} \delta_{\mathbf{n'}_{lM},\mathbf{n}_{lM}}\right)$} \\ \times \left[ (\mathbf{n}_{iM})_{\beta}(\mathbf{n}_{jM})_{\beta} \left(\delta_{\mathbf{n}_{iM}, \mathbf{n'}_{iM}+k\mathbf{e}_{\gamma}}\delta_{\mathbf{n}_{jM}, \mathbf{n'}_{jM}+k\mathbf{e}_{\gamma}} + \delta_{\mathbf{n}_{iM}, \mathbf{n'}_{iM}-k\mathbf{e}_{\gamma}}\delta_{\mathbf{n}_{jM}, \mathbf{n'}_{jM}-k\mathbf{e}_{\gamma}} \right. \right. \\ \left. \left.  -\delta_{\mathbf{n}_{iM}, \mathbf{n'}_{iM}-k\mathbf{e}_{\gamma}}\delta_{\mathbf{n}_{jM}, \mathbf{n'}_{jM}+k\mathbf{e}_{\gamma}} -\delta_{\mathbf{n}_{iM}, \mathbf{n'}_{iM}+k\mathbf{e}_{\gamma}}\delta_{\mathbf{n}_{jM}, \mathbf{n'}_{jM}-k\mathbf{e}_{\gamma}}\right) \right. \\ \left. -  (\mathbf{n}_{iM})_{\beta}(\mathbf{n}_{jM})_{\gamma} \left(\delta_{\mathbf{n}_{iM}, \mathbf{n'}_{iM}+k\mathbf{e}_{\gamma}}\delta_{\mathbf{n}_{jM}, \mathbf{n'}_{jM}+k\mathbf{e}_{\beta}} +  \delta_{\mathbf{n}_{iM}, \mathbf{n'}_{iM}-k\mathbf{e}_{\gamma}}\delta_{\mathbf{n}_{jM}, \mathbf{n'}_{jM}-k\mathbf{e}_{\beta}} \right. \right. \\ \left. \left. - \delta_{\mathbf{n}_{iM}, \mathbf{n'}_{iM}-k\mathbf{e}_{\gamma}}\delta_{\mathbf{n}_{jM}, \mathbf{n'}_{jM}+k\mathbf{e}_{\beta}} - \delta_{\mathbf{n}_{iM}, \mathbf{n'}_{iM}+k\mathbf{e}_{\gamma}}\delta_{\mathbf{n}_{jM}, \mathbf{n'}_{jM}-k\mathbf{e}_{\beta}}\right)\right]~. \label{3.3-08}
\end{gathered}
\end{equation}
Like in the previous case, the application of the discretized version of this operator in absoulte and relative (\textit{i.e.} primed) coordinates to the relative basis, even if followed by the cancelation of the center of mass momentum, gives rise to two unequal results, namely
\begin{equation}
\mathcal{L}_{\mathbf{n'},\mathbf{n}}^{2\hspace{0.5mm}(r,0)} \neq \mathcal{L}_{\mathbf{n'},\mathbf{n}}^{'2\hspace{0.5mm}(r,0)}~.\label{3.3-09}
\end{equation}
respectively. This is a consequence of the discretization of the one-body terms containing second mixed and pure derivatives (cf. Eq.~\eqref{3.3-02}), that transform together under linear coordinate changes. As observed, also the cancellation of diagonal terms in the Greek indexes in the summations for the one-body terms of $L^2_{\rm tot}$ (\textit{i.e.} the ones with $\beta=\gamma$ in the third line of Eq.~\eqref{3.3-02}), that are straightforward in the continuum, does not occur in the lattice. Nevertheless, in the large volume and small lattice spacing limit, the average values of the squared collective angular momentum operator calculated in the two approaches (cf. Eq.~\eqref{3.3-09}) coincide, as expected in the case of the kinetic energy operator.\\
Besides this inequality, another feature of the discretized version of $L_{\rm tot}^2$ is the loss of hermiticity, due again to the last two terms of the one-body part (cf. Eq.~\eqref{3.3-02}) whose sum is self-adjoint only in the continuum.

\section{\textsf{Symmetries}}\label{S-4.0}

Let us begin the analysis of the transformation properties of the Hamiltonian under spacetime symmetries. Since the potentials depend only on interparticle distances, Eq.~\eqref{2.1-01} is invariant under parity, $\mathscr{P}$, 
\begin{equation}
[H, \mathscr{P}] = 0~,\label{4.0-01}
\end{equation}  
a feature that is preserved by its realization on the cubic lattice. This invariance allows for the construction of projectors to the two irreducible representations, $+$ and $-$, of the parity group ($\approx \mathcal{C}_2$), 
\begin{equation}
P_{\pm} = \mathbbm{1}\pm \mathscr{P} \label{4.0-02}
\end{equation}
acting on continuum (and lattice) eigenfunctions of $H$ (resp. $\mathcal{H}$), that can thus bear the two irrep labels. Moreover, the implementation of the reducible $3M-3$ dimensional representation of the inversion operator on the lattice, $\mathscr{P}$, omitted in the last section, depends on the choice of the map between lattice points $\mathbf{n}_{iM}$ and the physical points on $\mathbb{R}^3$.\\
Furthermore, the Hamiltonian of a system of $M$ particles interacting with central forces is rotationally invariant,
\begin{equation}
[H, \mathbf{L}_{\rm tot}] = 0   \hspace{0.2cm}\mathrm{and}\hspace{0.2cm} [H, L_i^2] = 0\label{4.0-03}
\end{equation} 
with $i=1,2,...M$. Switching to the relative reference frame, cf. Eq.~\eqref{3.1-09}, and setting the center of mass momentum to zero, $H \lvert_{\mathbf{p}_{CM}=0} \equiv H_r$, this invariance is naturally preserved, but the relative squared angular momentum operator $L^2_{iM} \equiv (L_i^{'})^2$ of each of the particles no longer commutes with the relative Hamiltonian, due to the non-orthogonality of the linear transformation, $\mathbb{J}$, to the relative reference frame, cf. Eq.~\eqref{3.1-16},
\begin{equation}
 [H_r, (L_i^{'})^2] \neq 0\label{4.0-04} 
\end{equation}
where $i=1,2,...M-1$. Therefore, continuum eigenstates of $H_r$ can be labeled with the eigenvalues of the (squared) collective angular momentum, quadratic Casimir operator of SO(3), and by the ones of its third component, $L_{\rm tot,z}$, Casimir of the group of rotations on the plane, 
\begin{equation}
\begin{array}{ccc}
\mathrm{SO}(3) & \supset & \mathrm{SO}(2)\\
\downarrow & & \downarrow \\
 \ell & & m~, \\
\end{array}\label{4.0-05}
\end{equation}
\textit{i.e.} as basis of the $2\ell+1$ dimensional irreducible representation of SO(3) and eigenstates of rotations about the $z$ axis. However, the discretized Hamiltonian on the cubic lattice does not inherit this symmetry, being left invariant only by a subset of SO(3), forming the cubic group, $\mathcal{O}$, of order 24 and isomorphic to the permutation group of four elements, $\mathcal{S}_{4}$.  Equivalently, the dependence of the collective angular momentum on spatial derivatives and, therefore, the necessisity of resorting to an approximation scheme, prevents its commutation with the lattice Hamiltonian.\\  Nevertheless, like in the previous case, the basis vectors of each irrep of $\mathcal{O}$ can be chosen to be simultaneously diagonal with respect to a subset of its operations. Considering again the z axis, the set generated by a counterclockwise rotation of $\pi/2$, $\mathscr{R}_{z}^{\pi/2}$, forms an abelian group, isomorphic to the cyclic group of order four, $\mathcal{C}_4$ \footnote{Like SO(2) with SO(3), also $\mathcal{C}_4$ is not a normal subgroup of $\mathcal{O}$, as the conjugacy classes $3C_4^2(\pi)$ and $6C_4(\pi/2)$ of the latter are only partially included in the cyclic group.}. Since its complex 1-dimensional inequivalent irreps are four and the distinct eigenvalues of $\mathscr{R}_{z}^{\pi/2}$ are $\pm 1$ and $\pm i$, we can label the irreducible representations of $\mathcal{C}_4$ with the integers $I_z$ ranging from $0$ to three,  
\begin{equation}
R_{z}^{\pi/2} = \exp\left(-i\frac{\pi}{2}I_z\right)~.\label{4.0-06}
\end{equation}
Diagonalizing the lattice Hamiltonian together with $\mathscr{R}_{z}^{\pi/2}$, 
\begin{equation}
(\mathcal{H}+\mathscr{R}_z^{\pi/2})\Psi = (E+R_z^{\pi/2})\Psi~,\label{4.0-07}
\end{equation}
the simultaneous eigenstates $\Psi$ can be denoted, thus, with the irreducible representations of $\mathcal{O}$ and $\mathcal{C}_4$ (\textit{i.e.} \textit{quantum numbers}) 
\begin{equation}
\begin{array}{ccc}
\mathcal{O} & \supset & \mathcal{C}_4\\
\downarrow & & \downarrow \\
 \Gamma & & I_z~, \\
\end{array}\label{4.0-08}
\end{equation}
where $\Gamma \in A_1$, $A_2$, $E$, $T_1$ and $T_2$.
Due to this descent in symmetry, each of the original $2\ell+1$ degenerate eigenstates of $\mathrm{H}$ is split into smaller multiplets, their dimension ranging from one to three (cf. Tab.~\ref{T-4.0-01}).\\
\begin{table}[h!]
\begin{center}
\vspace{-0.2cm}
\begin{small}
\begin{tabular}{c|ccccccccc}
\toprule
$\Gamma$ & $D^0$ & $D^1$ & $D^2$ & $D^3$ & $D^4$ & $D^5$ & $D^6$ & $D^7$ & $D^8$\\
\midrule
$A_1$ & 1 & 0 & 0 & 0 & 1 & 0 & 1 & 0 & 1\\
$A_2$ & 0 & 0 & 0 & 1 & 0 & 0 & 1 & 1 & 0\\
$E$ & 0 & 0 & 1 & 0 & 1 & 1 & 1 & 1 & 2\\
$T_1$ & 0 & 1 & 0 & 1 & 1 & 2 & 1 & 2 & 2\\
$T_2$ & 0 & 0 & 1 & 1 & 1 & 1 & 2 & 2 & 2\\
\bottomrule
\end{tabular}
\end{small}
\caption{Coefficients of the decomposition of the representations of the spherical tensors of rank $2\ell+1$, $D^{\ell}$ into irreps of the cubic group. These can be obtained by repeated application of the Great Orthogonality Theorem for characters to the $2\ell+1$-dimensional representations of SO(3) and the irreps of $\mathcal{O}$.} \label{T-4.0-01}
\vspace{-0.6cm}
\end{center}
\end{table}

As in the case of parity, by expressing the cubic group elements $g$ as terns of Euler angles, $(\alpha,\beta,\gamma)$, it is possible to construct projectors on the irreps of $\mathcal{O}$ for spherical tensors of rank $2\ell+1$ \cite{Alt57},
\begin{equation}
P_{\Gamma}^{2\ell+1} = \sum_{g \in \mathcal{O}}\chi_{\Gamma}(g)D^{\ell}(g)~,\label{4.0-10}
\end{equation}
where the $D^{\ell}(g)$ are Wigner D-matrices, $D_{mk}^{\ell}(\alpha,\beta,\gamma)$, and $\chi_{\Gamma}(g)$  are characters of the irrep $\Gamma$ of the cubic group. It is exactly from the columns (resp. rows) of the projector matrix that cubic basis vectors (resp. tensors) from spherical basis vectors (resp. tensors) can be constructed \cite{Wig31}. Nevertheless, when the same irrep of $\mathcal{O}$ appears more then once in the decomposition of $D^{\ell}$ (cf. Tab.~\ref{T-4.0-01}) further rearrangement on the outcoming linear combinations is needed (cf. \ref{S-10.2}). Moreover, only tensors or basis vectors having the same projection of the angular momentum along the z axis, $m$, \textit{modulo} $4$ mix among themselves when projected to any cubic group irrep.\\
Eventually, we conclude the paragraph with particle space symmetries. Since both the relative and the full Hamiltonian commute with the permutation operators of $M$ parti\-cles,
\begin{equation}
[H,\mathscr{S}_g] = [H_r,\mathscr{S}_g] = 0~, \label{4.0-11}
\end{equation} 
where $g\in \mathcal{S}_M$, the permutation group of $M$ elements represents a symmetry for the system. Since the representatives of the sequences of transpositions, $\mathscr{S}_g$ does not affect the configuration space on which $\mathcal{O}$ and $\mathscr{P}$ act, they naturally commute with the elements of the space-time symmetry groups. In the \ce{^8Be} case, where two particle transposition $(12)$ coincides with parity, the latter assertion is ensured by means of commutation between rotations and space inversion.
As a consequence, whenever the states does not transform according to the bosonic representations, 
\begin{equation}
{\tiny\yng(2)...\tiny\yng(1)} \sim [\mathrm{M}]~,\label{4.0-12}
\end{equation}
 or the fermionic ones, 
\begin{equation}
\begin{array}{c}
\tiny\yng(1,1)\\
\vdots\\ 
\tiny\yng(1) \\ 
\end{array}
\sim [1^{\mathrm{M}}]~,\label{4.0-13}
\end{equation}
 they appear in the energy spectrum as repeated degenerate cubic group multiplets, their multiplicity being equal to the dimension of the irrep of $S_{M}$ to which they belong. It follows that Young diagrams or partitions can be included among the labels of the simultaneous eigenstates $\Psi$ (cf. Eq.~\eqref{4.0-07}).  Due to the bosonic nature of the $\alpha$-particles, the construction of the projector on the completely symmetric irrep of the permutation group,
\begin{equation}
P_{_{\tiny\yng(2)\dots\tiny\yng(1)}} = \sum_{g \in \mathscr{S}_M}\chi_{_{\tiny\yng(2)\dots\tiny\yng(1)}}(g)\mathscr{S}_{g} = \sum_{g \in \mathscr{S}_M}\mathscr{S}_{g}~,\label{4.0-14}
\end{equation}
turns out to be useful in the computation of the numerical eigenstates of the lattice Hamiltonian $\mathcal{H}_r$, see Sec.~\ref{S-5.1}, since unphysical eigenstates of parastatistic or fermionic nature are filtered out. In analogous way the projectors to all the other irreducible representations of $\mathcal{S}_M$ can be constructed.

\section{\textsf{Physical Observables}}\label{S-5.0}

\subsection{\textsf{Space coordinates}}\label{S-5.1}

The computation of matrix elements of lattice operators in the configuration-space representation requires the replacement of the lattice coordinates $n_{nM, \alpha}$  introduced in Sec.~\ref{S-3.1} by their physical counterpart $(r_{nM, \alpha})_{\mathrm{phys}}$. This is the case of the collective squared angular momentum operator (cf. Eqs.~\eqref{3.3-06}-\eqref{3.3-08}) and $V^{\rm II}$ and $V^{\rm III}$ terms of the Hamiltonian which are diagonal in the $3M-3$ dimensional configuration space, due to the absence of velocity-dependent potentials. \\
 Therefore, it is necessary to define a map between lattice points and the physical coordinates. If we encode the former by an unique positive integer index $r$, ranging from $0$ to $N^{3M-3}-1$, the lattice coordinates $n_{nM, \alpha}$ are can be extracted from $r$ via the \textit{modulo} function,
{\renewcommand\arraystretch{1.3}
\begin{equation}
\begin{gathered}
n_{nM, x} = \mathrm{mod}\left(\floor*{\frac{r}{N^{n}}}, N\right)\\
n_{nM,y} = \mathrm{mod}\left(\floor*{\frac{r}{N^{n+1}}},N\right)\\
n_{nM,z} = \mathrm{mod}\left(\floor*{\frac{r}{N^{n+2}}},N\right)\\
\end{gathered}
\label{5.1-01}
\end{equation}}
with $n \in 1,2,\dots M-1$. An invertible map from the latter to physical coordinates is provided by
{\renewcommand\arraystretch{1.0}
\begin{equation}
(r_{nM, \alpha})_{\mathrm{phys}} = 
\begin{dcases}
a\hspace{0.5mm}n_{nM, \alpha} \hspace{6.5mm}\mathrm{if}\hspace{1mm} n_{nM, \alpha} < N/2\\
a\hspace{0.5mm}(n_{nM, \alpha} - N) \hspace{3mm}\mathrm{if}\hspace{1mm} n_{nM, \alpha} \geq N/2\\
\end{dcases}\label{5.1-02}
\end{equation}}
where the lattice spacing $a$ is treated here as a dimensional parameter, expressed in femtometres. The three-dimensional configuration space is, thus, reduced to a cubic finite set of points encompassing the origin, which is centered on the latter only when the number of points per dimension $N$ is odd. However, the cubic region can be centered in the origin of the axes by considering the following definition of the physical coordinates \cite{BaH84}
{\renewcommand\arraystretch{1.0}
\begin{equation}
(r_{nM, \alpha})_{\mathrm{phys}} = a\left(\hspace{0.5mm}n_{nM, \alpha} - \frac{N-1}{2}\right)~.
\label{5.1-03}
\end{equation}}
As a consequence, when $N$ is even the physical points $(r_{nM, \alpha})_{\mathrm{phys}}$ do not include the origin any more and assume only half-integer values. This second map between lattice and physical coordinates, that had been already adopted in a study on rotational invariance restoration of lattice eigenfunctions in ref.~\cite{BaH84}, is preferable for plotting the discretized wavefunctions. \\
Finally, it is worth remarking that, if  the lattice configuration space is restricted to the first octant of the three-dimensional space (\textit{e.g.} Eq.~\eqref{5.1-02} with a sign reversal in the argument of the second row) the average values of $\mathcal{L}^2$ on states with good angular momentum converge to incorrect values in the continuum and infinite volume limit, due to the exclusion of physical points bearing negative entries.\\

\subsection{\textsf{Binding energy}}\label{S-5.2}

Another physical quantity of interest for our analysis is the binding energy $BE(Z,N)$ that can be obtained from the energy of the lattice Hamiltonian $\mathcal{H}$ ground state, $E_{0^+}$, via the relation
\begin{equation}
BE(2M,2M) =  2M m_{_{^1\mathrm{H}}} c^2 + 2M m_nc^2 - M m_{_{^4\mathrm{He}}}c^2 - E_{0^+}~.\label{5.2-01}
\end{equation}
Since the parameters of the Ali-Bodmer potential are fitted to the $\alpha-\alpha$ scattering lengths, the experimental value of the binding energy of \ce{^8Be} from Eq.~\eqref{5.2-01} differs from the observational one, even in the large boxes limit. On the other hand, for \ce{^{12}C} the addition of a 3-body potential permitted to fix the ground state energy to the $3\alpha$ decay threshold, thus yielding binding energies consistent with their experimental counterparts, provided the experimental energy gap between the Hoyle state and the former breakup threshold is added to $E_{0^+}$. \\

\subsection{\textsf{Multiplet averaging}}\label{S-5.3}

The multiplet averaged value of energy the is defined as 
\begin{equation}
E(\ell_A^P) = \sum_{\Gamma \in \mathcal{O}}\frac{\chi^{\Gamma}(E)}{2\ell+1}E(\ell_{\Gamma}^P)~,\label{5.3-01}
\end{equation}
where $\Gamma$ is an irreducible representation of the cubic group (cf. Tab.~\ref{T-10.2-02}), $\chi^{\Gamma}(E)$ is its character with respect to the conjugacy class of the identity and $P$ is the eigenvalue of the inversion operator, $\mathscr{P}$. The same operation can be performed for average values of operators representing physical observables $\mathcal{Q}$ on lattice eigenstates, 
\begin{equation}
\langle \mathcal{Q} \rangle (\ell_A^P) = \sum_{\Gamma \in \mathcal{O}}\frac{\chi^{\Gamma}(E)}{2\ell+1}\langle \mathcal{Q} \rangle(\ell_{\Gamma}^P)~.\label{5.3-02}
\end{equation}
In particular, the latter formula that has been extensively applied for the squared angular momentum operator, $\mathcal{L}^2$, in the analysis of finite-volume and discretization effects. 

\section{\textsf{Implementation of the method}}\label{S-6.0}

As it can be inferred from Sec.~\ref{S-5.1}, the extent of configuration space of \ce{^{12}C} on the cubic lattice would require the storage of vectors and matrices with a huge amount of entries. For instance, any eigenvector of the lattice Hamiltonian with $N = 31$ for the latter nucleus implies the storage of almost nine hundred millions of entries, a number that rises to circa 32 $10^9$ double precision items if all the meaningful operators involved in the diagonalization and eigenspace analysis stored as sparse matrices are considered. Although in the previous literature on the subject (cf. Refs.~\cite{BNL14} and \cite{BNL15}) pre-built numerical diagonalization functions for the Hamiltonian matrix were considered, the increased dimension of the lattice operators acting on the eigenvectors led us to the choice of the memory-saving \textit{Lanczos algorithm} (cf. Sec.~\ref{S-6.1}), an iterative method reducing the overall storage cost to the one of subset of eigenvectors of interest and making extensive use of indexing.

\subsection{\textsf{The Lanczos algorithm}}\label{S-6.1}

The algorithm chosen for the simultaneous diagonalization of $\mathcal{H}_r$ and $\mathscr{R}_z^{\pi/2}$ is an implementation of the \textit{Lanczos algorithm} and is based on the repeated multiplication of the matrix of interest on a vector followed by its subsequent normalization, like the \textit{power} or \textit{Von Mises} \textit{iteration}. Once a suitable initial state, $\Psi_0$, is constructed, our method produces a c-number and a vector, that reproduce the lowest signed eigenvalue of the matrix and the relevant eigenvector respectively with increasing precision after an increasing number of iterations.  \\
Before the beginning of the iteration loop, the trial eigenvector, $\Psi_0$, is defined. Although also random states could be used for attaining the task, the construction of trial states that reflect the symmetries of the Hamiltonian often reduces the number of necessary iterations. Besides, an initial value for the eigenenergy, $E_0$, is entered together with $\Psi_0$ and the \textit{pivot energy}, $E_p$, a c-number that ensures the convergence of the desired eigenvector to the one corresponding to the lowest signed eigenvalue. Once $\Psi_0$ is passed into the loop, the updated vector in the beginning of the $k+1$-th iteration, $\Psi_{k+1}^{new}$, is related to the resulting state from the previous iteration,$\Psi_{k}$, via the following realtion 
\begin{equation}
\Psi_{k+1}^{new} = (\mathcal{H}_r+\mathscr{R}_z^{\pi/2}-E_p)\Psi_{k}~,\label{6.1-01}
\end{equation}
\textit{i.e.} a multiplication of $\Psi_{k}$ by the matrices to be simultaneously diagonalized followed by the subtraction of the same vector multiplied by $E_p$. Then, the updated value of the energy eigenvalue is drawn from the updated state by taking the scalar product of $\Psi_{k+1}^{new}$ with $\Psi_{k}$, 
\begin{equation}
E_{k+1} = (\Psi_{k},\Psi_{k+1}^{new}) + E_p~.\label{6.1-02} 
\end{equation}
Immediately after, also the pivot energy undergoes an update. If $E_{k+1}-E_{k}$ turns out to be positive (resp. negative), in fact, $E_p$ is incremented (resp. decremented) by a positive integer, whose magnitude is usually different in the two cases,   
\begin{equation}
E_p^{new} = E_p + \Delta [\sign(E_{k+1}-E_k)]\label{6.1-03}
\end{equation}
where $\Delta [+1] > \Delta [-1]$, in order to make the series $\{E_k\}$ converge to $E_r$. More precisely, in all the computations that follow, $\Delta [+1]$ is tuned to be approximately ten times larger than $\Delta [-1]$, even if further adjustment of these two parameters depending on the $\mathcal{O}$ irreps of the eigenstates of interest leads to faster convergence. At this point, it is worth observing that, if the pivot energy is set equal to zero and its update loop, cf. Eq.~\eqref{6.1-03}, is suppressed, the body of this version of the Lanczos algorithm would exactly coincide with the one of the \textit{power iteration}. Finally, as in the \textit{Von Mises iteration}, the normalization of the $k+1$-times improved eigenfunction,
\begin{equation}
\Psi_{k+1} = \frac{\Psi_{k+1}^{new}}{\norm{\Psi_{k+1}^{new}}}~,\label{6.1-04}
\end{equation}
ends the body of the iteration loop, that runs until the absolute value of the difference between the updated energy eigenvalue and $E_k$ falls below a given value of precision, $\delta_C$, customarily set equal to $10^{-9}$ or $10^{-10}~\mathrm{MeV}$.
The convergence of the outcoming state vector to the actual eigenfunction of $\mathcal{H}_r$ and $\mathcal{R}_z^{\pi/2}$ is ensured by both the non-degeneracy of the common eigenvalues of the two matrices and by the construction of a trial state with a nonzero component in the direction of the eigenvector associated to the ground state: in case one of these two conditions is not satisfied, convergence of the $\{\Psi_k\}$ series is no longer guaranteed. \\
Moreover, the number of iterations required to attain the given precision, $\delta_C$, in the extraction of the eigenvalues grows not only with the box size, $N$, (\textit{i.e.} with the dimension of the Hamiltonian matrix), but also with the inverse of lattice spacing. This is due to the fact that eigenenergies get closer in magnitude for small values of $a$ and the eigenvector under processing, $\Psi_k$, may \textit{oscillate} many times about the neighbouring eigenstates during the iterations before converging. Besides, a wise choice of the trial wavefunction turns out to reduce significantly the number of required iterations and can stabilize the process.\\
The bare Lanczos iteration just described, however, does not allow for the extraction of any other eigenvector than the ground state unless an orthogonalization scheme involving the already extracted states is introduced. In order to access a wider region of the spectrum (\textit{e.g.} $n+1$ eigenstates), \textit{Gram-Schmidt} orthogonalization has been introduced into the body of the iteration loop: if $\Psi^{(0)}$, $\Psi^{(1)}$, ... $\Psi^{(n-1)}$ is a set of $n$ converged states, the remaining eigenstate, $\Psi_{k+1}^{(n)}$, is finally orthogonalized in the end of each iteration with respect to the former eigensubspace. Is exactly this piece of the puzzle that prevents $\Psi_{k+1}^{(n)}$ to collapse into the ground state of the system, even when the initial trial function maximizes the overlap with the target eigenstate.\\
Furthermore, projectors upon cubic \footnote{For example, Eq.~\eqref{4.0-10} with the Wigner D matrix, $D^{J}(\alpha,\beta,\gamma)$, replaced by a representative of the element $(\alpha,\beta,\gamma)$ in the reducible $N^{3M-3}$-dimensional representation of the eigenstates of $\mathcal{H}_r$.} and permutation group irreps (cf. Eq.~\eqref{4.0-14}) have been applied to the $\Psi_{k+1}^{(n)}$ state just before orthonormalization, thus allowing for the investigation of specific regions of the spectrum of the two compatible operators.\\ 
Before concluding the paragraph, special attention has to be devoted to the $T_1$ and $T_2$ eigenstates of $\mathcal{H}_r+\mathscr{R}_z^{\pi/2}$. Even if the spectrum of the matrix is complex, the power method implemented in the space of real vectors of dimension $N^{3M-3}$, does not allow for the extraction of complex eigenvalues with nonzero imaginary part and the relevant eigenvectors, transforming as the $1$ and $3$ irreps of $C_4$. The outcoming vectors are real and orthogonal among themselves and remain associated to (almost) degenerate real energy eigenvalues. Since the remaing partner of the $T_1$ (resp. $T_2$) multiplet, with $I_z = 0$  (resp. $2$), transform in a separate block under the operations of $C_4$ and the exact eigenvectors are related by complex conjugation,
\begin{equation}
\Psi_{\Gamma,I_z = 3} = \left[\Psi_{\Gamma,I_z = 1}\right]^*\label{6.1-05}
\end{equation}
the true common eigenvectors of $\mathcal{H}_r$ and $\mathscr{R}_z^{\pi/2}$ can be drawn from the real degenerate ones, $\Psi_{\Gamma}^{(p)}$ and $\Psi_{\Gamma}^{(q)}$, by means of a $SU(2)$ transformation on the corresponding eigensubspace,
\begin{equation}
\left(
\begin{array}{c}
\Psi_{\Gamma,I_z = 1}\\
\Psi_{\Gamma,I_z = 3}\\
\end{array}
\right)
= \frac{1}{\sqrt{2}}
\left(
\begin{array}{c}
\Psi_{\Gamma}^{(p)} + i  \Psi_{\Gamma}^{(q)}\\
\Psi_{\Gamma}^{(p)} - i   \Psi_{\Gamma}^{(q)}
\end{array}
\right).\label{6.1-06}
\end{equation}
Since $\mathcal{C}_4$ is Abelian, made of four rotations about the same axis, any 2-dimensional representation of it can be reduced to a direct sum of 1-dimensional irreps, provided the similarity matrix is allowed to be complex.\\ As done with the cubic and the permutation group, projectors on the real ($I_z = 0,2$) irreducibles representation of  $\mathcal{C}_4$ can be constructed and introduced in the iteration loop, thus halving (resp. reducing to one third) the memory consumption for the storage of $E$ (resp. $T_1$ and $T_2$) states and extending the accessible region of the low-energy spectrum of the two nuclei considered here. \\

\subsection{\textsf{Parallel implementation}}\label{S-6.2}

The iteration code pointed out in the previous section has been written first in MATLAB and in Fortran 90 and, finally, in CUDA C++. Although devoid of the vector indexing conventions of MATLAB, Fortran 90 permitted us to perform parallel computations on the available clusters of CPU processors (cf. \textit{Acknowledgements}). The original MATLAB codes drafted for the first tests, in fact, have been rewritten in the latter language using the pre-built Message Passing Interface (MPI) routines. In particular, each of the converged eigenvectors has been assigned to a different processor (referred also as \textit{rank}) on the same node whereas, in the succeeding versions of the MPI codes, the eigenvectors themseleves have been split into different ranks, in order to achieve further speedup. Nevertheless, for the large-lattice ($ 25 \leq N \leq 31$) diagonalizations concerning \ce{^{12}C}, the exploitation of the graphic cards (GPUs) of the same cluster has been considered, thus leading to a significant reduction in the computational times (up to a factor of $5\cdot10^{-2}$) for the given box size interval. Accordingly, the Fortran MPI code has been rewritten in CUDA C++ in such a way that each of the vectors, assigned to a single CPU (\textit{host}), is copied, processed and analyzed entirely on a single GPU core (\textit{device}) and only finally copied back to the host, for the backup of the vector in the hard disk memory. This final rewriting of the codes for the diagonalization and the analysis of the state vectors allowed us to process vectors with $N=31$ of \ce{^{12}C} and a precision $\delta_C = 10^{-9}$ (cf. Sec.~\ref{S-6.1}) within six hours of running time. Finally, the use of more than one GPU node for the storage of each state vector is likely to extend the \ce{^{12}C} diagonalizations to $N \geq 32$ and to allow for the analysis of eigenvectors of mid-sized lattices ($10 \leq N \leq 12$) for the \ce{^{16}O} in the near future.

\subsection{\textsf{Boundary conditions}}\label{S-6.3}

So far, no reference to the way in which the Cauchy problem associated to the relative Hamiltonian $H_r$ (plus the cubic group operation) has been made. A customary choice in lattice realizations of Schr\"odinger equation is the imposition of periodic boundary conditions (PBC) on the eigenfunctions, 
\begin{equation}
\Psi^{(q)}(\mathbf{n}+\mathbf{m}N)=\Psi^{(q)}(\mathbf{n}), \label{6.3-01}
\end{equation}
where $\mathbf{m}$ and $\mathbf{m}$ are two vectors of integers. A practical realization of this constraint is provided by the application of the \textit{modulo} $N$ functions on the array indices corresponding to hopping terms of the lattice operators involved. This results the appearance of more entries in the matrix realizations of quantum mechanical operators, whose explicit storage has been wisely avoided.\\
Another choice of boundary conditions, subject of a recent investigation on three-body systems \cite{KoL16}, is given by the twisted boundary conditions (TBC),
\begin{equation}
\Psi^{(q)}(\mathbf{n}+\mathbf{m}N)=e^{i\mathbf{\theta}\cdot\mathbf{m}}\Psi^{(q)}(\mathbf{n})\label{6.3-02}~.
\end{equation}
Since for twisting angles equal to zero, $\theta_{\alpha} = 0$, the two constraints coincide, Eq.~\eqref{6.3-02} can be considered as a generalization to complex phases of the usual PBC. In particular, it has been proven that in two-body systems \textit{i-periodic} boundaries, \textit{i.e.} with $\theta_{\alpha} = \pi/2$, reduce significantly the leading order exponential dependence of the finite-volume energy corrections and that analogous suppressions of finite-volume effects for three-body systems can be achieved \cite{KoL16}. \\
Nevertheless, since our aim is the analysis of the breaking of rotational invariance in four $\alpha$ particle systems, we chose the computationally cheaper PBC.\\

\section{\textsf{The \ce{^{8}Be} nucleus}}\label{S-7.0}

It is firmly enstablished that the actual ground state of this nucleus lies 91.84~keV above the $\alpha-\alpha$ decay threshold, thus making it the only unbound $\alpha$-conjugate nucleus with $M \leq 10$. However, it remains of interest to dwell shortly on the behaviour of the binding energy (cf. Eq.~\eqref{5.2-01}) of this nucleus for different values of $N$ and lattice spacing kept fixed to $0.75~\mathrm{fm}$. As it can be inferred from Fig.~\ref{F-7-01}, the infinite volume value ($L \equiv Na = 40 $~fm) of the binding energy ($\approx$ 57.67 MeV) is inconsistent of about 1.2 MeV with the observational value ($\approx$ 56.50 MeV \cite{ABB97}), due to the choice of tuning the parameters of the Ali-Bodmer potential on the $0_1^+$ - $0_2^+$ gap of \ce{^{12}C}. \\ 
\begin{figure}[ht!]
\begin{tikzpicture}
\begin{axis}[
	width = 15.3cm,
          height = 6.9cm,
	xmin = 0,
	xmax = 45,
	xlabel={\small $N$},
	ylabel={\small $BE(4,4)$ $[\mathrm{MeV}]$},
	xtick={0,3,6,9,12,15,18,21,24,27,30,33,36,39,42,45},
	ytick={0,10,20,30,40,50,60},
	legend style ={at={(0.68,0.38)}, anchor=north west, draw=black,fill=white,align=left},
	legend entries ={\textcolor{Black}{\hspace{-0.2cm}$0_{A_1}^{+}$ - a=0.75 fm},\textcolor{OliveGreen}{Experimental},\textcolor{Violet}{Lattice}}
]
\addlegendimage{empty legend};
\addlegendimage{OliveGreen,dashed};
\addlegendimage{Violet, mark=square};
\addplot[dashed, color=OliveGreen, domain=0:45, samples=100] {56.4988217554};
\addplot[mark=square*, color=Violet,  mark size=1, mark options={fill=Violet}] table [y=BEexp, x=L] {8Be_A1_BE_075.txt};
\end{axis}
\end{tikzpicture}
\caption{Binding energy of the \ce{^8Be} as a function of  N, for a lattice with spacing $a =1.0$~fm.}\label{F-7-01}
\end{figure}

Nevertheless, the binding energy grows with the volume of the lattice, in accordance with the sign of the leading order finite volume correction for a $0^+$ $A_1$ state \cite{KLH12}. Besides, due to the choice of the $\mathcal{O}(a^8)$ approximation for the dispersion term, the smallest lattice of interest is the one with $N=K=4$, in which the binding energy turns out to be largely underestimated ($\approx 12$~MeV).\\ 
As discussed in Sec.~\ref{S-4.0}, the spectrum of the \ce{^8Be} Hamiltonian (cf. Eq.~\eqref{2.1-01} with $M=2$) on the lattice is made of simultaneous eigenstates of the cubic group, the cyclic group of order four generated by $\mathscr{R}_z^{\pi/2}$, spatial (and time) inversion and $\mathcal{S}_2$, the permutation group of two elements. In particular, being particle exchange equivalent to the reversal of the sign of the relative coordinate $\mathbf{r}_{12}$, bosonic (resp. fermionic) eigenstates possess even (resp. odd) parity. \\
In order to assess the capability of the model of describing the observed $\alpha$-cluster lines of this nucleus and receive some guidance for the subsequent choice of the multiplets of interest, we present a short excerpt of the low-energy spectrum of $\mathcal{H}_r$ for a box with $a=0.5$ fm and $N=36$ in the Tab.~\ref{T-7.0-01}. \\
\begin{table}[h!]
\begin{minipage}[c]{0.48\columnwidth}
\begin{footnotesize}
\begin{tabular}{c|cc|c|c|c}
\toprule
$E$ [MeV] & $\Gamma$ & $I_z$ & $\mathscr{P}$ & $\mathcal{S}_2$ & $\langle \mathcal{L}_{\rm tot}^2\rangle$ $[\hbar^2]$\\
\midrule
$-1.106778$ & $A_1$ & 0 & + & ${\tiny\yng(2)}$ & $-0.056$\\
\multirow{3}{1.25cm}{\centering{$0.353021$}} & \multirow{3}{0.75cm}{\centering{$T_1$}} & 0 & \multirow{3}{0.25cm}{\centering{-}} &  \multirow{3}{0.5cm}{\centering{${\tiny\yng(1,1)}$}} & \multirow{3}{1.25cm}{\centering{$2.086$}}\\
&  & 1 & & &\\
&  & 3 & & &\\
$0.948046$ & $A_1$ & 0 & + & ${\tiny\yng(2)}$ & $2.507$\\
\multirow{2}{1.25cm}{\centering{$1.721746$}} & \multirow{2}{0.75cm}{\centering{$E$}} & 0 & \multirow{2}{0.25cm}{\centering{+}} &  \multirow{2}{0.5cm}{\centering{${\tiny\yng(2)}$}} & \multirow{2}{1.25cm}{\centering{$6.899$}}\\
 &  & 2 & & &\\
\multirow{3}{1.25cm}{\centering{$2.261133$}} & \multirow{3}{0.75cm}{\centering{$T_1$}} & 0 & \multirow{3}{0.25cm}{\centering{-}} &  \multirow{3}{0.5cm}{\centering{${\tiny\yng(1,1)}$}} & \multirow{3}{0.75cm}{\centering{$10.029$}}\\
&  & 1 & & & \\
&  & 3 & & & \\
\multirow{3}{1.25cm}{\centering{$2.532701$}} & \multirow{3}{0.75cm}{\centering{$T_2$}} & 1 & \multirow{3}{0.25cm}{\centering{+}} &  \multirow{3}{0.5cm}{\centering{${\tiny\yng(2)}$}} & \multirow{3}{0.75cm}{\centering{$7.090$}}\\
&  & 2 & & &\\
&  & 3 & & &\\
$2.651441$ & $A_1$ & 0 & + & ${\tiny\yng(2)}$ & $18.908$\\
\bottomrule
\end{tabular}
\end{footnotesize}
\end{minipage}\hfill
\begin{minipage}[c]{0.48\columnwidth}
\begin{footnotesize}
\begin{tabular}{c|cc|c|c|c}
\toprule
$E$ [MeV] & $\Gamma$ & $I_z$ & $\mathscr{P}$ & $\mathcal{S}_2$ & $\langle \mathcal{L}_{\rm tot}^2\rangle$ $[\hbar^2]$\\
\midrule
\multirow{2}{1.25cm}{\centering{$2.834477$}} & \multirow{2}{0.75cm}{\centering{$E$}} & 0 & \multirow{2}{0.25cm}{\centering{+}} &  \multirow{2}{0.5cm}{\centering{${\tiny\yng(2)}$}} & \multirow{2}{0.75cm}{\centering{$15.332$}}\\
 &  & 2 & & &\\
\multirow{3}{1.25cm}{\centering{$3.133750$}} & \multirow{3}{0.75cm}{\centering{$T_2$}} & 1 & \multirow{3}{0.25cm}{\centering{-}} & \multirow{3}{0.5cm}{\centering{${\tiny\yng(1,1)}$}} & \multirow{3}{0.75cm}{\centering{$12.676$}}\\
&  & 2 & & &\\
&  & 3 & & &\\
\multirow{3}{1.25cm}{\centering{$3.868673$}} & \multirow{3}{0.75cm}{\centering{$T_2$}} & 1 & \multirow{3}{0.25cm}{\centering{+}} & \multirow{3}{0.5cm}{\centering{${\tiny\yng(2)}$}} & \multirow{3}{0.75cm}{\centering{$17.451$}}\\
&  & 2 & & &\\
&  & 3 & & &\\
\multirow{3}{1.25cm}{\centering{$3.960128$}} & \multirow{3}{0.75cm}{\centering{$T_1$}} & 0 & \multirow{3}{0.25cm}{\centering{-}} & \multirow{3}{0.5cm}{\centering{${\tiny\yng(1,1)}$}} & \multirow{3}{0.75cm}{\centering{$23.629$}}\\
&  & 1 & & &\\
&  & 3 & & &\\
$4.289695$ & $A_1$ & 0 & + & ${\tiny\yng(2)}$ & $30.743$\\
$4.302368$ & $A_2$ & 2 & - & ${\tiny\yng(1,1)}$ & $14.698$\\
\multirow{2}{1.25cm}{\centering{$4.308802$}} & \multirow{2}{0.75cm}{\centering{$E$}} & 0 & \multirow{2}{0.25cm}{\centering{+}} & \multirow{2}{0.5cm}{\centering{${\tiny\yng(2)}$}} & \multirow{2}{0.75cm}{\centering{$10.620$}}\\
 &  & 2 & & &\\
\bottomrule
\end{tabular}
\end{footnotesize}
\end{minipage}
\caption{The 14 lowest multiplets of eigenstates of the \ce{^8Be} lattice Hamiltonian with $N = 35$ and $a = 0.5~\mathrm{fm}$.} \label{T-7.0-01}
\end{table}

Noticeable are the discrepancies between the eigenvalues of the squared angular momentum operator and the average values of it reported in the table. Since the volume of the box ($Na = 17.5$ fm) is large enough to reduce finite-volume effects to the third decimal digit of the energy, these disagreements are due to discretization effects, whose magnitude increases with excitation energy and make the reconstruction of the infinite-volume angular momentum multiplet from the $\langle \mathcal{L}_{\rm tot}^2\rangle \equiv \mathcal{L}^2$ hardly reliable: for the first $2^+$ multiplet, consisting of an $E$ plus a $T_2$ state, $\Delta\mathcal{L}^2$ is already $15~\%$ of the expected angular momentum eigenvalue. The behaviour of the squared angular momentum, therefore, suggests that wavefunctions corresponding to states of increasing energy are also incrasingly position-dependent. \\
In addition the presence of an $A_1^+$ state at $0.948~\mathrm{MeV}$, that further diagonalizations of the lattice Hamiltonian indicate as $0^+$, appears to be in contrast with the present observational data, that position the first excited $0^+$ at $27.494~\mathrm{MeV}$ \cite{KPS17}. \\
In order to study a larger number of bound states as well as to test the results reported in Ref.~\cite{BNL14}, the strength parameter of the attactive part of the Ali-Bodmer potential, $V_0$, has been incremented by a $30$~\% with respect to its original value, see the dashed curve in Fig.~\ref{F-2.1-01}. Accordingly, the \textit{artificial} ground state lies approximately $10.70~\mathrm{MeV}$ below its observational counterpart. \\
Besides the fundamental state, the infinite-volume spectrum of the Hamiltonian includes also a $2^+$ multiplet, made of an $E$ and a $T_2$ state and another $0^+$ state, the closest to the $\alpha$-$\alpha$ decay threshold. Since the latter appears only for relatively large volumes ($Na \ge 25$), we focus the attention only on the $2^+$ multiplet, as in Ref.~\cite{BNL14}. Fixing the lattice spacing to $a = 0.25~\mathrm{fm}$ in order to reduce discretization effects and enlarge the samples of data, we investigate the finite-volume effects on the energy and the squared angular momentum of the three multiplets of states. \\

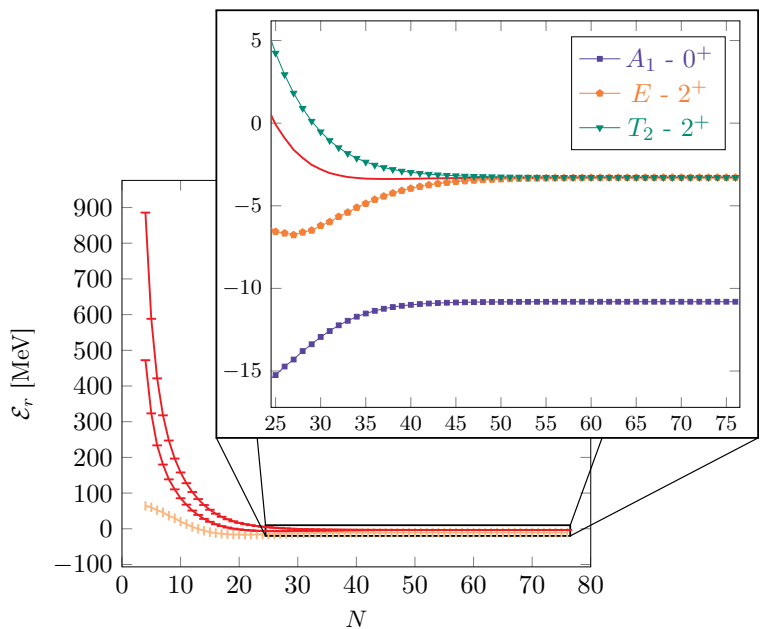
\begin{figure}[h!]
\begin{minipage}[l]{0.66\columnwidth}
\scalebox{0.9}{
\begin{tikzpicture}
\begin{axis}[
	name= bigEn,
	xmin = 0,
	xmax = 80,
	xlabel={\small $N$},
	ylabel={\small $\mathcal{E}_r$ [MeV]},
	xtick={0,10,20,30,40,50,60,70,80},
	ytick={-100,0,100,200,300,400,500,600,700,800,900}
]
\addplot[color=Apricot, mark=|, thick] table [y=E, x=L] {0+_I_A1_030.txt};
\addplot[color=Red, mark =-, thick] table [y=E, x=L] {2+_I_E_030.txt};
\addplot[color=Red, mark = -, thick] table [y=E, x=L] {2+_I_T2_030.txt};
\coordinate (c1) at (axis cs:24.5,-20);
\coordinate (c2) at (axis cs:76.5,10);
\coordinate (c3) at (axis cs:24.5,10);
\coordinate (c4) at (axis cs:76.5,-20);
\draw [line width = 0.75 pt] (c1) rectangle (axis cs:76.5,10);
\end{axis}
\coordinate (d1) at (1.38 cm,1.9 cm);
\coordinate (d2) at (9.3 cm, 8.2 cm);
\coordinate (d3) at (1.38 cm, 8.2 cm);
\coordinate (d4) at (9.3 cm, 1.9cm);
\draw [line width= 0.5 pt] (c1) -- (d1);
\draw [line width= 0.5 pt] (c2) -- (d2);
\draw [line width= 0.5 pt] (c3) -- (d3);
\draw [line width= 0.5 pt] (c4) -- (d4); 
\draw [line width = 0.75 pt, fill= white] (d1) rectangle (d2);
\begin{axis}[
	at = {($(2.18 cm,2.35 cm)$)},
	name= smaEn,
          axis background/.style={fill=white},
	xmin = 24.5,
	xmax = 76.5,
	xtick={25,30,35,40,45,50,55,60,65,70,75},
	xticklabel style = {font=\footnotesize},
	yticklabel style = {font=\footnotesize},
	legend style ={at={(0.64,0.96)}, anchor=north west, draw=black,fill=white,align=left},
	legend entries ={\textcolor{Violet}{$A_1$ - $0^{+}$},\textcolor{Orange}{$E$ - $2^{+}$},\textcolor{PineGreen}{$T_2$ - $2^{+}$}}
]
\addlegendimage{Violet,mark=square*, mark options={fill=Violet}, mark size = 1};
\addlegendimage{Orange, mark=pentagon*, mark options={fill=Orange}, mark size = 1.5};
\addlegendimage{PineGreen, mark=triangle*, mark options={fill=PineGreen,rotate=180}, mark size = 1.5};
\addplot[mark=square*, color=Violet,  mark size=1, mark options={fill=Violet}] table [y=E, x=L]{0+_I_A1_030.txt};
\addplot[mark=pentagon*, color=Orange,  mark size=1.5, mark options={fill=Orange} ] table [y=E, x=L] {2+_I_E_030.txt};
\addplot[mark=triangle*, color=PineGreen,  mark size=1.5, mark options={fill=PineGreen,rotate=180} ] table [y=E, x=L] {2+_I_T2_030.txt};
\addplot[color=Red,  solid, thick] table [y=E, x=L] {2+_I_MAVG_030.txt};
\end{axis}
\end{tikzpicture}}
\end{minipage}
\begin{minipage}[c]{0.32\columnwidth}
\caption{Behaviour of the energies of the lowest $0^+$ (vertical bars) and $2^+$ (horizontal bars) eigenstates as a function of the box size N for $a = 0.25$ fm.  As expected, the eigenenergies associated to states belonging to the same irrep of SO(3) but to different irreps of $\mathcal{O}$ become almost degenerate at the infinite-size limit, the remaining discrepancies owing to space discretization. Multiplet average of the energies between the $E$ and the $T_2$ states in the magnification has been denoted by a solid line.}\label{F-7.0-02}
\end{minipage}
\end{figure}
With this choice of the lattice spacing, the ground state energy reaches its infinite volume value within the third decimal digit for $Na = 13.25$, while the two multiplets, $E$ and $T_2$ become degenerate within the same precision only for $Na = 17$. Nevertheless, convergence for the latter can be boosted by considering the multiplet averaged energy \cite{BNL14}, $E(2^+_A)$, of the five states composing the $2^+$ continuum one, the third-digit accuracy is already achieved by $E(2^+_A)$ at $Na = 14.25$. The theoretical justification underlying this procedure resides in the cancellation of the polynomial dependence on $N$ of the lowest order finite-volume energy correction for the multiplet-averaged state. The main contribution to this energy shift is proportional to $\exp(\kappa N)$, where $\kappa = \sqrt{-2mE}$ is the binding momentum of the state, and turns out to be negative for all the values of $N$ (cf. Eq.~(19) of \cite{BNL14}) and \textit{even} angular momentum. \\
Even though we do not have an analytical formula for the finite-volume corrections to the average values of $\mathcal{L}^2$ at our disposal, we extend the use of the average on the dimensions of cubic group representations to the latter. As for the energies, an overall smoothing effect on the discrepancies between the average values and the eigenvalues of the squared angular momentum can be observed: a two digit accuracy in the estimates of the latter is reached at $N= 37$ by the multiplet-averaged $\mathcal{L}^2$ for the $2^+$ multiplet, see the red dashed line in Fig.~\ref{F-7.0-03}, while the individual members of the multiplet reach the same precision only at $N = 51$. Moreover, in the large volume limit ($N=72$) the $0^+$ state approaches the angular momentum eigenvalue within $2\times 10^{-5} \hbar^{-2}$, whereas for the $E$ and $T_2$ states of the $2^+$ multiplet the accuracy is poorer, \textit{i.e.} $2\times 10^{-3} \hbar^{-2}$ and $8\times 10^{-4} \hbar^{-2}$, in order.\\
\begin{figure}[h!]
\centering
\begin{minipage}[c]{0.49\columnwidth}
\scalebox{0.85}{
\begin{tikzpicture}
\begin{axis}[
	xmin = 0,
	xmax = 80,
	xlabel={\small $N$},
	ylabel={\small $\hbar^{-2}\mathcal{L}^2$},
	xtick={0,10,20,30,40,50,60,70,80},
	ytick={-3,0,3,6,9,12,15},
	legend style ={at={(0.64,0.96)}, anchor=north west, draw=black,fill=white,align=left},
	legend entries ={\textcolor{Violet}{$A_1$ - $0^{+}$},\textcolor{Orange}{$E$ - $2^{+}$},\textcolor{PineGreen}{$T_2$ - $2^{+}$}}
]
\addlegendimage{Violet,mark=square*,mark options={fill=Violet}, mark size = 1};
\addlegendimage{Orange,mark=pentagon*,mark options={fill=Orange}, mark size = 1.3};
\addlegendimage{PineGreen,mark=triangle*,mark options={fill=PineGreen,rotate=180}, mark size = 1.3};
\addplot[mark=square*, color=Violet,  mark size=1, mark options={fill=Violet}] table [y=J, x=L] {0+_I_A1_030.txt};
\addplot[mark=pentagon*, color=Orange,  mark size=1.3, mark options={fill=Orange}] table [y=J, x=L] {2+_I_E_030.txt};
\addplot[mark=triangle*, color=PineGreen,  mark size=1.3, mark options={fill=PineGreen,rotate=180}] table [y=J, x=L] {2+_I_T2_030.txt};
\addplot[color=Red, solid, thick] table [y=J, x=L] {2+_I_MAVG_030.txt};
\end{axis}
\end{tikzpicture}}
\caption{Average value of the squared angular momentum for the three bound state multiplets as a function of the lattice size.  As predicted, the average values of $\mathcal{L}^2$ for the $0_{A_1}^+$, $2_{E}^+$ and $2_{T_2}^+$ states smoothly converge to the eigenvalues equal to 0, 6 and 6 units of $\hbar^2$ respectively of the same operator, despite some oscillatory behaviour.}\label{F-7.0-03}
\end{minipage}\hfill
\begin{minipage}[c]{0.49\columnwidth}
\scalebox{0.85}{
\begin{tikzpicture}
\begin{axis}[
	xmin = 0,
	xmax = 80,
	xlabel={\small $N$},
	ylabel={\small $\hbar^{-2}|\Delta\mathcal{L}^2|$},
	xtick={0,10,20,30,40,50,60,70,80},
	ymode=log,
	log basis y = {2.718281828459},
	yticklabels = {$e^{-13}$,$e^{-9}$,$e^{-5}$,$e^{-1}$,$e^{3}$},
	legend style ={at={(0.09,0.32)}, anchor=north west, draw=black,fill=white,align=left},
	legend entries ={\textcolor{Violet}{$A_1$ - $0^{+}$},\textcolor{Orange}{$E$ - $2^{+}$},\textcolor{PineGreen}{$T_2$ - $2^{+}$}}
]
\addlegendimage{Violet,mark=square*,mark options={fill=Violet}, mark size = 1};
\addlegendimage{Orange,mark=pentagon*,mark options={fill=Orange}, mark size = 1.3};
\addlegendimage{PineGreen,mark=triangle*,mark options={fill=PineGreen,rotate=180}, mark size = 1.3};
\addplot[mark=square*, color=Violet,  mark size=1, mark options={fill=Violet} ] table [y=dJ, x=L] {0+_I_A1_030.txt};
\addplot[mark=pentagon*, color=Orange,  mark size=1.3, mark options={fill=Orange} ] table [y=dJ, x=L] {2+_I_E_030.txt};
\addplot[mark=triangle*, color=PineGreen,  mark size=1.3, mark options={fill=PineGreen,rotate=180}] table [y=dJ, x=L] {2+_I_T2_030.txt};
\addplot[color=Red,  solid, thick] table [y=dJ, x=L] {2+_I_MAVG_030.txt};
\end{axis}
\end{tikzpicture}}
\caption{Difference between the average value and the expected eigenvalue of the squared angular momentum for the three bound state multiplets as a function of the lattice size.  Logscale is set on the $y$ axis, thus unveiling a regular linear behaviour in the finite volume $\mathcal{L}^2$ corrections for boxes large enough, analogous to the well-known one of the energies of bound states \cite{KLH12}. Unlike the latter, the three spikes due to sign reversal of the $\Delta\mathcal{L}^2$ suggest that the finite volume corrections to this observable are not constant in sign.}\label{F-7.0-04}
\end{minipage}
\end{figure}

Plotting finally the discrepancies between the average values and the expected eigenvalues of the squared angular momentum of the three sets of degenerate energy eigenstates as function of the number of box sites per dimension, an exponential behavior of the former, $\Delta\mathcal{L}^2 = A\exp(mN)$ with $A$ and $m$ real parameters, can be recognized, cf. Fig.~\ref{F-7.0-04}. A linear regression with slope $m$ and intercept $\log A$ on the points with $N \gtrsim 35$ can be performed, highlighting a distinct descending behaviour for each of the multiplets: the $\Delta\mathcal{L}^2$ of the $2^+$  states decreases, in fact, with the same angular coefficient within three-digit precision. It follows that the precision with which the squared angular momentum average values agree with their expectation values is an increasing function of the the binding momentum: the more the state is bound, the greater is the reliability of the $\mathcal{L}^2$ estimation. Nevertheless, the derivation of an analytical formula for the finite volume corrections to the eigenvaues of the squared angular momentum operator remains a subject of interest for further publications.\\
Besides, once finite volume effects are reduced to the fourth decimal digit in the energies via the constraint $Na \geq 18$~fm, the effects of discretization for different values of $a$ can be inspected. As observed in~\cite{BNL14}, the energies as function of the lattice spacing display an oscillatory behaviour, whose amplitutes for the $A_1$ state are limited to the first decimal digit for $0.9 \lesssim a \lesssim 1.2$~fm, then second digit precision is achieved for $0.7 \lesssim a \lesssim 0.9$~fm. For the members of the $2^+$ multiplet the fluctuations about the continuum value of the energies become more pronounced, being the achievement of a three digit precision confined to $a \lesssim 0.5$~fm. Since only lattices with odd number of sites per dimension contain the origin of the axes, cf. the definition of the map between lattice sites and physical coordinates in Eq.~\eqref{5.1-03}, that is supposed to give important contribution to the lattice eigenenergies when the wavefunction is concentrated about the former point, only lattices with odd values of $N$ have been considered for the large ($a \gtrsim 1.25$~fm) lattice spacing analysis.\\
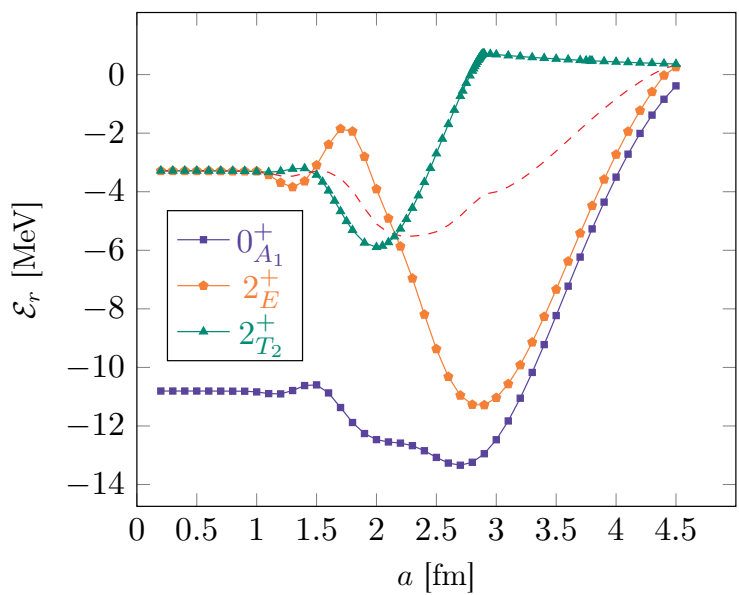
\begin{figure}[h!]
\begin{minipage}[c]{0.64\columnwidth}
\scalebox{1.15}{
\begin{tikzpicture}
\begin{axis}[
	name= bigEn,
	xmin = 0,
	xmax = 5,
	xlabel={\small $a$ [fm]},
	ylabel={\small $\mathcal{E}_r$ [MeV]},
	xtick={0,0.5,1,1.5,2,2.5,3,3.5,4,4.5},
	ytick={0,-2,-4,-6,-8,-10,-12,-14,-16},
           legend style ={at={(0.05,0.60)}, anchor=north west, draw=black,fill=white,align=left},
	legend entries ={\textcolor{Violet}{$0_{A_1}^{+}$},\textcolor{Orange}{$2_E^{+}$},\textcolor{PineGreen}{$2_{T_2}^{+}$}}
]
\addplot[mark=square*, color=Violet,  mark size=1, mark options={fill=Violet}] table [y=E, x=a] {0+_A1_I_18fm.txt};
\addplot[mark=pentagon*, color=Orange,  mark size=1.5, mark options={fill=Orange}] table [y=E, x=a] {2+_E_I_19fm.txt};
\addplot[mark=triangle*, color=PineGreen,  mark size=1.5, mark options={fill=PineGreen}] table [y=E, x=a] {2+_T2_I_19fm.txt};
\addplot[color=Red, dashed] table [y=E, x=a] {2+_I_MAVG_19fm.txt};
\end{axis}
\end{tikzpicture}}
\end{minipage}
\begin{minipage}[c]{0.32\columnwidth}
\caption{Behaviour of the energies of the bound eigenstates as a function of the lattice spacing a for $Na \geq 18$~fm ($\ell = 0$) and  $Na \geq 19$~fm ($\ell = 2$).  As expected, the eigenenergies associated to states belonging to the same irrep of SO(3) but to different irreps of $\mathcal{O}$ become almost degenerate in the zero-spacing limit. In the opposite direction, another level crossing is expected to occur at $a \approx 4.5$~fm. Multiplet-averaged energy of the $2^+$ states has been denoted by a dashed line.}
\end{minipage}
\label{F-7.0-05}
\end{figure}

Although a closed form for the leading order dirscretization corrections to the energy eigenvalues does not exist, it remains possible to associate some extrema of the latter, see Fig.~\ref{F-7.0-05} and Fig.~3 in Ref.~\cite{BNL14}, to the maxima of the squared modulus of the associated eigenstates. This interpretation rests on the assumption that $\mathcal{E}_r(a)$ reaches a local minimum for all the values of the spacing $a$ such that all the maxima of the squared modulus of the corresponding eigenfunction, $|\Psi_r(\mathbf{r})|^2$, are included in the lattice. This condition is satisfied when all the maxima lie along the symmetry axes of the cubic lattice. In case $|\Psi_r(\mathbf{r})|^2$ possesses only primary maxima, \textit{i.e.} points lying at distance $d^*$ from the origin such that the most probable $\alpha$-$\alpha$ separation, $\mathcal{R}^*$, coincides with $d^*$, the description of the behaviour of $\mathcal{E}_r(a)$ in terms of the spatial distribution of the associated wavefunction becomes more predictive. In particular, when all the maxima lie along the lattice axes and the decay of the probability density function (PDF) associated to $\Psi_r(\mathbf{r})$ with radial distance is fast enough, \textit{i.e.} $|\Psi_r(\mathbf{r})|^2_{\mathrm{Max}} \gg |\Psi_r(\mathbf{r})|^2$ for $|\mathbf{r}|=nd^*$ and $n \geq 2$, the average value of the interparticle distance coincides approximately with the most probable $\alpha$-$\alpha$ separation, $\mathcal{R} \approx d^*$, and the average value of the potential, $\mathcal{V}$, is minimized at the same time.\\

\begin{figure}[ht!]
\includegraphics[width=1.00\columnwidth]{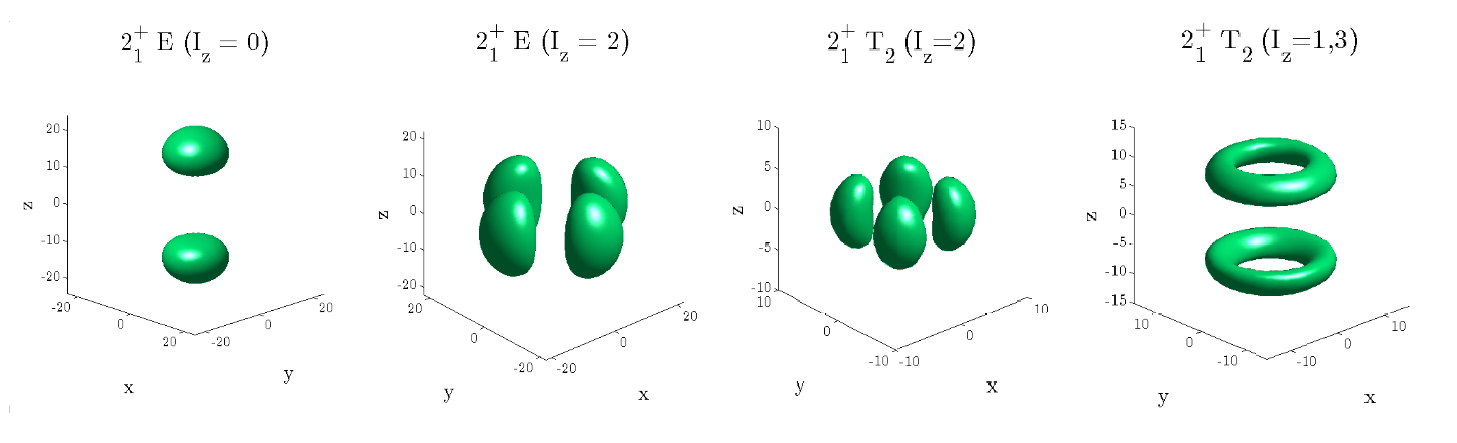}
\caption{The 3-d probability density distributions of the $\alpha$-$\alpha$ separation for the $2_1^+$ states. As in Figs.~\ref{F-7.0-07} and \ref{F-7.0-09}, the distances along the axes are measured in units of lattice spacing ($a = 0.2$ and  $0.5\hspace{1mm}\mathrm{fm}$ for the $E$ and $T_2$ states respectively).  In each subfigure the isohypses with 25\% of the maximal probability density are shown. Due to time-reversal symmetry the PDF corresponding to the $T_2$ $I_z=1$ and $3$ states exactly coincide.}
\label{F-7.0-10}
\end{figure}

Since the maxima of the eigenfunctions of both the $2_1^+$ $E$ states ($I_z =0,2$) lie on the lattice axes at distance $d^* \approx 2.83$~fm and no secondary maximum is found, cf. Fig.~\ref{F-7.0-10}, the energy eigenvalues of the two states are expected to display minima for $a = d^*/n$ with $n \in \mathbb{N}$, \textit{i.e.} for $a \approx 2.83, 1.42, 0.94, \ldots$~fm. Effectively, two energy minima at $a \approx 2.85$ and $1.36$~fm are detected (cf. Fig.~\ref{F-7.0-06}). In addition, for $a \approx d^*$ it is found that $\mathcal{R} \approx 2.88$~fm and $\mathcal{V} \approx -21.21$~MeV, both the values being in appreciable agreement with the minimum values of the two respective quantities, $2.70$~fm and $-21.40$~MeV, see Figs.~\ref{F-7.0-06}-\ref{F-7.0-07}: it follows that also the requirement on the decrease of the PDF with distance is approximately fulfilled.\\  
\begin{figure}[hb!]
\begin{minipage}[c]{0.48\columnwidth}
\includegraphics[width=0.90\columnwidth]{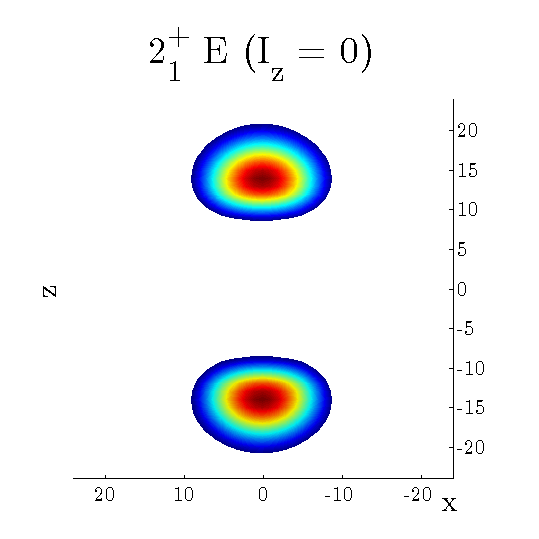}
\end{minipage}
\begin{minipage}[c]{0.48\columnwidth}
\scalebox{0.85}{
\begin{tikzpicture}
\begin{axis}[
	xmin = 0,
	xmax = 5,
           height = 8.2cm,
	width = 7.9cm,
	xlabel={\Large $a$ [fm]},
	ylabel={\Large $\mathcal{R}$ [fm]},
	xtick={0,0.5,1,1.5,2,2.5,3,3.5,4,4.5,5},
	ytick={2.5,2.8,3.1,3.4,3.7,4.0,4.3,4.6,4.9},
           legend style ={at={(0.10,0.75)}, anchor=north west, draw=black,fill=white,align=left},
	legend entries ={\hspace{-0.4cm}$2_1^+$ E $I_z=0$ State, $\mathcal{R}$},
]
\addlegendimage{empty legend};
\addlegendimage{mark=o,color=Orange};
\addplot[mark=o, color=Orange,  mark size=1, mark options={fill=Orange}] table [y=ravg, x=a] {2+_E_I_19fm.txt};
\end{axis}
\end{tikzpicture}}
\end{minipage}
\caption{Cross-sectional plot (xz plane) of the PDF of the $2_1^+$ $E$ $I_z=0$ state (left) and behaviour of the average value of the interparticle distance as a function of the lattice spacing for the same eigenstate (right). In particular, the outer isohypsic surfaces of the former plot correspond to a probability density equal to the 25\% (dark blue) of the maximum value of the PDF (dark red). Distances along the axes are measured in lattice spacing units ($a=0.2$~fm). In the other graph, two minima of $\mathcal{R}$ at $a \approx 1.4$ and $2.5$~fm are visible, implying that the condition on the decay of the wavefunction with increasing $\alpha$-$\alpha$ distance is only approximately fulfilled.}
\label{F-7.0-07}
\end{figure}
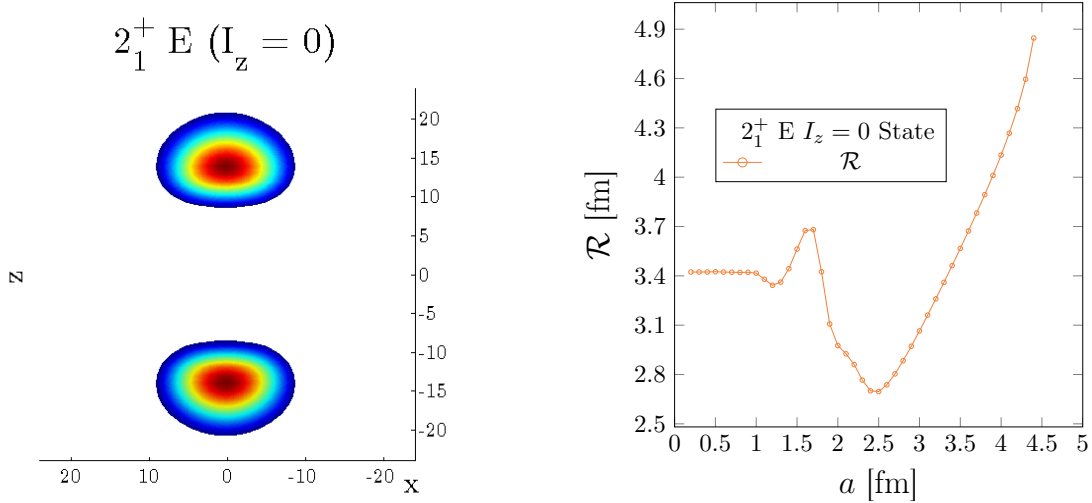

\begin{figure}[ht!]
\begin{center}
\scalebox{1.05}{
\begin{tikzpicture}
\begin{axis}[
	xmin = 0,
	xmax = 6.5,
           height = 6.5cm,
	width = 11.7cm,
	xlabel={\small $a$ [fm]},
	ylabel={\small [MeV]},
	xtick={0,0.5,1,1.5,2,2.5,3,3.5,4,4.5,5,5.5,6,6.5},
	ytick={-25,-20,-15,-10,-5,0,5,10,15,20,25},
          clip mode=individual,
           legend style ={at={(0.71,0.75)}, anchor=north west, draw=black,fill=white,align=left,/tikz/column 1/.style={
                column sep=4pt}},
          legend columns = 1,
	legend entries ={\hspace{-0.3cm}$2_1^+$ E States, $\mathcal{T}$,$\mathcal{V}$, $\mathcal{E}_r$}
]
\addlegendimage{empty legend};
\addlegendimage{color=Orange, dotted};
\addlegendimage{color=Orange, dash dot};
\addlegendimage{color=Orange, mark=o};
\addplot[color=Orange, dotted , thick] table [y=Kin, x=a] {2+_E_I_19fm.txt};
\addplot[color=Orange,  dash dot, thick] table [y=Pot, x=a] {2+_E_I_19fm.txt};
\addplot[mark=o, color=Orange,  mark size=1, mark options={fill=Orange}] table [y=E, x=a] {2+_E_I_19fm.txt};
\end{axis}
\end{tikzpicture}}
\end{center}
\caption{Behaviour of the average values of the kinetic energy, $\mathcal{T}$,  and the potential operator, $\mathcal{V}$, on the $2_1^+$ $E$ eigenstates as a function of the lattice spacing a for $Na \geq 19$~fm. The sum of the two average values produce the already displayed $\mathcal{E}_r$ curve (cf. Fig.~\ref{F-7.0-05}), that almost intercepts the dotted one of $\mathcal{T}$ as soon as the potential energy vanishes ($a\approx 4.5$~fm) and the two states of the multiplet become unbound.}
\label{F-7.0-06}
\end{figure}
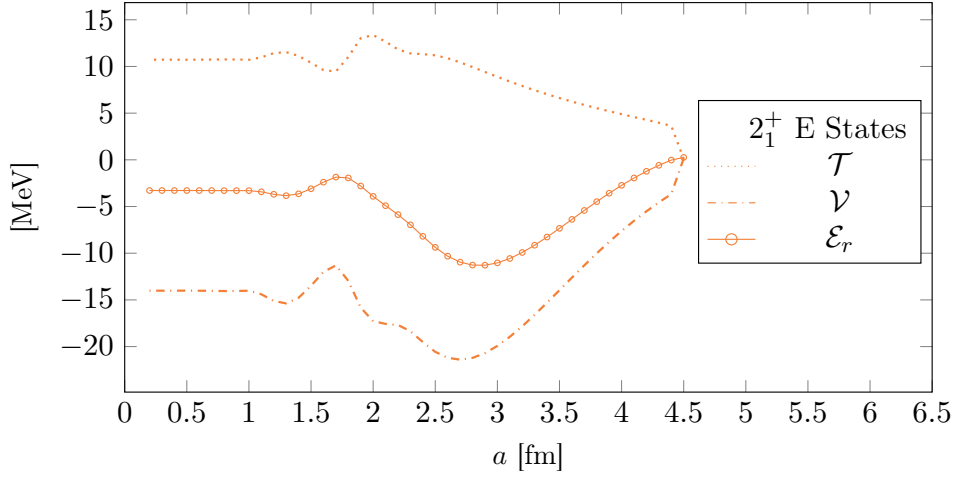

\begin{figure}[hb!]
\begin{minipage}{0.48\columnwidth}
\includegraphics[height=0.92\columnwidth]{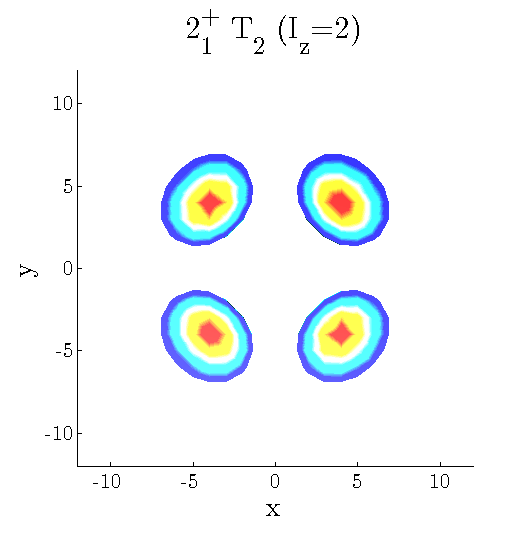}
\end{minipage}
\begin{minipage}{0.48\columnwidth}
\scalebox{0.90}{
\begin{tikzpicture}
\begin{axis}[
	xmin = 0,
	xmax = 3,
           height = 8.2cm,
	width = 7.9cm,
	xlabel={\Large $a$ [fm]},
	ylabel={\Large $\mathcal{R}$ [fm]},
           xtick={0,0.4,0.8,1.2,1.6,2.0,2.4,2.8,3.2},
	ytick={2.0,2.3,2.6,2.9,3.2,3.5,3.8,4.1,4.4,4.7,5},
           legend style ={at={(0.10,0.80)}, anchor=north west, draw=black,fill=white,align=left},
	legend entries ={\hspace{-0.4cm}$2_1^+$ $T_2$ $I_z=0$ State, $\mathcal{R}$},
]
\addlegendimage{empty legend};
\addlegendimage{mark=o,color=PineGreen};
\addplot[mark=o, color=PineGreen,  mark size=1, mark options={fill=PineGreen}, thick] table [y=ravg, x=a] {2+_T2_I_19fm.txt};
\end{axis}
\end{tikzpicture}}
\end{minipage}
\caption{Cross-sectional plot (xy plane) of the PDF of the $2_1^+$ $T_2$ $I_z=2$ state (left) and behaviour of the average value of the interparticle distance as a function of the lattice spacing for the same eigenstate (right). In particular, the outer isohypses of the former plot correspond to a probability density equal to the 25\% (dark blue) of the maximum value of the PDF (dark red). Distances along the axes are measured in lattice spacing units ($a=0.5$~fm). In the other graph, two minima of $\mathcal{R}$ at $a \approx 1.0$ and $1.8$~fm are visible, implying that the condition on the decay of the wavefunction with increasing $\alpha$-$\alpha$ distance is satisfied only to a first approximation.}
\label{F-7.0-09}
\end{figure}
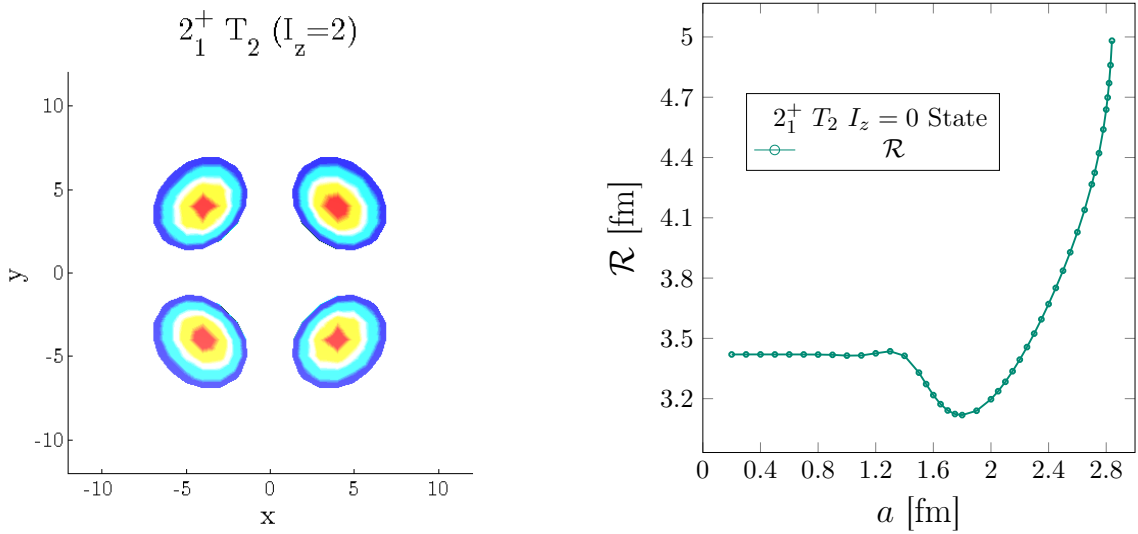

On the other hand, the PDF of the $2_1^+$ $T_2$ $I_z = 2$ state possesses four absolute maxima in the intersections between the xy plane and the $y = \pm x$ planes lying at the same distance $d^* \approx 2.83$~fm from the origin of the axes, whereas for the $I_z = 1,3$ states there are two circles of absolute maxima about the z axis, located at the same distance from the origin, cf. Figs.~\ref{F-7.0-10}-\ref{F-7.0-09}. \\

\begin{figure}[ht!]
\begin{center}
\scalebox{0.90}{
\begin{tikzpicture}
\begin{axis}[
	xmin = 0,
	xmax = 5,
           height = 6.5cm,
	width = 12.7cm,
	xlabel={\small $a$ [fm]},
	ylabel={\small [MeV]},
	xtick={0,0.5,1,1.5,2,2.5,3,3.5,4,4.5,5},
	ytick={-25,-20,-15,-10,-5,0,5,10,15,20,25},
           legend style ={at={(0.68,0.75)}, anchor=north west, draw=black,fill=white,align=left,/tikz/column 1/.style={
                column sep=4pt}},
          legend columns = 1,
	legend entries ={\hspace{-0.3cm}$2_1^+$ $T_2$ States, $\mathcal{T}$,$\mathcal{V}$,$E_B$}
]
\addlegendimage{empty legend};
\addlegendimage{color=PineGreen, dotted};
\addlegendimage{color=PineGreen, dash dot};
\addlegendimage{color=PineGreen, mark=o};
\addplot[color=PineGreen, dotted , ultra thick] table [y=Kin, x=a] {2+_T2_I_19fm.txt};
\addplot[color=PineGreen,  dash dot, ultra thick] table [y=Pot, x=a] {2+_T2_I_19fm.txt};
\addplot[mark=o, color=PineGreen,  mark size=1, mark options={fill=PineGreen}, thick] table [y=E, x=a] {2+_T2_I_19fm.txt};
\end{axis}
\end{tikzpicture}}
\end{center}
\caption{Behaviour of the average values of the kinetic energy, $\mathcal{T}$,  and the potential operator, $\mathcal{V}$, on the $2_1^+$ $T_2$ eigenstates as a function of the lattice spacing a for $Na \geq 19$~fm. The sum of the two average values produce the already displayed $\mathcal{E}_r$ curve (cf. Fig.~\ref{F-7.0-05}), that almost overlaps the dotted one of $\mathcal{T}$ when the potential energy is negligible ($a \gtrsim 2.8$~fm) and the three states of the multiplet are unbound.}
\label{F-7.0-08}
\end{figure}
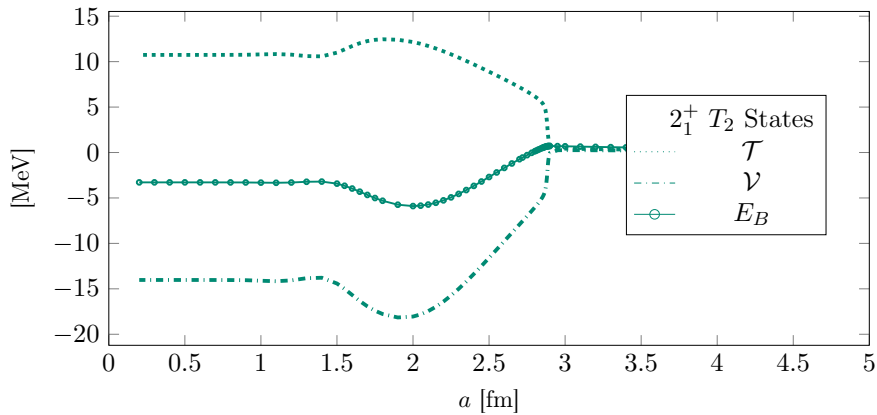

The two different patterns lead to the same inclusion conditions for the principal maxima, $a = d^*/\sqrt{2}n$ with $n \in \mathbb{N}$, \textit{i.e.} $a \approx 2.02, 1.01, 0.67, \ldots$~fm. In practice, two well-developed minima for $a \approx 2.02$ and $1.05$~fm are observed, still in agreement with the predictions. Moreover, two minima are detected in the potential at $a \approx 1.96$ and $1.05$~fm, whereas no extremum is found for around $a = d^*$, due to the absence of maxima along the lattice axes (cf. Fig.~\ref{F-7.0-08}).\\ 

Therefore, the interpretation of the behaviour of the eigenenergies of bound states based on the spatial configuration of the corresponding eigenfunctions and the average value of potential $\mathcal{V}$ on the latter reviewed also in Sec.~III~A of Ref.~\cite{BNL14} is further supported by our findings.\\
However, also the behaviour of the energy eigenvalue as a function of the lattice spacing for the ground state (cf. Fig.~\ref{F-7.0-05}) can find an interpretation if the extrema of the two-body potentials $V^{\rm II}$ are considered. Since the spatial distribution of the PDF of the $0_1^+$ state is spherical with a maximum when the two $\alpha$ particles completely overlap ($d^*=0$), minima of $\mathcal{E}_r$  may occur when the only minimum of $V^{\rm II}$ at $2.64$~fm is mapped in the cubic lattice, \textit{i.e.} for spacings equal to $2.64$, $1.32$, $0.85\ldots$~fm. Effectively, two minima at about $1.25$ and $2.70$~fm are found together with a quasi-stationary point at $2.35$~fm, perhaps due to the inclusion of the shallow maximum of the two-body potentials at $6.71$~fm, see Fig.~\ref{F-2.1-01}.\\
Concerning the angular momentum, similar considerations on fluctuations can be drawn: first decimal digit oscillations are associated to the region $0.96 \lesssim a \lesssim 1.55$~fm of the ground state, the  $1.05 \lesssim a \lesssim 1.58$~fm one of the $2_E^+$ state and the $0.96 \lesssim a \lesssim 1.7$~fm one of the $2_{T_2}^+$ , while third decimal digit accuracy is achieved for $a \lesssim 0.6$~fm by the $0^+$ and only at $a \lesssim 0.2$~fm and $a \lesssim 0.55$~fm for the two members of the $2^+$ multiplet, respectively. The overall behaviour of the angular momentum average values of the three states seems unaffected by level crossings and turns out to be smooth, with the noticeable exception of the evolution curve for the $2_{T_2}^+$ state. In correspondence with the local maximum of the energy eigenvalue at $a=2.9$~fm a rapid \textit{step} increase of the average value of the squared angular momentum of the latter eigenstate takes place, see Fig.~\ref{F-7.0-12}. This phase transition-like behaviour is perhaps related to the exclusion of a sharp extremum characterizing the wavefunction from the lattice, thus preventing an unambiguous determination of the angular momentum content of the $2_{T_2}^+$ state for $a \gtrsim 2.9$~fm. \\
Contrary to the finite-volume analysis, few conclusions can be drawn from the plot of the $\Delta\mathcal{L}^2$ average values (cf. Fig.~\ref{F-7.0-12}). Even if one keeps the logscale in the ordinate axis, the behaviour remains far from linear, due both to sign oscillations of the corrections and to a certain overall negative concavity. In addition, multiplet averaging seems to have little effect in smoothing these fluctuations. \\

\begin{figure}[h!]
\begin{minipage}[c]{0.48\columnwidth}
\scalebox{0.85}{
\begin{tikzpicture}
\begin{axis}[
	xmin = 0,
	xmax = 5,
	xlabel={\small $a$ [fm]},
	ylabel={\small $\hbar^{-2}\mathcal{L}^2$},
	xtick={0,0.5,1,1.5,2,2.5,3,3.5,4,4.5,5},
	ytick={0,2.5,5,7.5,10,12.5,15},
	legend style ={at={(0.12,0.90)}, anchor=north west, draw=black,fill=white,align=left},
	legend entries ={\textcolor{Blue}{$A_1$ - $0^{+}$},\textcolor{Orange}{$E$ - $2^{+}$},\textcolor{PineGreen}{$T_2$ - $2^{+}$}}
]
\addlegendimage{Violet,mark=square*,mark options={fill=Violet}, mark size = 1.0};
\addlegendimage{Orange,mark=pentagon*,mark options={fill=Orange}, mark size = 1.3};
\addlegendimage{PineGreen,mark=triangle*,mark options={fill=PineGreen,rotate=180}, mark size = 1.3};
\addplot[mark=square*, color=Violet,  mark size=1, mark options={fill=Violet}] table [y=J, x=a] {0+_A1_I_18fm.txt};
\addplot[mark=pentagon*, color=Orange,  mark size=1.3, mark options={fill=Orange}] table [y=J, x=a] {2+_E_I_19fm.txt};
\addplot[mark=triangle*, color=PineGreen,  mark size=1.3, mark options={fill=PineGreen,rotate=180}] table [y=J, x=a] {2+_T2_I_19fm.txt};
\addplot[color=Red, solid] table [y=J, x=a] {2+_I_MAVG_19fm.txt};
\end{axis}
\end{tikzpicture}}
\caption{Average value of the squared angular momentum for the six bound states as a function of the lattice spacing when $Na \geq 18$~fm ($\ell = 0$) and  $Na \geq 19$~fm ($\ell = 2$).  As previously, convergence of the average values of $\mathcal{J}^2$ to its expected eigenvalues is attained in the zero-spacing limit.}\label{F-7.0-11}
\end{minipage}\hfill
\begin{minipage}[c]{0.48\columnwidth}
\scalebox{0.85}{
\begin{tikzpicture}
\begin{axis}[
	xmin = 0,
	xmax = 5,
	xlabel={\small $a$ [fm]},
	ylabel={\small $\hbar^{-2}|\Delta\mathcal{L}^2|$},
	xtick={0,0.5,1,1.5,2,2.5,3,3.5,4,4.5,5},
	ymode=log,
	log basis y = {2.718281828459},
	yticklabels = {$e^{-16}$,$e^{-11}$,$e^{-6}$,$e^{-1}$},
	legend style ={at={(0.55,0.40)}, anchor=north west, draw=black,fill=white,align=left},
	legend entries ={\textcolor{Blue}{$A_1$ - $0^{+}$},\textcolor{Orange}{$E$ - $2^{+}$},\textcolor{PineGreen}{$T_2$ - $2^{+}$}}
]
\addlegendimage{Violet,mark=square*,mark options={fill=Violet}, mark size = 1.0};
\addlegendimage{Orange,mark=pentagon*,mark options={fill=Orange}, mark size = 1.3};
\addlegendimage{PineGreen,mark=triangle*,mark options={fill=PineGreen,rotate=180}, mark size = 1.3};
\addplot[mark=square*, color=Violet,  mark size=1, mark options={fill=Violet} ] table [y=dJ, x=a] {0+_A1_I_18fm.txt};
\addplot[mark=pentagon*, color=Orange,  mark size=1.3, mark options={fill=Orange} ] table [y=dJ, x=a] {2+_E_I_19fm.txt};
\addplot[mark=triangle*, color=PineGreen,  mark size=1.3, mark options={fill=PineGreen, rotate=180} ] table [y=dJ, x=a] {2+_T2_I_19fm.txt};
\end{axis}
\end{tikzpicture}}
\caption{Difference between the average value and the expected eigenvalue of the squared angular momentum for the six bound states as a function of the lattice spacing for $Na \geq 18$~fm ($\ell = 0$) and  $Na \geq 19$~fm ($\ell = 2$).  Even if a logscale is set on the $y$ axis, no regular behaviour in the finite volume $\mathcal{L}^2$ corrections can be detected, apart from an overall negative concavity and piecewise linearity of the $0_{A_1}^+$ and $2_{T_2}^+$ curves.}\label{F-7.0-12}
\end{minipage}
\end{figure}

With the aim of extending the previous analysis to higher angular momentum states and assessing the effectivity of multiplet averaging, we increase artificially the stength parameter of the attractive part of the Ali-Bodmer potential up to the 150~\% of its original value, see the dotted curve in Fig.~\ref{F-2.1-01}. By means of this artifact, the wavefunctions of the \ce{^4He} nuclei become more localized about the origin, a consequence of the enhanced attraction of the $\alpha-\alpha$ potential. Moreover, finite volume effects in lattices with size $Na = 12$~fm are already limited to the third decimal digit for the energies of the bound states, a precision that, in the previous case, was attained by the $2^+$ multiplet only at $17$~fm. \\
Besides the latter states and the fundamental one, the bound region of the spectrum now contains four $0^+$ and two further $2^+$ multiplets, together with two $4^+$ and the expected $6^+$, in whose decomposition into irreps of the cubic group all the representations appear at least once. \\
\begin{figure*}[ht!]
\scalebox{0.90}{
\begin{tikzpicture}
\begin{axis}[
	name= bigEn,
	xmin = 5,
	xmax = 55,
	width = 17 cm,
	height = 11 cm,
	xlabel={\small $N$},
	ylabel={\small $\mathcal{E}_r$ [MeV]},
	xtick={0,5,10,15,20,25,30,35,40,45,50,55},
	ytick={-225,-150,-75,0,75,150,225,300,375,450,525,600,675,750,825}
]
\addplot[mark=-, color=Apricot,  mark size=1.5, mark options={fill=Apricot} ] table [y=E, x=L] {0+_A1_I_150.txt};
\addplot[mark=-, color=Apricot,  mark size=1.5, mark options={fill=Apricot} ] table [y=E, x=L] {0+_A1_II_150.txt};
\addplot[mark=-, color=Apricot,  mark size=1.5, mark options={fill=Apricot} ] table [y=E, x=L] {0+_A1_III_150.txt};
\addplot[mark=|, color=Red,  mark size=1.5, mark options={fill=Red} ] table [y=E, x=L] {2+_E_I_150.txt};
\addplot[mark=|, color=Red,  mark size=1.5, mark options={fill=Red} ] table [y=E, x=L] {2+_T2_I_150.txt};
\addplot[mark=|, color=Red,  mark size=1.5, mark options={fill=Red} ] table [y=E, x=L] {2+_E_II_150.txt};
\addplot[mark=|, color=Red,  mark size=1.5, mark options={fill=Red} ] table [y=E, x=L] {2+_T2_II_150.txt};
\addplot[mark=|, color=Red,  mark size=1.5, mark options={fill=Red} ] table [y=E, x=L] {2+_E_III_150.txt};
\addplot[mark=|, color=Red,  mark size=1.5, mark options={fill=Red} ] table [y=E, x=L] {2+_T2_III_150.txt};
\addplot[mark=+, color=Green,  mark size=1.5, mark options={fill=Green} ] table [y=E, x=L] {4+_A1_II_150.txt};
\addplot[mark=+, color=Green,  mark size=1.5, mark options={fill=Green} ] table [y=E, x=L] {4+_E_I_150.txt};
\addplot[mark=+, color=Green,  mark size=1.5, mark options={fill=Green} ] table [y=E, x=L] {4+_T1_I_150.txt};
\addplot[mark=+, color=Green,  mark size=1.5, mark options={fill=Green} ] table [y=E, x=L] {4+_T2_I_150.txt};
\addplot[mark=+, color=Green,  mark size=1.5, mark options={fill=Green} ] table [y=E, x=L] {4+_A1_II_150.txt};
\addplot[mark=+, color=Green,  mark size=1.5, mark options={fill=Green} ] table [y=E, x=L] {4+_E_II_150.txt};
\addplot[mark=+, color=Green,  mark size=1.5, mark options={fill=Green} ] table [y=E, x=L] {4+_T1_II_150.txt};
\addplot[mark=+, color=Green,  mark size=1.5, mark options={fill=Green} ] table [y=E, x=L] {4+_T2_II_150.txt};
\addplot[mark=asterisk, color=Fuchsia,  mark size=1.5, mark options={fill=Fuchsia} ] table [y=E, x=L] {6+_A1_I_150.txt};
\addplot[mark=asterisk, color=Fuchsia,  mark size=1.5, mark options={fill=Fuchsia} ] table [y=E, x=L] {6+_A2_I_150.txt};
\addplot[mark=asterisk, color=Fuchsia,  mark size=1.5, mark options={fill=Fuchsia} ] table [y=E, x=L] {6+_E_I_150.txt};
\addplot[mark=asterisk, color=Fuchsia,  mark size=1.5, mark options={fill=Fuchsia} ] table [y=E, x=L] {6+_T1_I_150.txt};
\addplot[mark=asterisk, color=Fuchsia,  mark size=1.5, mark options={fill=Fuchsia} ] table [y=E, x=L] {6+_T2_I_1_150.txt};
\addplot[mark=asterisk, color=Fuchsia,  mark size=1.5, mark options={fill=Fuchsia} ] table [y=E, x=L] {6+_T2_I_2_150.txt};
\coordinate (c1) at (axis cs:14.5,-40);
\coordinate (c2) at (axis cs:40.5,20);
\coordinate (c3) at (axis cs:14.5,20);
\coordinate (c4) at (axis cs:40.5,-40);
\draw [line width = 0.9 pt] (c1) rectangle (axis cs:40.5,20);
\end{axis}
\coordinate (d1) at (4.58 cm,2.7 cm);
\coordinate (d2) at (15.0 cm, 9.0 cm);
\coordinate (d3) at (4.58 cm, 9.0 cm);
\coordinate (d4) at (15.0 cm, 2.7 cm);
\draw [line width= 0.3 pt] (c1) -- (d1);
\draw [line width= 0.3 pt] (c2) -- (d2);
\draw [line width= 0.3 pt] (c3) -- (d3);
\draw [line width= 0.3 pt] (c4) -- (d4); 
\draw [line width = 0.9 pt, fill= white] (d1) rectangle (d2);
\begin{axis}[
	at = {($(5.44 cm,3.15 cm)$)},
	name= smaEn,
          axis background/.style={fill=white},
	xmin = 14.5,
	xmax = 40.5,
	ymin = -40,
	ymax = 20,
	height = 7.09 cm,
	width = 11.03 cm,
	xtick={15,20,25,30,35,40},
	ytick={-40,-30,-20,-10,0,10,20},
	xticklabel style = {font=\footnotesize},
	yticklabel style = {font=\footnotesize},
	legend columns=5,
	legend style ={at={(0.21,0.21)}, anchor=north west, draw=black,fill=white,align=left,/tikz/column 5/.style={
                column sep=4pt}},
	legend entries ={\textcolor{Violet}{$A_1$},\textcolor{Turquoise}{$A_2$},\textcolor{Orange}{$E$},\textcolor{Magenta}{$T_1$},\textcolor{PineGreen}{$T_2$}}
]
\addlegendimage{Violet,mark=square*, mark size=1.2, mark options={fill=Violet}};
\addlegendimage{Turquoise, mark=diamond*, mark size=1.5, mark options={fill=Turquoise}};
\addlegendimage{Orange, mark=pentagon*, mark size=1.5, mark options={fill=Orange}};
\addlegendimage{Magenta, mark=triangle*, mark size=1.5, mark options={fill=Magenta}};
\addlegendimage{PineGreen, mark=triangle*, mark size=1.5, mark options={fill=PineGreen,rotate=180}};
\addplot[mark=square*, color=Violet,  mark size=1.2, mark options={fill=Violet}] table [y=E, x=L] {4+_A1_II_150.txt};
\addplot[mark=pentagon*, color=Orange,  mark size=1.5, mark options={fill=Orange}] table [y=E, x=L] {4+_E_II_150.txt};
\addplot[mark=triangle*, color=Magenta,  mark size=1.5, mark options={fill=Magenta}] table [y=E, x=L] {4+_T1_II_150.txt};
\addplot[mark=triangle*, color=PineGreen,  mark size=1.5, mark options={fill=PineGreen,rotate=180}] table [y=E, x=L] {4+_T2_II_150.txt};
\addplot[color=Green, dashed, thick] table [y=E, x=L] {4+_II_MAVG.txt};
\addplot[mark=square*, color=Violet,  mark size=1.2, mark options={fill=Violet}] table [y=E, x=L] {6+_A1_I_150.txt};
\addplot[mark=diamond*, color=Turquoise,  mark size=1.5, mark options={fill=Turquoise}] table [y=E, x=L] {6+_A2_I_150.txt};
\addplot[mark=pentagon*, color=Orange,  mark size=1.5, mark options={fill=Orange}] table [y=E, x=L] {6+_E_I_150.txt};
\addplot[mark=triangle*, color=Magenta,  mark size=1.5, mark options={fill=Magenta}] table [y=E, x=L] {6+_T1_I_150.txt};
\addplot[mark=triangle*, color=PineGreen,  mark size=1.5, mark options={fill=PineGreen,rotate=180} ] table [y=E, x=L] {6+_T2_I_1_150.txt};
\addplot[mark=triangle*, color=PineGreen,  mark size=1.5, mark options={fill=PineGreen,rotate=180} ] table [y=E, x=L] {6+_T2_I_2_150.txt};
\addplot[color=Fuchsia, dotted, thick] table [y=E, x=L] {6+_I_MAVG.txt};
\end{axis}
\end{tikzpicture}}
\caption{Behaviour of the energies of the bound eigenstates as a function of the box size N for $a = 0.25$~fm. In the background graph, lines marked by horizontal bars are associated to $0^+$ states, lines marked by vertical bars with $2^+$ states, lines marked by crosses with $4^+$ and lines marked by asterisks with $6^+$. As expected, rotational symmetry is almost restored in the lage box size limit ($N=52$), the remaining discrepancies $\mathcal{O}(10^{-4})\hspace{1mm}\mathrm{MeV}$ being essentially due to space discretization. The magnification resolves the $4_2^+$ and $6_1^+$ states in terms of the underlying cubic group multiplets. Multiplet-averaged eigenenergies of the two are denoted by dashed and dotted lines, in order.}\label{F-7.0-13}
\end{figure*}
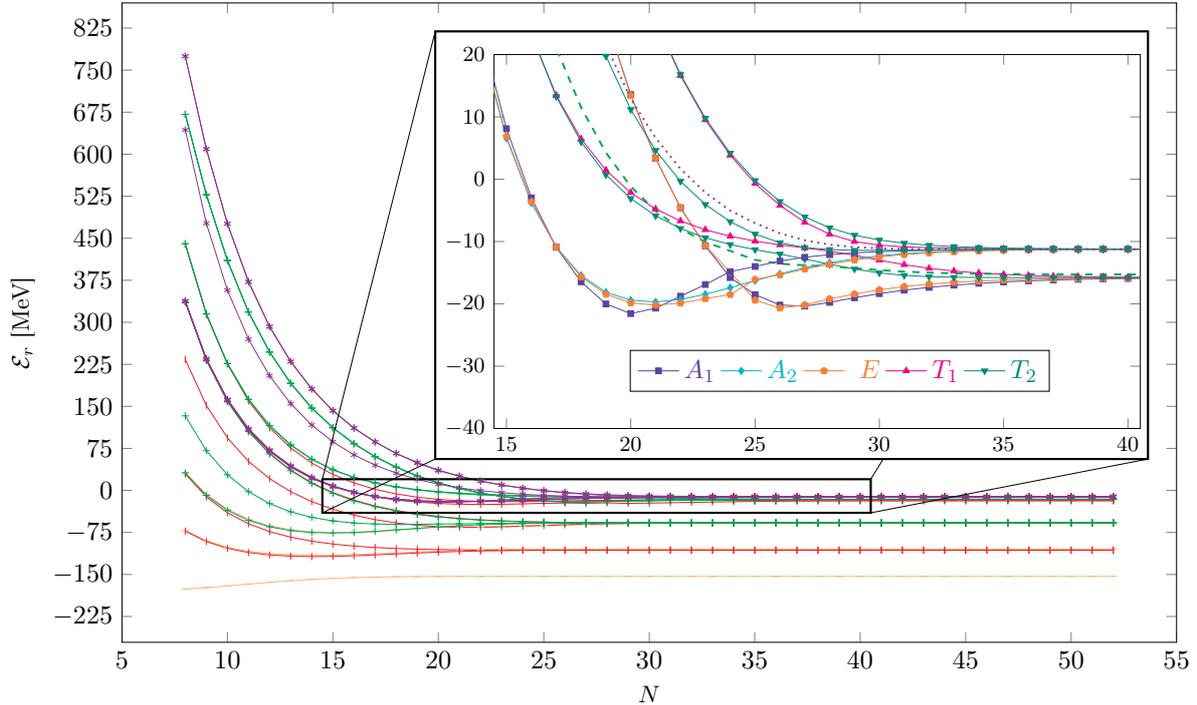

As in the previous case, multiplet averaging of the energies of the $4^+$ and $6^+$ multiplets finds further justification with the cancellation of the polynomial dependence on the lattice size $N$ in the lowest order finite-volume energy corrections (FVEC). More precisely, the leading order correction for the multiplet averaged energies with angular momentum $\ell$ and parity $P$ assumes the universal form \cite{BNL14}
\begin{equation}
E_{\infty}(\ell_A^P) - E_N(\ell_A^P) \lvert^{LO} = (-1)^{\ell+1} 3|\gamma|^2\frac{e^{-\kappa N}}{m N} \label{7.0-01},
\end{equation}
as its magnitude is independent on the particular SO(3) irrep according to which the energy eigenstate transforms. Keeping the lattice spacing invariant with respect to the previous case, we repeat the finite-volume analysis for all the bound states, but dedicating a special attention to the two uppermost SO(3) multiplets, $4_2^+$ and $6_1^+$. Even if the extraction of a greater number of bound states increases the runtime of the numerical computations, the faster decay of the wavefunctions with distance allows to keep the same lattice spacings. Due to the changes in the spatial distribution, the cubic group multiplets composing the SO(3) ones become degenerate with a minimum precision of $10^{-3}$~MeV already at $N=52$, while the average values of the squared angular momentum reach the expectation values with a four decimal digit minimum accuracy. \\ 
As it can be inferred from the magnification in Fig.~\ref{F-7.0-13}, at least two level crossings between states with the same transofmation properties under the operations of the cubic group take place: the involved states are the $A_1$ and the $E$ ones of the two SO(3) multiplets. These intersections are at the origin of sudden \textit{spikes} in the evolution curves of the average values of the squared angular momentum with $N$ for the latter states. As this is presumably due to the insufficient sampling in these regions limited by the lattice spacing constraint, these points have been accurately removed from the plots in Figs.~\ref{F-7.0-14} and \ref{F-7.0-15}. Therefore, better estimations of the intrinsic behaviour of these angular momentum evolution lines can be drawn from $\mathcal{O}$ multiplets that never experience level crossings with states having the same transformation properties under the cubic group. Optimal candidates for such curves are the smooth ones associated to the $6_{A_2}^+$, $4_{T_1}^+$, $4_{T_2}^+$ and $6_{T_2}^{+}~\mathrm{I}$ \footnote{With $6_{T_2}^{+}~\mathrm{I}$ has been denoted the $T_2$ multiplet lying always below in energy with respect to the $J = 6$ partner bearing the same cubic irrep.} levels.\\
\begin{figure}[ht!]
\begin{minipage}[c]{0.48\columnwidth}
\scalebox{0.90}{
\begin{tikzpicture}
\begin{axis}[
	xmin = 5,
	xmax = 55,
	xlabel={\small $N$},
	ylabel={\small $\hbar^{-2}\mathcal{L}^2$},
	xtick={5,10,15,20,25,30,35,40,45,50,55},
	ytick={10,15,20,25,30,35,40,45,50,55},
	legend columns=2,
	legend style ={at={(0.55,0.91)}, anchor=north west, draw=black,fill=white,align=left,/tikz/column 2/.style={
                column sep=4pt}},
	legend entries ={\textcolor{Violet}{$A_1$},\textcolor{Orange}{$E$},\textcolor{Magenta}{$T_1$},\textcolor{PineGreen}{$T_2$}}
]
\addlegendimage{Violet,mark=square*, mark size=1.0, mark options={fill=Violet}};
\addlegendimage{Orange, mark=pentagon*, mark size=1.3, mark options={fill=Orange}};
\addlegendimage{Magenta, mark=triangle*, mark size=1.3, mark options={fill=Magenta}};
\addlegendimage{PineGreen, mark=triangle*, mark size=1.3, mark options={fill=PineGreen, rotate=180}};
\addplot[mark=square*, color=Violet,  mark size=1, mark options={fill=Violet}] table [y=J, x=L] {4+_A1_II_150.txt};
\addplot[mark=pentagon*, color=Orange,  mark size=1.3, mark options={fill=Orange}] table [y=J, x=L] {4+_E_II_150.txt};
\addplot[mark=triangle*, color=Magenta,  mark size=1.3, mark options={fill=Magenta}] table [y=J, x=L] {4+_T1_II_150.txt};
\addplot[mark=triangle*, color=PineGreen,  mark size=1.3, mark options={fill=PineGreen, rotate=180}] table [y=J, x=L] {4+_T2_II_150.txt};
\addplot[color=Green, dashed] table [y=J, x=L] {4+_II_MAVG.txt};
\end{axis}
\end{tikzpicture}}
\caption{Average value of the squared angular momentum for the $4^+$ states as a function of the lattice size.  As predicted, the average values of $\mathcal{L}^2$ for the cubic group partners of the SO(3) multiplet converge to the eigenvalue of 20 units of $\hbar^2$ of the same operator, even if a well-pronounced oscillatory behaviour for relatively small lattices ($N \lesssim 32$).}\label{F-7.0-14}
\end{minipage}\hfill
\begin{minipage}[c!]{0.48\columnwidth}
\scalebox{0.90}{
\begin{tikzpicture}
\begin{axis}[
	xmin = 5,
	xmax = 55,
	xlabel={\small $N$},
	ylabel={\small $\hbar^{-2}\mathcal{L}^2$},
	xtick={5,10,15,20,25,30,35,40,45,50,55},
	ytick={20,25,30,35,40,45,50,55},
	legend columns=2,
	legend style ={at={(0.55,0.31)}, anchor=north west, draw=black,fill=white,align=left,/tikz/column 2/.style={
                column sep=4pt}},
	legend entries ={\textcolor{Violet}{$A_1$},\textcolor{Turquoise}{$A_2$},\textcolor{Orange}{$E$},\textcolor{Magenta}{$T_1$},\textcolor{PineGreen}{$T_2$}}
]
\addlegendimage{Violet,mark=square*, mark size=1.0, mark options={fill=Violet}};
\addlegendimage{Turquoise, mark=diamond*, mark size=1.3, mark options={fill=Turquoise}};
\addlegendimage{Orange, mark=pentagon*, mark size=1.3, mark options={fill=Orange}};
\addlegendimage{Magenta, mark=triangle*, mark size=1.3, mark options={fill=Magenta}};
\addlegendimage{PineGreen, mark=triangle*, mark size=1.3, mark options={fill=PineGreen,rotate=180}};
\addplot[mark=square*, color=Violet,  mark size=1, mark options={fill=Violet}] table [y=J, x=L] {6+_A1_I_150.txt};
\addplot[mark=diamond*, color=Turquoise,  mark size=1.3, mark options={fill=Turquoise}] table [y=J, x=L] {6+_A2_I_150.txt};
\addplot[mark=pentagon*, color=Orange,  mark size=1.3, mark options={fill=Orange}] table [y=J, x=L] {6+_E_I_150.txt};
\addplot[mark=triangle*, color=Magenta,  mark size=1.3, mark options={fill=Magenta}] table [y=J, x=L] {6+_T1_I_150.txt};
\addplot[mark=triangle*, color=PineGreen,  mark size=1.3, mark options={fill=PineGreen, rotate=180}] table [y=J, x=L] {6+_T2_I_1_150.txt};
\addplot[mark=triangle*, color=PineGreen,  mark size=1.3, mark options={fill=PineGreen, rotate=180}] table [y=J, x=L] {6+_T2_I_2_150.txt};
\addplot[color=Fuchsia, dotted] table [y=J, x=L] {6+_I_MAVG.txt};
\end{axis}
\end{tikzpicture}}
\caption{Average value of the squared angular momentum for the $6^+$ states as a function of the lattice size.  As predicted, the average values of $\mathcal{L}^2$ for the cubic group partners of the SO(3) multiplet converge to the eigenvalue of 42 units of $\hbar^2$ of the same operator, even if a well-pronounced oscillatory behaviour for relatively small lattices ($N \lesssim 32$) is observed.}\label{F-7.0-15}
\end{minipage}
\end{figure}

The plot of the differences between the average values and the expected values of $\mathcal{L}^2$ with the number of lattice sites per dimension enables us to confirm the qualitative observations on the finite volume corrections for the squared angular momentum. For lattices large enough ($N \gtrsim 26$), the latter decreases exponentially with $N$, the decay constant being approximately shared by all the members of the same SO(3) multiplet. Besides, convergence to the expected angular momentum is faster for more tightly bound states, suggesting again a dependence of the decay constants on the energies of the spectral lines. Moreover, the chosen value of the lattice spacing is responsible of the \textit{saturation} behaviour of the lines for the $6_{T_1}^+$ and $6_{T_2}^+~\mathrm{II}$ for $N \geq 37$: as observed in Fig.~\ref{F-7.0-12}, discretization affects states belonging to different SO(3) and $\mathcal{O}$ irreps in different extent. \\
\begin{figure}[ht!]
\begin{minipage}[c]{0.57\columnwidth}
\begin{tikzpicture}
\begin{axis}[
	xmin = 20,
	xmax = 55,
	xlabel={\small $N$},
	ylabel={\small $\hbar^{-2}|\Delta\mathcal{L}^2|$},
	legend columns=5,
	xtick={25,35,45,55},
	ymode=log,
	log basis y = {2.718281828459},
	yticklabels = {$e^{-9}$,$e^{-6}$,$e^{-3}$,$e^0$,$e^3$},
	width = 4.5 cm,
	height = 7 cm,
]
\addplot[mark=square*, color=Violet,  mark size=1, mark options={fill=Violet}] table [y=dJ, x=L] {4+_A1_II_150.txt};
\addplot[mark=pentagon*, color=Orange,  mark size=1.3, mark options={fill=Orange}] table [y=dJ, x=L] {4+_E_II_150.txt};
\addplot[mark=triangle*, color=Magenta,  mark size=1.3, mark options={fill=Magenta}] table [y=dJ, x=L] {4+_T1_II_150.txt};
\addplot[mark=triangle*, color=PineGreen,  mark size=1.3, mark options={fill=PineGreen, rotate=180}] table [y=dJ, x=L] {4+_T2_II_150.txt};
\addplot[color=Green, dashed] table [y=dJ, x=L] {4+_II_MAVG.txt};
\end{axis}
\end{tikzpicture}
\begin{tikzpicture}
\begin{axis}[
	xmin = 20,
	xmax = 55,
	width = 4.5 cm,
	height = 7 cm,
	xlabel={\small $N$},
	ytick = {},
	xtick={25,35,45,55},
	ymode=log,
	log basis y = {2.718281828459},
	yticklabels = {$e^{-10}$,$e^{-7}$,$e^{-4}$,$e^{-1}$,$e^2$}
]
\addplot[mark=square*, color=Violet,  mark size=1, mark options={fill=Violet} ] table [y=dJ, x=L] {6+_A1_I_150.txt};
\addplot[mark=diamond*, color=Turquoise,  mark size=1.3, mark options={fill=Turquoise}] table [y=dJ, x=L] {6+_A2_I_150.txt};
\addplot[mark=pentagon*, color=Orange,  mark size=1.3, mark options={fill=Orange}] table [y=dJ, x=L] {6+_E_I_150.txt};
\addplot[mark=triangle*, color=Magenta,  mark size=1.3, mark options={fill=Magenta}] table [y=dJ, x=L] {6+_T1_I_150.txt};
\addplot[mark=triangle*, color=PineGreen,  mark size=1.3, mark options={fill=PineGreen, rotate=180}] table [y=dJ, x=L] {6+_T2_I_1_150.txt};
\addplot[mark=triangle*, color=PineGreen,  mark size=1.3, mark options={fill=PineGreen, rotate=180}] table [y=dJ, x=L] {6+_T2_I_2_150.txt};
\addplot[color=Fuchsia, dotted] table [y=dJ, x=L] {6+_I_MAVG.txt};
\end{axis}
\end{tikzpicture}
\end{minipage}\hfill
\begin{minipage}[c]{0.38\columnwidth}
\caption{Difference between the average value and the expected eigenvalue of the squared angular momentum for the $4_2^+$ (left) and $6_1^+$ (right) states as a function of the lattice size.  A regular linear behaviour in the finite volume $\mathcal{L}^2$ corrections for boxes large enough neatly emerges by setting the logscale on the y axis. The same convention on the markers for the cubic group irreps of Figs.~\ref{F-7.0-08}-\ref{F-7.0-10} is used.}\label{F-7.0-16}
\end{minipage}
\end{figure}
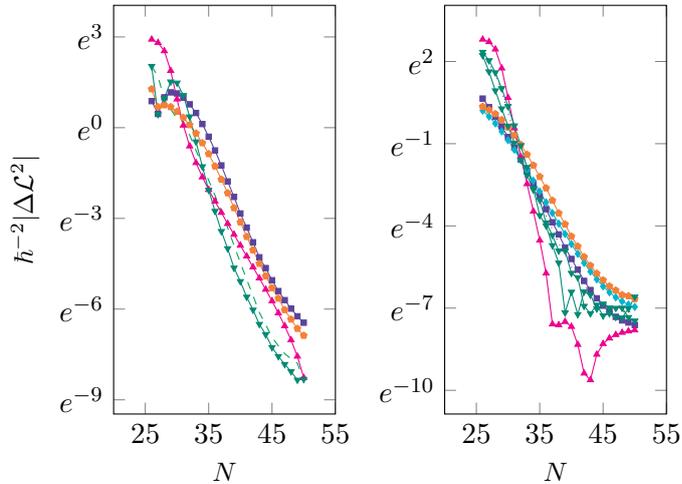

Setting a box size equal to $12$~fm, we can now concentrate on discretization effects. As expected, here the consequences of a more localized distribution of the wavefunctions about the origin become even more evident. Discretization errors for the energies remain large in a wide range of lattice spacing, dropping to the first decimal digit for most of the bound states only at $a \approx 0.60$~fm and then reaching third digit precision only at $0.25$~fm. Nevertheless, the behaviour of the $4_2^+$ and $6_1^+$ eigenenergies as function of the lattice spacing appears smooth in the interval of interest, $0.24 \leq a \leq 1.8$~fm.
In particular, the curves for the $4_2^+$ $E$, $A_1$ and $T_2$ multiplets display a deep minimum located around $0.95$~fm, cf. Fig.~\ref{F-7.0-17}, whereas the one of the $T_1$ levels possesses a shallower pocket, closer to the origin ($a\approx 0.75$~fm). Similarily, the energy curves of the $6_{A_1}^+$, $6_{A_2}^+$ and $6_{E}^+$ states possess a well developed first minimum about $1.38$, $1.02$ and $0.91$~fm, respectively, while $T_1$ and $T_2$ states are characterized by a first shallow minimum at about $0.9$~fm followed by a second even less-developed one around $1.5$~fm. \\
\begin{figure}[ht!]
\begin{minipage}[c]{0.49\columnwidth}
\scalebox{0.85}{
\begin{tikzpicture}
\begin{axis}[
	xmin = 0,
	xmax = 2.0,
	xlabel={\small $a$ [fm]},
	ylabel={\small $\mathcal{E}_r$ [MeV]},
	xtick={0,0.3,0.6,0.9,1.2,1.5,1.8},
	ytick={-35,-30,-25,-20,-15,-10,-5,0,5},
	legend columns=1,
	legend style ={at={(0.19,0.91)}, anchor=north west, draw=black,fill=white,align=left,/tikz/column 1/.style={
                column sep=4pt}},
	legend entries ={\textcolor{Violet}{$A_1$},\textcolor{Orange}{$E$},\textcolor{Magenta}{$T_1$},\textcolor{PineGreen}{$T_2$}},
]
\addlegendimage{Violet, mark=square*, mark size=1.0, mark options={fill=Violet}};
\addlegendimage{Orange, mark=pentagon*, mark size=1.3, mark options={fill=Orange}};
\addlegendimage{Magenta, mark=triangle*, mark size=1.3, mark options={fill=Magenta}};
\addlegendimage{PineGreen, mark=triangle*, mark size=1.3, mark options={fill=PineGreen, rotate=180}};
\addplot[mark=square*, color=Violet,  mark size=1, mark options={fill=Violet} ] table [y=E, x=a] {4+_A1_II_12fm.txt};
\addplot[mark=pentagon*, color=Orange,  mark size=1.3, mark options={fill=Orange} ] table [y=E, x=a] {4+_E_II_12fm.txt};
\addplot[mark=triangle*, color=Magenta,  mark size=1.3, mark options={fill=Magenta} ] table [y=E, x=a] {4+_T1_II_12fm.txt};
\addplot[mark=triangle*, color=PineGreen,  mark size=1.3, mark options={fill=PineGreen,rotate=180} ] table [y=E, x=a] {4+_T2_II_12fm.txt};
\addplot[color=Green, dashed] table [y=E, x=a] {4+_II_MAVG_12fm.txt};
\end{axis}
\end{tikzpicture}}
\caption{Behaviour of the energies of the  $4_2^+$ eigenstates as a function of the lattice spacing for $Na \geq 12$~fm. }\label{F-7.0-17}
\end{minipage}
\hfill
\begin{minipage}[c]{0.49\columnwidth}
\scalebox{0.85}{
\hspace{0.2cm}
\begin{tikzpicture}
\begin{axis}[
	name= bigEn,
	xmin = 0,
	xmax = 2.0,
	xlabel={\small $a$ [fm]},
	ylabel={\small $\mathcal{E}_r$ [MeV]},
	xtick={0,0.3,0.6,0.9,1.2,1.5,1.8},
	ytick={-60,-55,-50,-45,-40,-35,-30,-25,-20,-15,-10,-5,0,5,10},
	legend columns=1,
	legend style ={at={(0.12,0.49)}, anchor=north west, draw=black,fill=white,align=left,/tikz/column 1/.style={
                column sep=4pt}},
	legend entries ={\textcolor{Violet}{$A_1$},\textcolor{Turquoise}{$A_2$},\textcolor{Orange}{$E$},\textcolor{Magenta}{$T_1$},\textcolor{PineGreen}{$T_2$}},
]
\addlegendimage{Violet,mark=square*, mark size=1.0, mark options={fill=Violet}};
\addlegendimage{Turquoise, mark=diamond*, mark size=1.3, mark options={fill=Turquoise}};
\addlegendimage{Orange, mark=pentagon*, mark size=1.3, mark options={fill=Orange}};
\addlegendimage{Magenta, mark=triangle*, mark size=1.3, mark options={fill=Magenta}};
\addlegendimage{PineGreen, mark=triangle*, mark size=1.3, mark options={fill=PineGreen,rotate=180}};
\addplot[mark=square*, color=Violet,  mark size=1, mark options={fill=Violet} ] table [y=E, x=a] {6+_A1_I_12fm.txt};
\addplot[mark=diamond*, color=Turquoise,  mark size=1.3, mark options={fill=Turquoise} ] table [y=E, x=a] {6+_A2_I_12fm.txt};
\addplot[mark=pentagon*, color=Orange,  mark size=1.3, mark options={fill=Orange} ] table [y=E, x=a] {6+_E_I_12fm.txt};
\addplot[mark=triangle*, color=Magenta,  mark size=1.3, mark options={fill=Magenta} ] table [y=E, x=a] {6+_T1_I_12fm.txt};
\addplot[mark=triangle*, color=PineGreen,  mark size=1.3, mark options={fill=PineGreen} ] table [y=E, x=a] {6+_T2_I_1_12fm.txt};
\addplot[mark=triangle*, color=PineGreen,  mark size=1.3, mark options={fill=PineGreen} ] table [y=E, x=a] {6+_T2_I_2_12fm.txt};
\addplot[color=Fuchsia, dotted] table [y=E, x=a] {6+_I_MAVG_12fm.txt};
\end{axis}
\end{tikzpicture}}
\caption{Behaviour of the energies of the $6_1^+$ eigenstates as a function of the lattice spacing for $Na \geq 12$~fm. }\label{F-7.0-18}
\end{minipage}
\end{figure}

As illustrated for $2_1^+$ states and in Ref.~\cite{BNL14}, the position of these minima can find an interpretation via the analysis of the spatial distribution of the PDFs associated to the relevant states. However, the presence of secondary maxima and of absolute maxima off the lattice symmetry axes in the $4_2^+$ and $6_1^+$ PDFs make these predictions less effective than in the previous case. Nevertheless, the inclusion conditions for the maxima of the $6_{A_2}^+$ $I_z=2$ state are satisfied in good approximation for a relatively large value of the spacing, $a$, leading to a successful description of the behaviour of the turquoise curve in Fig.~\ref{F-7.0-18}.\\
The probability density function for this $6^+$ state is characterized by four equidistant couples of principal maxima separated by an angle $\gamma \approx 34.2^{\circ}$ and located at a distance $d^*\approx 2.31$~fm from the origin in the $x,y$ and $z=0$ planes.\\ 

\begin{figure}[ht!]
\begin{minipage}[c]{0.48\columnwidth}
\includegraphics[width=0.82\columnwidth]{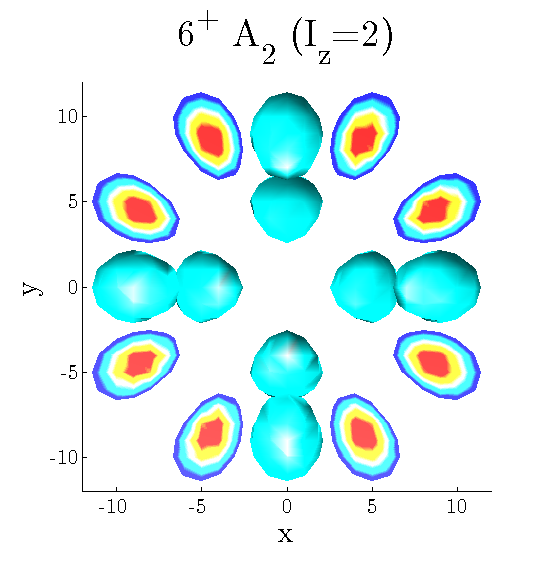}
\end{minipage}\hfill
\begin{minipage}[c]{0.48\columnwidth}
\includegraphics[width=0.84\columnwidth]{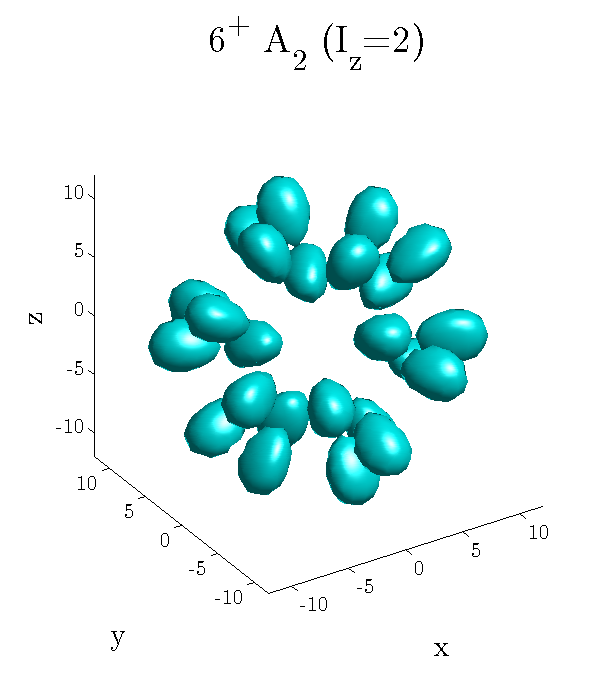}
\end{minipage}
\caption{Three-dimensional probability density distribution of the $\alpha$-$\alpha$ separation (left) and cross-sectional plot (xy plane) of the PDF (right) of the $6_1^+$ $A_2$ state. In particular, the outer isohypsic surfaces of the two plots correspond to a probability density equal to the 25\% (dark blue) of the maximum value of the PDF (dark red). Distances along the axes are measured in lattice spacing units ($a=0.24$~fm). Despite the strong resemblance, the arrangement of the maxima in the xy, xz and yz planes is not octagonal.}
\label{F-7.0-20}
\end{figure}

Even if the 24 maxima cannot be simultaneously included in the cubic lattice, the inclusion conditions on the lattice spacing approximately match for $1.02 \lesssim a \lesssim 1.08$~fm. From the inclusion conditions of a pair of maxima in the first quadrant of the xy plane, see Fig.~\ref{F-7.0-20}, in fact, it follows that
\begin{equation}
a_x = \frac{d^*}{n}\cos\left(\frac{\pi}{4}-\frac{\gamma}{2}\right)\label{7.0-02},
\end{equation}
\textit{i.e.} $a_x \approx 2.04, 1.02, 0.68...$ for the x-axis and
\begin{equation}
a_y = \frac{d^*}{n}\sin\left(\frac{\pi}{4}-\frac{\gamma}{2}\right)\label{7.0-03},
\end{equation}
\textit{i.e.} $a_y \approx 1.08, 0.54, 0.36...$ for the y-axis. Effectively, a sharp minimum of the total energy (cf. Figs.~\ref{F-7.0-18} and \ref{F-7.0-19}) is detected, confirming the predictions. On the other hand, the minimum of the average value of the potential, cf. Fig.~\ref{F-7.0-19}, and the $\alpha$-$\alpha$ distance see Fig.~\ref{F-7.0-21}, is shifted towards smaller spacings ($\approx 0.85$~fm), due to a slow decrease of the associated probability density function in the vicinity the maxima. \\ 
\begin{figure}[ht!]
\begin{minipage}[c]{0.54\columnwidth}
\scalebox{0.71}{
\begin{tikzpicture}
\begin{axis}[
	xmin = 0,
	xmax = 2.5,
           height = 8.5cm,
	width = 11.9cm,
	xlabel={\small $a$ [fm]},
	ylabel={\small [MeV]},
	xtick={0,0.2,0.4,0.6,0.8,1.0,1.2,1.4,1.6,1.8,2.0,2.2,2.4},
	ytick={-150,-125,-100,-75,-50,-25,0,25,50,75,100,125,150},
           legend style ={at={(0.69,0.75)}, anchor=north west, draw=black,fill=white,align=left,/tikz/column 1/.style={
                column sep=4pt}},
          legend columns = 1,
	legend entries ={\hspace{-0.3cm}$6_1^+$ $A_2$ State, $\mathcal{T}$,$\mathcal{V}$,$E_B$}
]
\addlegendimage{empty legend};
\addlegendimage{color=Turquoise, dotted};
\addlegendimage{color=Turquoise, dash dot};
\addlegendimage{color=Turquoise, mark=o};
\addplot[color=Turquoise, dotted , ultra thick] table [y=Kin, x=a] {6+_A2_I_12fm_Suppl.txt};
\addplot[color=Turquoise,  dash dot, ultra thick] table [y=Pot, x=a] {6+_A2_I_12fm_Suppl.txt};
\addplot[mark=o, color=Turquoise,  mark size=1, mark options={fill=Turquoise}, thick] table [y=E, x=a] {6+_A2_I_12fm_Suppl.txt};
\end{axis}
\end{tikzpicture}}
\caption{Behaviour of the average values of the kinetic energy, $\mathcal{T}$,  and the potential operator, $\mathcal{V}$, on the $6_1^+$ $A_2$ eigenstate as a function of the lattice spacing a for $Na \geq 12$~fm. The sum of the two average values produce the already displayed $\mathcal{E}_r$ curve, see Fig.~\ref{F-7.0-18}. }\label{F-7.0-19}
\end{minipage}\hfill
\begin{minipage}[c]{0.42\columnwidth}
\begin{center}
\scalebox{0.63}{
\begin{tikzpicture}
\begin{axis}[
	xmin = 0,
	xmax = 2.1,
           height = 8.5cm,
	width = 8.3cm,
	xlabel={\Large $a$ [fm]},
	ylabel={\Large $\mathcal{R}$ [fm]},
	xtick={0,0.3,0.6,0.9,1.2,1.5,1.8,2.1},
	ytick={1.0,1.25,1.5,1.75,2.0,2.25,2.5,2.75,3.0,3.25,3.5,3.75,4.0,4.25},
           legend style ={at={(0.10,0.88)}, anchor=north west, draw=black,fill=white,align=left},
	legend entries ={\hspace{-0.4cm}$2_1^+$ $T_2$ $I_z=0$ State, $\mathcal{R}$},
]
\addlegendimage{empty legend};
\addlegendimage{mark=o,color=Turquoise};
\addplot[mark=o, color=Turquoise,  mark size=1, mark options={fill=Turquoise}, thick] table [y=ravg, x=a] {6+_A2_I_12fm_Suppl.txt};
\end{axis}
\end{tikzpicture}}
\end{center}
\caption{Behaviour of the average value of the  interparticle distance as a function of the lattice spacing for the $6_1^+$ $A_2$ eigenstate. A minimum in $\mathcal{R}$ at $a \approx 0.88$~fm is visible, implying that the condition on the decay of the wavefunction with increasing $\alpha$-$\alpha$ distance is appreciably satisfied.}\label{F-7.0-21}
\end{minipage}
\end{figure}

Concerning the angular momentum, the fluctuations of the average values of $\mathcal{L}^2$ about the corresponding expectation values for $a \gtrsim 0.6$~fm are even larger than the ones of the energies. The effect is even amplified for the $0_3^+$ and the $A_1$ member of the lowest $4_1^+$ state due to their quasi-degeneracy and the many level crossings they undergo before reaching their continuum eigenvalue. \\ 
\begin{figure}[hb!]
\begin{minipage}[c]{0.48\columnwidth}
\scalebox{0.85}{
\begin{tikzpicture}
\begin{axis}[
	xmin = 0,
	xmax = 2.0,
	xlabel={\small $a$ [fm]},
	ylabel={\small $\hbar^{-2}\mathcal{L}^2$},
	xtick={0,0.3,0.6,0.9,1.2,1.5,1.8},
	ytick={-5,0,5,10,15,20,25,30,35,40,45,50},
	legend columns=1,
	legend style ={at={(0.09,0.89)}, anchor=north west, draw=black,fill=white,align=left,/tikz/column 1/.style={
                column sep=4pt}},
	legend entries ={\textcolor{Violet}{$A_1$},\textcolor{Orange}{$E$},\textcolor{Magenta}{$T_1$},\textcolor{PineGreen}{$T_2$}},
]
\addlegendimage{Violet, mark=square*, mark size=1.0, mark options={fill=Violet}};
\addlegendimage{Orange, mark=pentagon*, mark size=1.3, mark options={fill=Orange}};
\addlegendimage{Magenta, mark=triangle*, mark size=1.3, mark options={fill=Magenta}};
\addlegendimage{PineGreen, mark=triangle*, mark size=1.3, mark options={fill=PineGreen, rotate=180}};
\addplot[mark=square*, color=Violet,  mark size=1, mark options={fill=Violet} ] table [y=J, x=a] {4+_A1_II_12fm.txt};
\addplot[mark=pentagon*, color=Orange,  mark size=1.3, mark options={fill=Orange} ] table [y=J, x=a] {4+_E_II_12fm.txt};
\addplot[mark=triangle*, color=Magenta,  mark size=1.3, mark options={fill=Magenta} ] table [y=J, x=a] {4+_T1_II_12fm.txt};
\addplot[mark=triangle*, color=PineGreen,  mark size=1.3, mark options={fill=PineGreen, rotate=180} ] table [y=J, x=a] {4+_T2_II_12fm.txt};
\addplot[color=Green, dashed] table [y=J, x=a] {4+_II_MAVG_12fm.txt};
\end{axis}
\end{tikzpicture}}
\caption{Behaviour of the squared angular momentum of the  $4_2^+$ eigenstates as a function of the lattice spacing for $Na \geq 12$~fm. As before, convergence of the average values of $\mathcal{L}^2$ to its expected eigenvalues is achieved in the zero-spacing limit.}\label{F-7.0-22}
\end{minipage}\hfill
\begin{minipage}[c]{0.48\columnwidth}
\scalebox{0.85}{
\begin{tikzpicture}
\begin{axis}[
	name= bigEn,
	xmin = 0,
	xmax = 2.0,
	xlabel={\small $a$ [fm]},
	ylabel={\small $\hbar^{-2}\mathcal{L}^2$},
	xtick={0,0.3,0.6,0.9,1.2,1.5,1.8},
	ytick={-5,0,5,10,15,20,25,30,35,40,45,50},
	legend columns=1,
	legend style ={at={(0.12,0.62)}, anchor=north west, draw=black,fill=white,align=left,/tikz/column 2/.style={
                column sep=4pt}},
	legend entries ={\textcolor{Violet}{$A_1$},\textcolor{Turquoise}{$A_2$},\textcolor{Orange}{$E$},\textcolor{Magenta}{$T_1$},\textcolor{PineGreen}{$T_2$}},
]
\addlegendimage{Violet,mark=square*, mark size=1.0, mark options={fill=Violet}};
\addlegendimage{Turquoise, mark=diamond*, mark size=1.3, mark options={fill=Turquoise}};
\addlegendimage{Orange, mark=pentagon*, mark size=1.3, mark options={fill=Orange}};
\addlegendimage{Magenta, mark=triangle*, mark size=1.3, mark options={fill=Magenta}};
\addlegendimage{PineGreen, mark=triangle*, mark size=1.3, mark options={fill=PineGreen, rotate=180}};
\addplot[mark=square*, color=Violet,  mark size=1, mark options={fill=Violet} ] table [y=J, x=a] {6+_A1_I_12fm.txt};
\addplot[mark=diamond*, color=Turquoise,  mark size=1.3, mark options={fill=Turquoise} ] table [y=J, x=a] {6+_A2_I_12fm.txt};
\addplot[mark=pentagon*, color=Orange,  mark size=1.3, mark options={fill=Orange} ] table [y=J, x=a] {6+_E_I_12fm.txt};
\addplot[mark=triangle*, color=Magenta,  mark size=1.3, mark options={fill=Magenta} ] table [y=J, x=a] {6+_T1_I_12fm.txt};
\addplot[mark=triangle*, color=PineGreen,  mark size=1.3, mark options={fill=PineGreen,rotate=180} ] table [y=J, x=a] {6+_T2_I_1_12fm.txt};
\addplot[mark=triangle*, color=PineGreen,  mark size=1.3, mark options={fill=PineGreen,rotate=180} ] table [y=J, x=a] {6+_T2_I_2_12fm.txt};
\addplot[color=Fuchsia, dotted] table [y=J, x=a] {6+_I_MAVG_12fm.txt};
\end{axis}
\end{tikzpicture}}
\caption{Behaviour of the squared angular momentum of the  $6_1^+$ eigenstates as a function of the lattice spacing for $Na \geq 12$~fm. Convergence of the average values of $\mathcal{L}^2$ to its expected eigenvalues is attained in the zero-spacing limit.}\label{F-7.0-23}
\end{minipage}
\end{figure}

Due both to the absence of nearby levels with the same transformation properties under $\mathcal{O}$ and the smaller number of these crossings, the $4_2^+$ and the $6_1^+$ multiplets converge sensibly faster to their expected squared angular momentum eigenvalue in the zero spacing limit. Nevertheless, the appreciable continuity of the evolution curves of $\mathcal{L}^2$ with $a$ remains seldom interrupted by sharp spikes and wells, withnessing level crossings of the aforementioned kind. \\ 
Because of the presence of many low-lying $0^+$ and $2^+$ states, the $A_1$ and, to a smaller extent, $T_2$ and $E$ lines are more heavily affected by cusps than $T_1$ and $A_2$ states, whose behaviour exhibits the transition-like features already observed in Fig.~\ref{F-7.0-12}. The onset point of these step-growing and falling parts marks the upper bound of the lattice spacing interval in which the observed levels can be classified as partners of a SO(3) multiplet. Beyond $a \approx 0.9$~fm, the characterizing part of all the wavefunctions composing the $4_2^+$ and $6_1^+$ multiplets in not sampled any more by the lattice, thus making angular momentum classification of the states almost unreliable. \\
Since the $|\Delta\mathcal{L}^2|(a)$ curve for the $2_E^+$ state in the above is heavily affected by the sign inversions of the angular momentum correction, no particular conclusion was drawn from the graph in Fig.~\ref{F-7.0-12}. In this case, a part from a spike in the $4_{T_1}^+$ curve around $0.3$~fm and some disturbance in the $4_{T_2}^+$ one around $0.75$~fm, an appreciable quasi-linear behaviour of the $\log|\Delta\mathcal{L}^2|$'s can be inferred from $0.7$~fm towards the continuum limit. Consequently, the corrections to the squared angular momentum average values for lattice cubic group eigenstates can be reproduced by a positive exponential of $a$,
\begin{equation}
|\Delta\mathcal{L}^2(\ell)| \underset{a\rightarrow 0}{\approx} \mathcal{A_{\ell}}\exp(a\cdot\kappa_{\ell})~.\label{7.0-02}
\end{equation}
in the small-spacing region. In particular, the constant in the argument of the exponential, $\kappa_{\ell}$, is approximately independent on the cubic group irrep $\Gamma$ according to which each state of a given angular momentum multiplet $\ell$ transform. Moreover, the proportionality constant $\mathcal{A_{\ell}}$ in Eq.~\eqref{7.0-02} vanishes exactly for infinite-volume lattices and is expected to decrease with increasing box size $Na$.\\
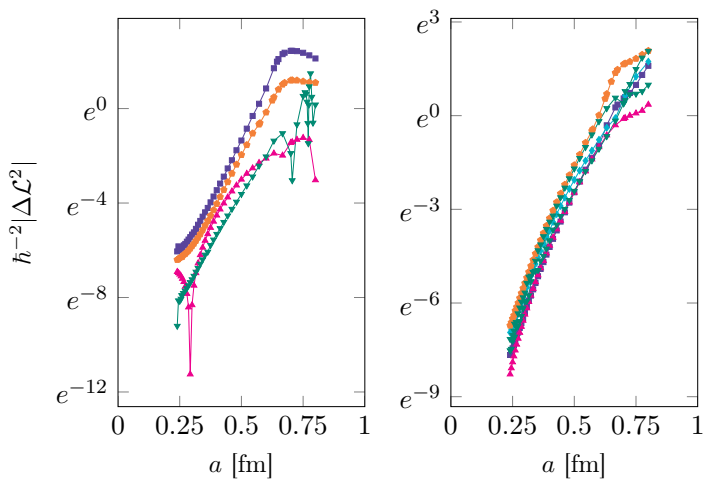
\begin{figure}[ht!]
\begin{minipage}[c]{0.60\columnwidth}
\scalebox{0.95}{
\begin{tikzpicture}
\begin{axis}[
	xmin = 0,
	xmax = 1.0,
	width = 5.0 cm,
	height = 7.0 cm,
	xlabel={\small $a$ [fm]},
	ylabel={\small $\hbar^{-2}|\Delta\mathcal{L}^2|$},
	xtick={0,0.25,0.5,0.75,1},
	ymode=log,
	log basis y = {2.718281828459},
	yticklabels = {$e^{-12}$,$e^{-8}$,$e^{-4}$,$e^0$}
]
\addplot[mark=square*, color=Violet,  mark size=1, mark options={fill=Violet} ] table [y=dmJ, x=a] {4+_A1_II_12fm.txt};
\addplot[mark=pentagon*, color=Orange,  mark size=1.3, mark options={fill=Orange} ] table [y=dmJ, x=a] {4+_E_II_12fm.txt};
\addplot[mark=triangle*, color=Magenta,  mark size=1.3, mark options={fill=Magenta} ] table [y=dmJ, x=a] {4+_T1_II_12fm.txt};
\addplot[mark=triangle*, color=PineGreen,  mark size=1.3, mark options={fill=PineGreen, rotate=180} ] table [y=dmJ, x=a] {4+_T2_II_12fm.txt};
\end{axis}
\end{tikzpicture}
\begin{tikzpicture}
\begin{axis}[
	xmin = 0,
	xmax = 1.0,
	width = 5.0 cm,
	height = 7.0 cm,
	xlabel={\small $a$ [fm]},
	xtick={0,0.25,0.5,0.75,1},
	ymode=log,
	log basis y = {2.718281828459},
	yticklabels = {$e^{-9}$,$e^{-6}$,$e^{-3}$,$e^0$,$e^3$}
]
\addplot[mark=square*, color=Violet,  mark size=1, mark options={fill=Violet} ] table [y=dmJ, x=a] {6+_A1_I_12fm.txt};
\addplot[mark=diamond*, color=Turquoise,  mark size=1.3, mark options={fill=Turquoise} ] table [y=dmJ, x=a] {6+_A2_I_12fm.txt};
\addplot[mark=pentagon*, color=Orange,  mark size=1.3, mark options={fill=Orange} ] table [y=dmJ, x=a] {6+_E_I_12fm.txt};
\addplot[mark=triangle*, color=Magenta,  mark size=1.3, mark options={fill=Magenta} ] table [y=dmJ, x=a] {6+_T1_I_12fm.txt};
\addplot[mark=triangle*, color=PineGreen,  mark size=1.3, mark options={fill=PineGreen, rotate=180} ] table [y=dmJ, x=a] {6+_T2_I_1_12fm.txt};
\addplot[mark=triangle*, color=PineGreen,  mark size=1.3, mark options={fill=PineGreen, rotate=180} ] table [y=dmJ, x=a] {6+_T2_I_2_12fm.txt};
\end{axis}
\end{tikzpicture}}
\end{minipage}\hfill
\begin{minipage}[c]{0.37\columnwidth}
\caption{Difference between the average value and the expected eigenvalue of the squared angular momentum for the $4_2^+$ (left) and the $6_1^+$ states (right) as a function of the lattice spacing. The same convention on the markers for the cubic group irreps of Figs.~\ref{F-7.0-13}-\ref{F-7.0-19} is used.}\label{F-7.0-24}
\end{minipage}
\end{figure}

However, the extent of the region where this approximation can be successfully applied depends on the onset point of the step growing or falling parts of the squared angular momentum curves. Since the $\alpha$-$\alpha$ average distance is larger for the $6_1^+$ than for the $4_2^+$, this interval is wider for the former and the positive exponential behaviour more evident.\\

\section{\textsf{The \ce{^{12}C} nucleus}}\label{S-8.0}

After having investigated finite volume and discretization effects in the low-lying spectrum of \ce{^8Be}, we now focus on the analysis of the bound states of a system three interacting $\alpha$ particles in the same framework, the \ce{^{12}C} nucleus. Due to the particular choice of the parameters of $V_{AB}$, the addition of the attractive phenomenologic three-body potential  in Eq.~\eqref{2.1-04} permits us to reproduce the binding energy of this nucleus. Although the ground state is tuned on the energy of the Hoyle state rather than on the $3\alpha$ decay threshold, in fact, the binding energy can be still recovered, provided the well-established positive gap between the latter two is added to the ground state energy, $E_{0^+}$ in Eq.~\eqref{5.2-01}. \\
\begin{figure}[hb!]
\scalebox{0.99}{
\begin{tikzpicture}
\begin{axis}[
	width = 15.3cm,
	height = 6.9cm,
	xmin = 0,
	xmax = 30,
	ymin = -50,
	ymax = 125,
	xlabel={\small $N$},
	ylabel={\small $BE(6,6)$ $[\mathrm{MeV}]$},
	xtick={0,2,4,6,8,10,12,14,16,18,20,22,24,26,28,30},
          ytick={-50,-25,0,25,50,75,100,125},
	legend style ={at={(0.66,0.45)}, anchor=north west, draw=black,fill=white,align=left},
	legend entries ={\textcolor{Black}{ $0_{A_1}^{+}$ - a = 0.75 fm},\textcolor{OliveGreen}{Experimental},\textcolor{Blue}{Lattice}}
]
\addlegendimage{empty legend};
\addlegendimage{OliveGreen,dashed};
\addlegendimage{Blue, mark=square*, mark size = 1, mark options={fill=Blue}};
\addplot[dashed, color=OliveGreen, domain=0:30, samples=100] {92.1598590065378};
\addplot[mark=square*, color=Blue,  mark size=1, mark options={fill=Blue} ] table [y=BEexp, x=L] {12C_A1_BE_075.txt};
\end{axis}
\end{tikzpicture}}
\caption{Binding energy of the \ce{^{12}C} as a function of N, for lattices with spacing $a = 0.75$~fm.}\label{F-8.0-01}
\end{figure}

Even if the behaviour of lattice binding energy (cf. Eq.~\eqref{5.2-01}) with the box size $N$ is all in all analogous to the one of Beryllium, two digit accuracy with the observational counterpart ($\approx 92.16$~MeV) of the former is finally reached at $N = 24$ and spacing equal to $0.75$~fm. Therefore, finite volume effects can be reasonably neglected for our purpose in lattices with size $Na \geq 18$~fm. \\
Differently from the preceeding case, there is no more isomorphism between parity and particle permutation group, $\mathcal{S}_3$, a six element non-abelian group bearing also a 2-dimensional irreducible representation (diagrammatically ${\tiny\yng(2,1)}$). As a consequence, besides bosonic and fermionic symmetry, the eigenstates of the lattice Hamiltonian $\mathcal{H}_r$ can be now symmetric with respect to the exchange of a pair of particles and antisymmetric with respect to the transposition of another couple of them, resulting in the appearance of \textit{unphysical} parastatistic eigenstates.  \\
Given the duration and memory consumption of the eigenvector extraction process and being parity itself uncorrelated with particle exchange symmetry, also projectors on parity and $\mathcal{C}_4$ irreps have been incorporated in the iteration loop, thus minimizing the number of eigenvectors involved in the Gram-Schmidt othogonalization. On the other hand, the matrix $\mathscr{R}_z^{\pi/2}$, to be simultaneously diagonalized together with the Hamiltonian (cf. Eq.~\eqref{6.1-01}), has been excluded from the iteration loop.\\
\begin{table}[ht!]
\begin{minipage}[c]{0.48\columnwidth}
\begin{footnotesize}
\begin{tabular}{c|cc|c|c|c}
\toprule
$E$ [MeV] & $\Gamma$ & $I_z$ & $\mathscr{P}$ & $\mathcal{S}_3$ & $\langle \mathcal{L}_{\rm tot}^2\rangle$ $[\hbar^2]$\\
\midrule
$\mathbf{-7.698420}$ & $\mathbf{A_1}$ & $\mathbf{0}$ & $\boldsymbol{+}$ & $\mathbf{{\tiny\yng(3)}}$ & $\mathbf{0.373}$\\
\multirow{3}{1.5cm}{\centering{$-6.306062$}} & \multirow{3}{0.75cm}{\centering{$T_1$}} & 0 & \multirow{3}{0.25cm}{\centering{$-$}} & \multirow{3}{0.5cm}{\centering{${\tiny\yng(2,1)}$}} & \multirow{3}{0.75cm}{\centering{$2.429$}}\\
&  & 1 & & &\\
&  & 3 & & &\\
\multirow{3}{1.5cm}{\centering{$-5.457046$}} & \multirow{3}{0.75cm}{\centering{$T_1$}} & 0 & \multirow{3}{0.25cm}{\centering{$+$}} & \multirow{3}{0.5cm}{\centering{${\tiny\yng(1,1,1)}$}} & \multirow{3}{0.75cm}{\centering{2.466}}\\
&  & 1 & & &\\
&  & 3 & & &\\
\multirow{3}{1.5cm}{\centering{$-4.550694$}} & \multirow{3}{0.75cm}{\centering{$T_2$}} & 1 & \multirow{3}{0.25cm}{\centering{$+$}} & \multirow{3}{0.5cm}{\centering{${\tiny\yng(2,1)}$}} & \multirow{3}{0.75cm}{\centering{6.612 }}\\
&  & 2 & & &\\
&  & 3 & & &\\
\multirow{2}{1.5cm}{\centering{$-4.470975$}} & \multirow{2}{0.75cm}{\centering{$E$}} & 0 & \multirow{2}{0.25cm}{\centering{$+$}} & \multirow{2}{0.5cm}{\centering{${\tiny\yng(2,1)}$}} & \multirow{2}{0.75cm}{\centering{$6.175$}}\\
&  & 2 & & &\\
\multirow{2}{1.5cm}{\centering{$\mathbf{-3.420394}$}} & \multirow{2}{0.75cm}{\centering{$\mathbf{E}$}} & $\mathbf{0}$ & \multirow{2}{0.25cm}{\centering{$\boldsymbol{+}$}} & \multirow{2}{0.5cm}{\centering{$\mathbf{{\tiny\yng(3)}}$}} & \multirow{2}{0.75cm}{\centering{$\mathbf{6.729}$}}\\
&  & $\mathbf{2}$ & & &\\
\multirow{3}{1.5cm}{\centering{$\mathbf{-3.177991}$}} & \multirow{3}{0.75cm}{\centering{$\mathbf{T_2}$}} & $\mathbf{1}$ & \multirow{3}{0.25cm}{\centering{$\boldsymbol{+}$}} & \multirow{3}{0.5cm}{\centering{$\mathbf{{\tiny\yng(3)}}$}} & \multirow{3}{0.75cm}{\centering{$\mathbf{6.824}$}}\\
&  & $\mathbf{2}$ & & &\\
&  & $\mathbf{3}$ & & &\\
\multirow{3}{1.5cm}{\centering{$-2.873875$}} & \multirow{3}{0.75cm}{\centering{$T_2$}} & 1 & \multirow{3}{0.25cm}{\centering{$-$}} & \multirow{3}{0.5cm}{\centering{${\tiny\yng(2,1)}$}} & \multirow{3}{0.75cm}{\centering{7.086}}\\
&  & 2 & & &\\
&  & 3 & & &\\
$-2.862931$ & $A_1$ & 0 & $+$ & ${\tiny\yng(2,1)}$ & $2.074$\\
\bottomrule
\end{tabular}
\end{footnotesize}
\end{minipage}\hfill
\begin{minipage}[c]{0.48\columnwidth}
\begin{footnotesize}
\begin{tabular}{c|cc|c|c|c}
\toprule
$E$ [MeV] & $\Gamma$ & $I_z$ & $\mathscr{P}$ & $\mathcal{S}_3$ & $\langle \mathcal{L}_{\rm tot}^2\rangle$ $[\hbar^2]$\\
\midrule
$-2.686463$ & $A_1$ & 0 & $+$ & ${\tiny\yng(3)}$ & $1.690$\\
\multirow{3}{1.5cm}{\centering{$-2.637041$}} & \multirow{3}{0.75cm}{\centering{$T_1$}} & 0 & \multirow{3}{0.25cm}{\centering{$-$}} & \multirow{3}{0.5cm}{\centering{${\tiny\yng(1,1,1)}$}} & \multirow{3}{0.75cm}{\centering{8.320}}\\
&  & 1 & & &\\
&  & 3 & & &\\
\multirow{3}{1.5cm}{\centering{$-2.483865$}} & \multirow{3}{0.75cm}{\centering{$T_2$}} & 1 & \multirow{3}{0.25cm}{\centering{$-$}} & \multirow{3}{0.5cm}{\centering{${\tiny\yng(1,1,1)}$}} & \multirow{3}{0.75cm}{\centering{$12.603$}}\\
&  & 2 & & &\\
&  & 3 & & &\\
$-2.297536$ & $A_2$ & 2 & $-$ & ${\tiny\yng(1,1,1)}$ & $12.493$\\
\multirow{3}{1.5cm}{\centering{$-2.281911$}} & \multirow{3}{0.75cm}{\centering{$T_1$}} & 0 & \multirow{3}{0.25cm}{\centering{$-$}} & \multirow{3}{0.5cm}{\centering{${\tiny\yng(1,1,1)}$}} & \multirow{3}{0.75cm}{\centering{7.943}}\\
&  & 1 & & &\\
&  & 3 & & &\\
\multirow{3}{1.5cm}{\centering{$\mathbf{-1.981923}$}} & \multirow{3}{0.75cm}{\centering{$\mathbf{T_2}$}} & $\mathbf{1}$ & \multirow{3}{0.25cm}{\centering{$\boldsymbol{-}$}} & \multirow{3}{0.5cm}{\centering{$\mathbf{{\tiny\yng(3)}}$}} & \multirow{3}{0.75cm}{\centering{$\mathbf{12.536}$}}\\
&  & $\mathbf{2}$ & & &\\
&  & $\mathbf{3}$ & & &\\
\multirow{3}{1.5cm}{\centering{$\mathbf{-1.797457}$}} & \multirow{3}{0.75cm}{\centering{$\mathbf{T_1}$}} & $\mathbf{0}$ & \multirow{3}{0.25cm}{\centering{$\boldsymbol{-}$}} & \multirow{3}{0.5cm}{\centering{$\mathbf{{\tiny\yng(3)}}$}} & \multirow{3}{0.75cm}{\centering{$\mathbf{12.360}$}}\\
&  & $\mathbf{1}$ & & &\\
&  & $\mathbf{3}$ & & &\\
$\mathbf{-1.779066}$ & $\mathbf{A_2}$ & $\mathbf{2}$ & $\boldsymbol{-}$ & $\mathbf{{\tiny\yng(3)}}$ & $\mathbf{12.384}$\\
\multirow{3}{1.5cm}{\centering{$-1.706789$}} & \multirow{3}{0.75cm}{\centering{$T_1$}} & 0 & \multirow{3}{0.25cm}{\centering{$-$}} & \multirow{3}{0.5cm}{\centering{${\tiny\yng(2,1)}$}} & \multirow{3}{0.75cm}{\centering{$4.441$}}\\
&  & 1 & & &\\
&  & 3 & & &\\
\bottomrule
\end{tabular}
\end{footnotesize}
\end{minipage}
\caption{Sample of the spectrum of the \ce{^{12}C} lattice Hamiltonian with $N = 15$ and $a = 1.00$~fm, consisting of the 17 lowest degenerate energy multiplets. The three angular momentum multiplets of interest, $0_1^+$, $2_1^+$ and $3_1^-$ are highlighted in bold. Cubic group multiplets labeled by the Young Tableau with three unaligned boxes appear twice in the spectrum, since the irrep of the permutation group $\mathcal{S}_3$ according to which they transform is 2-dimensional.} \label{T-8.0-01}
\end{table}

Since the actual nucleus is naturally bound, no artificial increase of the Ali-Bodmer potential attractive parameter is needed for the investigation of finite-volume and discretization effects in the lowest bound eigenstates. By sampling the sprectrum of the relative Hamiltonian with $N = 15$ and $a = 1.0$~fm, see Tab.~\ref{T-8.0-01}, and the one with $N=20$ and $a = 0.9$~fm to a smaller extent, it turns out that this nucleus possesses seven SO(3) multiplets of completely-symmetric bound states, namely three $0^+$, a $1^-$, two $2^+$ and a $3^-$,  in the continuum and infinite-volume limit. Experimentally, only a $2^+$ line at $4.44$~MeV is found to lie below the $3\alpha$ decay threshold \cite{ASK90}, while the lowest $3^-$ and $1^-$ observed excitations result to be unbound by circa $1.9$ and $2.2$~MeV respectively.\\
Starting from this set of bound eigenstates, we choose to restrict our analysis to the ground state at $-7.65$~MeV, the $2_1^+$ state at $-3.31$~MeV and the $3_1^-$ multiplet at $-1.80$~MeV, decomposing into an $A_2$, a $T_1$ and a $T_2$ multiplet with respect to the cubic group. \\
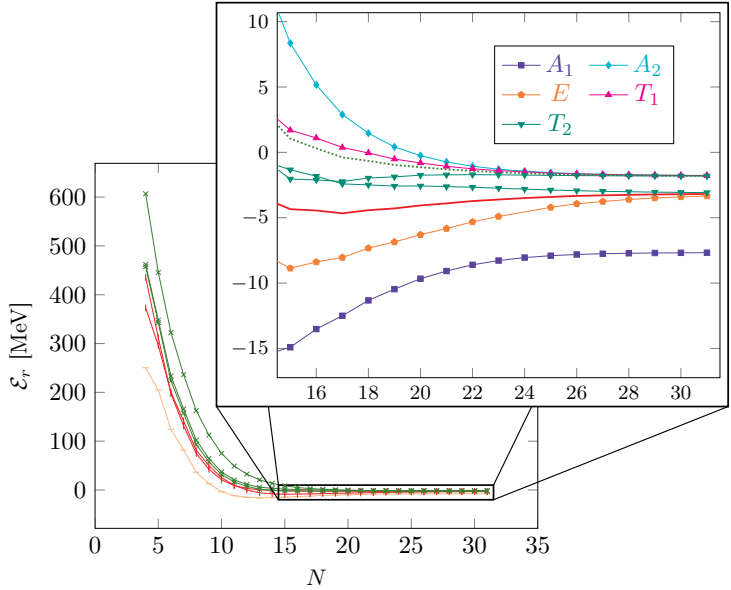
\begin{figure}[ht!]
\begin{center}
\begin{minipage}[c]{0.62\columnwidth}
\scalebox{0.85}{
\begin{tikzpicture}
\begin{axis}[
	name= bigEn,
	xmin = 0,
	xmax = 35,
	xlabel={\small $N$},
	ylabel={\small $\mathcal{E}_r$ [MeV]},
	xtick={0,5,10,15,20,25,30,35},
	ytick={0,100,200,300,400,500,600}
]
\addplot[mark=-, color=Apricot,  mark size=1.5, mark options={fill=Apricot}] table [y=E, x=L] {12C_0+_I_A1_050.txt};
\addplot[mark=|, color=Red,  mark size=1.5, mark options={fill=Red}] table [y=E, x=L] {12C_2+_I_E_050.txt};
\addplot[mark=|, color=Red,  mark size=1.5, mark options={fill=Red}] table [y=E, x=L] {12C_2+_I_T2_050.txt};
\addplot[mark=x, color=OliveGreen,  mark size=1.5, mark options={fill=OliveGreen}] table [y=E, x=L] {12C_3-_I_A2_050.txt};
\addplot[mark=x, color=OliveGreen,  mark size=1.5, mark options={fill=OliveGreen}] table [y=E, x=L] {12C_3-_I_T1_050.txt};
\addplot[mark=x, color=OliveGreen,  mark size=1.5, mark options={fill=OliveGreen}] table [y=E, x=L] {12C_3-_I_T2_050.txt};
\coordinate (c1) at (axis cs:14.5,-20);
\coordinate (c2) at (axis cs:31.5,10);
\coordinate (c3) at (axis cs:14.5,10);
\coordinate (c4) at (axis cs:31.5,-20);
\draw [line width = 0.75 pt] (c1) rectangle (axis cs:31.5,10);
\end{axis}
\coordinate (d1) at (1.88 cm,1.9 cm);
\coordinate (d2) at (9.8 cm, 8.2 cm);
\coordinate (d3) at (1.88 cm, 8.2 cm);
\coordinate (d4) at (9.8 cm, 1.9cm);
\draw [line width= 0.5 pt] (c1) -- (d1);
\draw [line width= 0.5 pt] (c2) -- (d2);
\draw [line width= 0.5 pt] (c3) -- (d3);
\draw [line width= 0.5 pt] (c4) -- (d4); 
\draw [line width = 0.75 pt, fill= white] (d1) rectangle (d2);
\begin{axis}[
	at = {($(2.82 cm,2.35 cm)$)},
	name= smaEn,
          axis background/.style={fill=white},
	xmin = 14.5,
	xmax = 31.5,
	xtick={16,18,20,22,24,26,28,30},
	xticklabel style = {font=\footnotesize},
	yticklabel style = {font=\footnotesize},
	legend columns = 2,
	legend style ={at={(0.49,0.91)}, anchor=north west, draw=black,fill=white,align=left,/tikz/column 2/.style={
	column sep=4pt}},
	legend entries ={\textcolor{Violet}{$A_1$},\textcolor{Turquoise}{$A_2$},\textcolor{Orange}{$E$},\textcolor{Magenta}{$T_1$},\textcolor{PineGreen}{$T_2$}},
]
\addlegendimage{Violet,mark=square*, mark size=1.2, mark options={fill=Violet}};
\addlegendimage{Turquoise, mark=diamond*, mark size=1.5, mark options={fill=Turquoise}};
\addlegendimage{Orange, mark=pentagon*, mark size=1.5, mark options={fill=Orange}};
\addlegendimage{Magenta, mark=triangle*, mark size=1.5, mark options={fill=Magenta}};
\addlegendimage{PineGreen, mark=triangle*, mark size=1.5, mark options={fill=PineGreen, rotate=180}};
\addplot[mark=square*, color=Violet,  mark size=1.2, mark options={fill=Violet}] table [y=E, x=L] {12C_0+_I_A1_050.txt};
\addplot[mark=pentagon*, color=Orange,  mark size=1.5, mark options={fill=Orange}] table [y=E, x=L] {12C_2+_I_E_050.txt};
\addplot[mark=triangle*, color=PineGreen,  mark size=1.5, mark options={fill=PineGreen,rotate=180}] table [y=E, x=L] {12C_2+_I_T2_050.txt};
\addplot[mark=diamond*, color=Turquoise,  mark size=1.5, mark options={fill=Turquoise}] table [y=E, x=L] {12C_3-_I_A2_050.txt};
\addplot[mark=triangle*, color=Magenta,  mark size=1.5, mark options={fill=Magenta} ] table [y=E, x=L] {12C_3-_I_T1_050.txt};
\addplot[mark=triangle*, color=PineGreen,  mark size=1.5, mark options={fill=PineGreen,rotate=180} ] table [y=E, x=L] {12C_3-_I_T2_050.txt};
\addplot[color=Red,  solid, thick] table [y=E, x=L] {12C_2+_I_MAVG_050.txt};
\addplot[color=OliveGreen,  densely dotted, thick] table [y=E, x=L] {12C_3-_I_MAVG_050.txt};
\end{axis}
\end{tikzpicture}}
\end{minipage}
\begin{minipage}[c]{0.29\columnwidth}
\caption{Behaviour of the energies of the lowest $0^+$ (horizontal bars), $2^+$ (vertical bars) and $3^-$ (crosses) bound eigenstates as a function of the box size N for $a = 0.50$~fm.  As expected, the eigenenergies associated to states belonging to the same irrep of SO(3) but to different irreps of $\mathcal{O}$ become almost degenerate at the infinite-size limit. The same convention on the markers for the cubic group irreps adopted in the Figures of Sec.~\ref{S-7.0} is understood.}\label{F-8.0-02}
\end{minipage}
\end{center}
\end{figure}

Analogously to the Beryllium case, we fix the lattice spacing in such a way to reduce the discretization errors to less than two decimal digits in the infinite-volume limit ($Na \gtrsim 19$~fm) for all the multiplets of interest and plot the behaviour of the energy as a function of the lattice size $N$ (cf. Fig.~\ref{F-8.0-02}). The evolution curve for the energy of the ground state follows a similar path to the one of the $0^+$ states of \ce{^8Be}: after a minimum at $Na \approx 6$~fm, the continuum and infinite-volume eigenvalue is reached asymptotically from below, as prescribed by the FVEC formulas from Ref.~\cite{KLH11} for a two-body system. \\
\begin{figure}[ht!]
\begin{minipage}[c]{0.48\columnwidth}
\scalebox{0.85}{
\begin{tikzpicture}
\begin{axis}[
	xmin = 0,
	xmax = 32.5,
	xlabel={\small $N$},
	ylabel={\small $\mathcal{R}$ $[\mathrm{fm}]$},
	xtick={0,5,10,15,20,25,30},
	ytick={0.5,1,1.5,2,2.5,3,3.5,4,4.5,5},
	legend style ={at={(0.69,0.38)}, anchor=north west, draw=black,fill=white,align=left},
	legend entries ={\textcolor{Violet}{$A_1$},\textcolor{Orange}{$E$},\textcolor{PineGreen}{$T_2$}},
]
\addlegendimage{Violet,mark=square*,mark size=1,mark options={fill=Violet}};
\addlegendimage{Orange,mark=pentagon*,mark size=1.3,mark options={fill=Orange}};
\addlegendimage{PineGreen,mark=triangle*,mark size=1.3,mark options={fill=PineGreen,rotate=180}};
\addplot[mark=square*, color=Violet, mark size=1, mark options={fill=Violet}] table [y=ravg, x=L] {12C_0+_I_A1_050.txt};
\addplot[mark=pentagon*, color=Orange,  mark size=1.3, mark options={fill=Orange} ] table [y=ravg, x=L] {12C_2+_I_E_050.txt};
\addplot[mark=triangle*, color=PineGreen,  mark size=1.3, mark options={fill=PineGreen,rotate=180} ] table [y=ravg, x=L] {12C_2+_I_T2_050.txt};
\addplot[color=Red, solid] table [y=ravg, x=L] {12C_2+_I_MAVG_050.txt};
\end{axis}
\end{tikzpicture}}
\caption{Behaviour of the average interparticle distance for the $0_1^+$ and $2_1^+$ multiplets as a function of the lattice size. Due to the broader spatial distribution of the $2_E^+$ and $T_2^+$ wavefunctions, the finite-volume effects on the average values of the $\alpha-\alpha$ separation distance remain sensitive ($\approx 0.24$~fm at $N=31$).}\label{F-8.0-03}
\end{minipage}\hfill
\begin{minipage}[c]{0.48\columnwidth}
\scalebox{0.85}{
\begin{tikzpicture}
\begin{axis}[
	xmin = 0,
	xmax = 32.5,
	xlabel={\small $N$},
	ylabel={\small $\mathcal{R}$ $[\mathrm{fm}]$},
	xtick={0,5,10,15,20,25,30},
	ytick={0.5,1,1.5,2,2.5,3,3.5,4,4.5,5},
	legend style ={at={(0.69,0.38)}, anchor=north west, draw=black,fill=white,align=left},
	legend entries ={\textcolor{Turquoise}{$A_2$},\textcolor{Magenta}{$T_1$},\textcolor{PineGreen}{$T_2$}},
]
\addlegendimage{Turquoise,mark=diamond*,mark size=1.3,mark options={fill=Turquoise}};
\addlegendimage{Magenta,mark=triangle*,mark size=1.3,mark options={fill=Magenta}};
\addlegendimage{PineGreen, mark=triangle*,mark size=1.3,mark options={fill=PineGreen,rotate=180}};
\addplot[mark=diamond*, color=Turquoise,  mark size=1.3, mark options={fill=Turquoise} ] table [y=ravg, x=L] {12C_3-_I_A2_050.txt};
\addplot[mark=triangle*, color=Magenta,  mark size=1.3, mark options={fill=Magenta} ] table [y=ravg, x=L] {12C_3-_I_T1_050.txt};
\addplot[mark=triangle*, color=PineGreen,  mark size=1.3, mark options={fill=PineGreen, rotate=180} ] table [y=ravg, x=L] {12C_3-_I_T2_050.txt};
\addplot[color=OliveGreen, densely dotted] table [y=ravg, x=L] {12C_3-_I_MAVG_050.txt};
\end{axis}
\end{tikzpicture}}
\caption{Behaviour of the average interparticle distance for the $3_1^-$ multiplet of states as a function of the lattice size. As expected, both the three members of this SO(3) multiplet converge to same average values of the $\alpha$-$\alpha$ separation distance, that at $N=31$ coincide within $0.05$~fm accuracy.}\label{F-8.0-04}
\end{minipage}
\end{figure}

In particular, an agreement within one decimal digit with the fitted value of $-7.65$~MeV is already reached at $Na \approx 13$~fm, whereas the overlap with all the meaningful digits is going to be achieved at $Na \approx 16.5$~fm.  However, the $2^+$ doublet is expected to become degenerate within one-digit precision only at $Na \approx 16$~fm, due to a broader spatial distribution of the $E$ and $T_2$ eigenfunctions. The average separation between the $\alpha$ particles in the equilateral triangular equilibrium configuration, in fact, amounts approximately to $4.65$~fm  for the latter states and to $4.05$~fm  for the $0_1^+$ state, see Fig.~\ref{F-8.0-03}. Furthermore, in the $3^-$ energy multiplet the $T_1$ and the $A_2$ states approach the continuum and infinite-volume energy from above, whereas the $T_2$ multiplet requires corrections of opposite sign, see Fig.~\ref{F-8.0-02}. \\
Although analytical formulas for the leading order FVEC for three body systems are still unknown, the sign of these corrections for the $\ell = 3$ multiplet seem coincide with the one of the FVECs for a multiplet of bound eigenstates with the same angular momentum of a two-body system. Besides, rotational symmetry for this multiplet is already restored within one decimal digit accuracy for $Na \approx 14$~fm, due to the more localized spatial distribution of the wavefunctions, see Fig.~\ref{F-8.0-04}. The infinite-volume value of the average $\alpha-\alpha$ distance for the states of these multiplets is $4.40$~fm, in between the one of the $0_1^+$ and the $3_1^-$ multiplets.\\
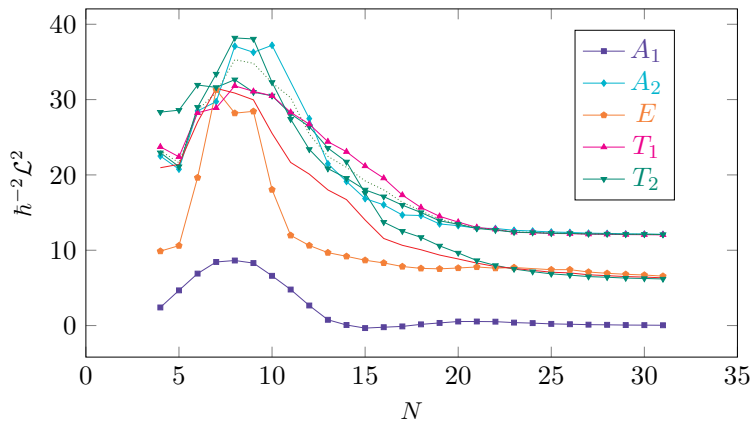
\begin{figure}[hb!]
\begin{minipage}[c]{0.60\columnwidth}
\scalebox{0.92}{
\begin{tikzpicture}
\begin{axis}[
	width = 10.9cm,
	height = 6.6cm,
	xmin = 0,
	xmax = 35,
	xlabel={\small $N$},
	ylabel={\small $\hbar^{-2}\mathcal{L}^2$},
	xtick={0,5,10,15,20,25,30,35},
	ytick={0,10,20,30,40},
	legend style ={at={(0.75,0.94)}, anchor=north west, draw=black,fill=white,align=left},
	legend entries ={\textcolor{Violet}{$A_1$},\textcolor{Turquoise}{$A_2$},\textcolor{Orange}{$E$},\textcolor{Magenta}{$T_1$},\textcolor{PineGreen}{$T_2$}},
]
\addlegendimage{Violet,mark=square*,mark size=1,mark options={fill=Violet}};
\addlegendimage{Turquoise, mark=diamond*,mark size=1.3,mark options={fill=Turquoise}};
\addlegendimage{Orange, mark=pentagon*,mark size=1.3,mark options={fill=Orange}};
\addlegendimage{Magenta, mark=triangle*,mark size=1.3,mark options={fill=Magenta}};
\addlegendimage{PineGreen, mark=triangle*,mark size=1.3,mark options={fill=PineGreen,rotate=180}};
\addplot[mark=square*, color=Violet,  mark size=1, mark options={fill=Violet} ] table [y=J, x=L] {12C_0+_I_A1_050.txt};
\addplot[mark=pentagon*, color=Orange,  mark size=1.3, mark options={fill=Orange} ] table [y=J, x=L] {12C_2+_I_E_050.txt};
\addplot[mark=triangle*, color=PineGreen,  mark size=1.3, mark options={fill=PineGreen,rotate=180}] table [y=J, x=L] {12C_2+_I_T2_050.txt};
\addplot[mark=diamond*, color=Turquoise,  mark size=1.3, mark options={fill=Turquoise} ] table [y=J, x=L] {12C_3-_I_A2_050.txt};
\addplot[mark=triangle*, color=Magenta,  mark size=1.3, mark options={fill=Magenta} ] table [y=J, x=L] {12C_3-_I_T1_050.txt};
\addplot[mark=triangle*, color=PineGreen,  mark size=1.3, mark options={fill=PineGreen,rotate=180} ] table [y=J, x=L] {12C_3-_I_T2_050.txt};
\addplot[color=Red, solid] table [y=J, x=L] {12C_2+_I_MAVG_050.txt};
\addplot[color=OliveGreen, densely dotted] table [y=J, x=L] {12C_3-_I_MAVG_050.txt};
\end{axis}
\end{tikzpicture}}
\end{minipage}\hfill
\begin{minipage}[c]{0.35\columnwidth}
\caption{Average value of the squared angular momentum for the six bound states as a function of the lattice size.  After displaying a peak in the interaction region, the average values of the squared angular momentum for the $0_1^+$, $2_1^+$ and $3_1^-$ states converge to the eigenvalues of $\mathcal{L}^2$ equal to $0$, $6$ and $12$  units of $\hbar^2$ respectively in the infinite-volume limit. The multiplet averages of the $2_1^+$ and $3_1^-$ states are denoted by solid and densely dotted lines.}\label{F-8.0-05}
\end{minipage}
\end{figure}

The average values of the angular momentum as function of the lattice size $N$ for both the three SO(3) multiplets considered display a well-developed maximum at about $N=7$, eventually followed by a shallow minimum lying between $N=15 $ and $N=20$. In particular, the angular momentum of the $0_1^+$ state reaches the expected asymptotic value from below, as observed in the beryllium case (cf. Fig.~\ref{F-7.0-03}), while the $2_E^+$ and $2_{T_2}$ multiplets approach the continuum and infinite volume limit from below and above, respectively. This suggests the sign of the leading order finite volume corrections for the eigenvalues of the $\mathcal{L}^2$ operator. Although the $\mathcal{L}^2$ evolution curves for the three SO(3) multiplets resemble the ones of the $0_1^+$ and $2_1^+$ states of the \ce{^8Be} nucleus, cf. Fig.~\ref{F-7.0-03}, the $E$ and the $A_2$ levels for $N \lesssim 14$ seem to be heavily affected by level crossings with adiacent energy states (note that a spike marking the $2_E^+$ evolution curve at $N=11$  has been omitted).\\
Next, we concentrate the attention to the systematic errors due to finite lattice spacing. By fixing the size of the lattice at $Na \geq 19$~fm in order to reduce finite-volume errors to the third decimal digit, we inspect the behaviour of the energy eigenvalues of the aforementioned $0_1^+$, $2_1^+$ and $3_1^-$ multiplets for lattice spacings $a$ ranging from $0.65$ to $3.50$~fm. From the plot in Fig.~\ref{F-8.0-06}, the $0_{A_1}^+$ state already equates the fitted energy eigenvalue of $-7.65$~MeV within one and two decimal digit precision at $a \approx 1.15$ and $1.00$~fm, whereas the two members of the $2_1^+$ multiplet become degenerate within the same accuracy for $a=1.30$ and $0.75$~fm respectively. \\
\begin{figure}[ht!]
\begin{minipage}[c]{0.47\columnwidth}
\scalebox{0.85}{
\begin{tikzpicture}
\begin{axis}[
	height= 8.9cm,
	width = 8.5cm,
	xmin = 0,
	xmax = 4.0,
	ymin = -16,
	ymax = 0,
	xlabel={\small $a$ [fm]},
	ylabel={\small $\mathcal{E}_r$ [MeV]},
	xtick={0,0.5,1.0,1.5,2.0,2.5,3.0,3.5,4.0},
	ytick={-16,-14,-12,-10,-8,-6,-4,-2,0},
	legend columns=1,
	legend style ={at={(0.19,0.35)}, anchor=north west, draw=black,fill=white,align=left,/tikz/column 1/.style={
                column sep=4pt}},
	legend entries ={\textcolor{Violet}{$A_1$},\textcolor{Orange}{$E$},\textcolor{PineGreen}{$T_2$}},
]
\addlegendimage{Violet,mark=square*,mark size=1,mark options={fill=Violet}};
\addlegendimage{Orange, mark=pentagon*,mark size=1.3, mark options={fill=Orange}};
\addlegendimage{PineGreen, mark=triangle*,mark size =1.3,mark options={fill=PineGreen,rotate=180}};
\addplot[mark=square*, color=Violet,  mark size=1, mark options={fill=Violet} ] table [y=E, x=a] {12C_0+_I_A1_19fm.txt};
\addplot[mark=pentagon*, color=Orange,  mark size=1.3, mark options={fill=Orange} ] table [y=E, x=a] {12C_2+_I_E_19fm.txt};
\addplot[mark=triangle*, color=PineGreen,  mark size=1.3, mark options={fill=PineGreen,rotate=180} ] table [y=E, x=a] {12C_2+_I_T2_19fm.txt};
\addplot[color=Red, solid] table [y=E, x=a] {12C_2+_I_MAVG_19fm.txt};
\end{axis}
\end{tikzpicture}}
\caption{Behaviour of the energies of the $0_1^+$ and $2_1^+$ eigenstates as a function of the lattice spacing for $Na \geq 19$~fm. Although the multiplet-averaged $2_1^+$ energy (solid line) improves the convergence rate to the continuum and infinite-volume counterpart, for $a \gtrsim 2.0$~fm discretization corrections amount to more than 33\% of the asymptotic energy eigenvalue.}\label{F-8.0-06}
\end{minipage}\hfill
\begin{minipage}[c]{0.47\columnwidth}
\scalebox{0.85}{
\begin{tikzpicture}
\begin{axis}[
	xmin = 0,
	xmax = 4.0,
	xlabel={\small $a$ [fm]},
	ylabel={\small $\mathcal{R}$ [fm]},
	xtick={0,0.5,1.0,1.5,2.0,2.5,3.0,3.5,4.0},
	ytick={3.0,3.2,3.4,3.6,3.8,4.0,4.2,4.4,4.6,4.8},
	legend columns=1,
	legend style ={at={(0.15,0.44)}, anchor=north west, draw=black,fill=white,align=left,/tikz/column 1/.style={
                column sep=4pt}},
	legend entries ={\textcolor{Violet}{$A_1$},\textcolor{Orange}{$E$},\textcolor{PineGreen}{$T_2$}},
]
\addlegendimage{Violet,mark=square*,mark size=1,mark options={fill=Violet}};
\addlegendimage{Orange, mark=pentagon*,mark size=1.3,mark options={fill=Orange}};
\addlegendimage{PineGreen, mark=triangle*,mark size=1.3,mark options={fill=PineGreen, rotate=180}};
\addplot[mark=square*, color=Violet,  mark size=1, mark options={fill=Violet} ] table [y=ravg, x=a] {12C_0+_I_A1_19fm.txt};
\addplot[mark=pentagon*, color=Orange,  mark size=1.3, mark options={fill=Orange} ] table [y=ravg, x=a] {12C_2+_I_E_19fm.txt};
\addplot[mark=triangle*, color=PineGreen,  mark size=1.3, mark options={fill=PineGreen,rotate=180} ] table [y=ravg, x=a] {12C_2+_I_T2_19fm.txt};
\addplot[color=Red, solid] table [y=ravg, x=a] {12C_2+_I_MAVG_19fm.txt};
\end{axis}
\end{tikzpicture}}
\caption{Behaviour of the average $\alpha-\alpha$ distance of the $0_1^+$ and $2_1^+$ eigenstates as a function of the lattice spacing for $Na \geq 19$~fm. It is worth observing that the values of $\mathcal{R}$ to which the $2_E^+$ and the $2_{T_2}^+$ states seem to converge do not coincide by an amount of $0.06$~fm. It is possible that this small bias is due to residual finite-volume effects, since, as noticed in Fig.~\ref{F-8.0-03}, for $a=0.5$~fm and $N=31$ the two average interparticle distances differ still by $0.24$~fm. Nevertheless, the other observables concerning this angular momentum multiplet, cf. Figs~\ref{F-8.0-06} and \ref{F-8.0-10}, perhaps less sensitive to finite-volume effects, do not display this behaviour in the small-spacing limit.}\label{F-8.0-07}
\end{minipage}
\end{figure}

\begin{figure}[ht!]
\hspace{1.7cm}
\begin{minipage}{0.46\columnwidth}
\includegraphics[height=0.83\columnwidth]{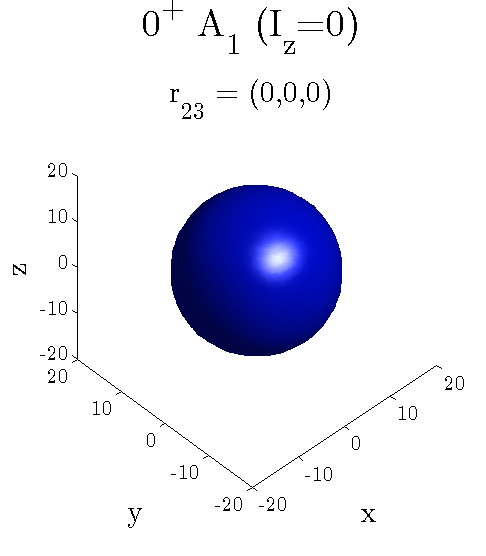}
\end{minipage}
\begin{minipage}{0.49\columnwidth}
\includegraphics[height=0.81\columnwidth]{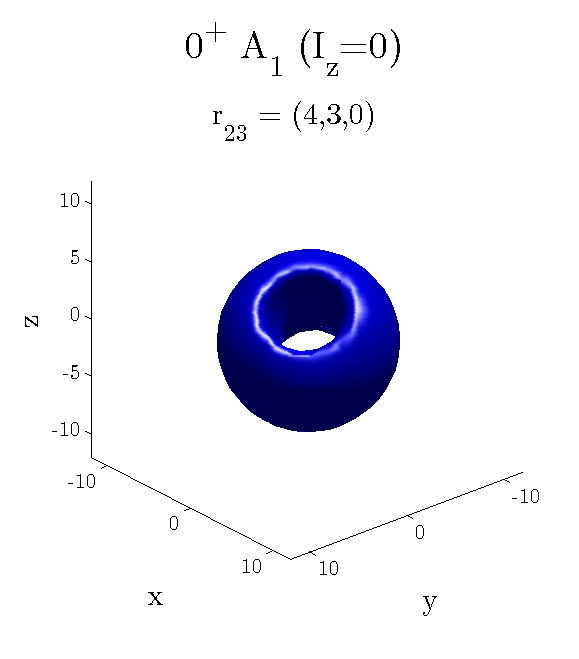}
\end{minipage}
\caption{Spatial distribution of the PDF of the $0_1^+$ $A_1$ state with $I_z=0$ in the configuration space slices with $\mathbf{r}_{23} = (0,0,0)$ (left) and $\mathbf{r}_{23} = (4,3,0)$ (right). The outer isohypsic surfaces of the former plot correspond to a probability density equal to 50 times the local minimum value of the PDF ($\approx 2.6\cdot 10^{-16}$~fm$^{-6}$), whereas the one of the latter is equal to $10\%$ the probability density of the absolute maximum ($\approx 1.7\cdot 10^{-9}$~fm$^{-6}$). Distances along the axes are measured in lattice spacing units ($a=0.65$~fm). In particular, the toroidal region in the right panel encompasses an entire circle of maxima, which correspond to principal extrema of the wavefunction. In the other plot, the probability density increases with the distance from the origin, until a saddle point consisting of a spherical shell is reached. Then the probability density decreases more slowly to zero. Finally, it is worth remarking that symmetry under particle exchange ensures that the two plots would remain unaffected if the two slices were kept from the $\mathbf{r}_{13}$ subspace.}
\label{F-8.0-06bis}
\end{figure}

\begin{figure}[ht!]
\hspace{1.6cm}
\begin{minipage}{0.48\columnwidth}
\includegraphics[height=0.83\columnwidth]{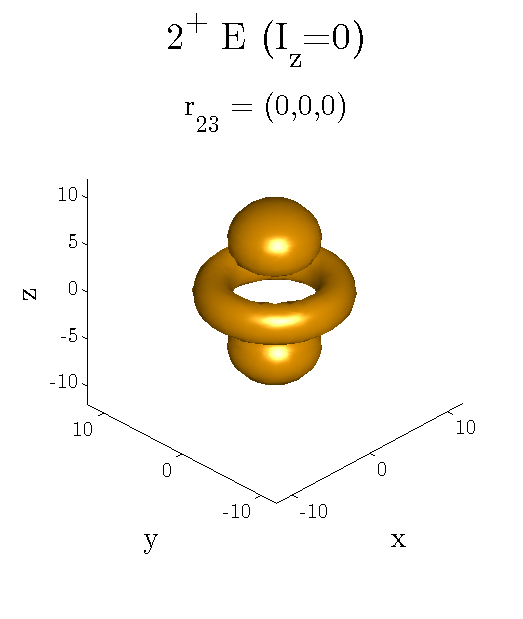}
\end{minipage}
\begin{minipage}{0.48\columnwidth}
\includegraphics[height=0.83\columnwidth]{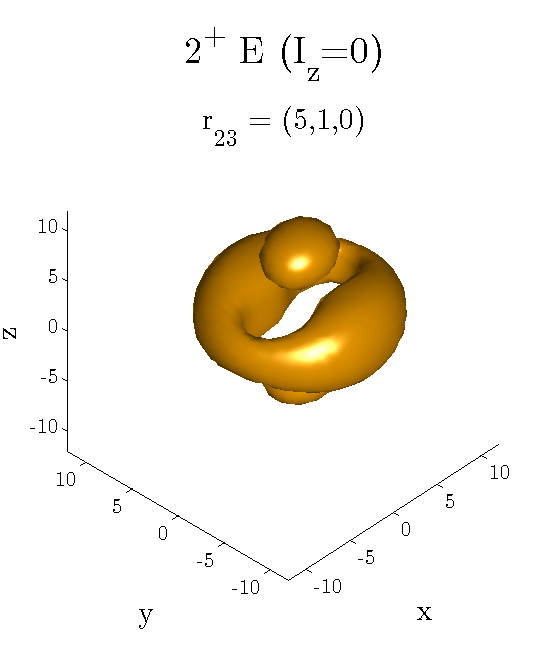}
\end{minipage}
\caption{Spatial distribution of the PDF of the $2_1^+$ $E$ state with $I_z=0$ in the configuration space slices with $\mathbf{r}_{23} = (0,0,0)$ (left) and $\mathbf{r}_{23} = (5,1,0)$ (right). The outer isohypsic surfaces of the two plots correspond to probability densities equal to the 15\%  and the 10\% of the largest extremal values of the squared modulus of the wavefunction enclosed by the surfaces. Distances along the axes are measured in lattice spacing units ($a=0.65$~fm). In particular, the bulges in the toroidal regions in the right plot encompass one single PDF extremum each, which correspond to principal maxima of the wavefunction ($\approx 3.2\cdot 10^{9}$~fm$^{-6}$). On the other hand, all the extrema in the $\mathbf{r}_{23} = (0,0,0)$ slice of the PDF are indeed saddle points. It follows that the configurations with two overlapping $\alpha$-particles and the third one lying in the centre of one of the spheres or in the inner circle of the regular torus are unstable.}
\label{F-8.0-06ter}
\end{figure}

As outlined in Sec.~\ref{S-7.0}, some of the minima of the energy curves can be associated to the values of the lattice spacing that permit the inclusion of relative maxima of the probability distribution functions of the states into the lattice. Differently to the two-body case, the \ce{^{12}C} eigenfunctions may possess a huge amount of local extrema and display rather complex spatial distributions, thus making the analysis of the PDF maxima by far more involved than in the beryllium case, see Figs.~\ref{F-8.0-06bis}-\ref{F-8.0-06quater}. Since the interactions are isotropic, the most probable separation distance between any of the pairs of $\alpha$ particles is expected to coincide exactly for all the eigenfunctions belonging to the same SO(3) multiplet in the zero-spacing limit. \\
Contrary to the beryllium case, the PDF of the ground state of this nucleus has a local non-zero minimum when $\mathbf{r}_{13}=\mathbf{r}_{23}=(0,0,0)$, meaning that configuration in which all the $\alpha$ particles completely overlap has become unstable. In addition, the squared modulus of the $0_1^+$ wavefunction possesses also maxima, the absolute ones corresponding to equilateral triangular equilibrium configurations in which $\alpha$-particles are separated by $ d^* \approx 3.3$~fm, see. Fig.~\ref{F-8.0-06bis}.  Even if none of these maxima can be exactly included in the lattice, both the three minima of the energy eigenvalue at $a \approx 1.40$, $2.35$ and $3.10$~fm are in good correspondence with the ones of the potential energy $\mathcal{V}$. In particular, for the latter two values of the spacing the average interparticle distance $\mathcal{R}$ differs from $d^*$ by only $0.3$~fm, cf. Fig~\ref{F-8.0-07}.\\

For what concerns the $2_E^+$ multiplet, its energy eigenvalue reaches a shallow minimum for $a \approx 2.30$~fm and two well-developed minima for $a \approx 1.45$ and $3.10$~fm (cf. Fig.~\ref{F-8.0-06} and Fig.~7 in Ref.~\cite{BNL14}). As before, these minima are found to be in correspondence with the ones of the average values of the potential energy. Although noone of the absolute maxima of the associated PDFs lies on the lattice axes (cf. Fig~\ref{F-8.0-06ter}), the average value of the interparticle distance at $a\approx 3.1$~fm is in reasonable agreement with the most probable $\alpha-\alpha$ separation distance $d^*$, equal to $\approx 3.3$~fm, see Fig.~\ref{F-8.0-07}. Conversely, for $a\approx 1.45$ and $2.30$~fm $\mathcal{R}$ appears far from $d^*$, due to the contributions of the tails of the wavefunction, certainly more significant than the ones of the ground state.\\
Analogous is the situation of the $2_{T_2}^+$ multiplet, for which the energy minima are in optimal agreement with the minima of the average values of the potential energy, and lie at spacings almost equal to the ones of the $2_E^+$ multiplet ($a \approx 1.40$,  $2.35$ and $3.20$~fm). Even if they do not lie on the lattice axes, the absolute maxima of the PDF can be exactly mapped in the cubic lattice and correspond to equilateral triangular configurations with side $d^*$ equal to $3.3$~fm, as in the previous case.\\

 Besides, the average values of $\mathcal{R}$ at a $\approx 2.35$ and $3.20$~fm roughly agree with the most probable $\alpha-\alpha$ separation distance $d^*$, although for the latter value of the interparticle distance the discrepancy is larger, see Fig.~\ref{F-8.0-07}.\\
\begin{figure}[hb!]
\begin{minipage}{0.48\columnwidth}
\scalebox{0.84}{
\begin{tikzpicture}
\begin{axis}[
	xmin = 0,
	xmax = 4.0,
	xlabel={\small $a$ [fm]},
	ylabel={\small $\mathcal{E}_r$ [MeV]},
	xtick={0,0.5,1.0,1.5,2.0,2.5,3.0,3.5,4.0},
	ytick={-12,-10,-8,-6,-4,-2,0},
	legend columns=1,
	legend style ={at={(0.19,0.35)}, anchor=north west, draw=black,fill=white,align=left,/tikz/column 1/.style={
                column sep=4pt}},
	legend entries ={\textcolor{Turquoise}{$A_2$},\textcolor{Magenta}{$T_1$},\textcolor{PineGreen}{$T_2$}},
]
\addlegendimage{Turquoise,mark=diamond*,mark size=1.3,mark options={fill=Turquoise}};
\addlegendimage{Magenta, mark=triangle*, mark size=1.3, mark options={fill=Magenta}};
\addlegendimage{PineGreen, mark=triangle*, mark size=1.3, mark options={fill=PineGreen,rotate=180}};
\addplot[mark=diamond*, color=Turquoise,  mark size=1.3, mark options={fill=Turquoise}] table [y=E, x=a] {12C_3-_I_A2_19fm.txt};
\addplot[mark=triangle*, color=Magenta,  mark size=1.3, mark options={fill=Magenta}] table [y=E, x=a] {12C_3-_I_T1_19fm.txt};
\addplot[mark=triangle*, color=PineGreen,  mark size=1.3, mark options={fill=PineGreen,rotate=180}] table [y=E, x=a] {12C_3-_I_T2_19fm.txt};
\addplot[color=OliveGreen, densely dotted] table [y=E, x=a] {12C_3-_I_MAVG_19fm.txt};
\end{axis}
\end{tikzpicture}}
\caption{Behaviour of the energies of the $3_1^-$ eigenstates as a function of the lattice spacing for $Na \geq 19$~fm. Even if the multiplet-averaged $3_1^-$ energy (densely dotted line) improves the convergence rate to the continuum and infinite-volume counterpart, for $a \gtrsim 2.0$~fm discretization corrections amount to more than 100\% of the asymptotic energy eigenvalue.}\label{F-8.0-08}
\end{minipage}\hfill
\begin{minipage}[c]{0.48\columnwidth}
\scalebox{0.84}{
\begin{tikzpicture}
\begin{axis}[
	xmin = 0,
	xmax = 4.0,
	xlabel={\small $a$ [fm]},
	ylabel={\small $\mathcal{R}$ [fm]},
	xtick={0,0.5,1.0,1.5,2.0,2.5,3.0,3.5,4.0},
	ytick={3.0,3.5,4.0,4.5,5.0,5.5,6.0,6.5,7.0,7.5,8.0,8.5,9.0,9.5},
	legend columns=1,
	legend style ={at={(0.15,0.71)}, anchor=north west, draw=black,fill=white,align=left,/tikz/column 1/.style={
                column sep=4pt}},
	legend entries ={\textcolor{Turquoise}{$A_2$},\textcolor{Magenta}{$T_1$},\textcolor{PineGreen}{$T_2$}},
]
\addlegendimage{Turquoise,mark=diamond*,mark size=1.3,mark options={fill=Turquoise}};
\addlegendimage{Magenta, mark=triangle*,mark size=1.3,mark options={fill=Magenta}};
\addlegendimage{PineGreen, mark=triangle*,mark size=1.3,mark options={fill=PineGreen,rotate=180}};
\addplot[mark=diamond*, color=Turquoise,  mark size=1.3, mark options={fill=Turquoise} ] table [y=ravg, x=a] {12C_3-_I_A2_19fm.txt};
\addplot[mark=triangle*, color=Magenta,  mark size=1.3, mark options={fill=Magenta} ] table [y=ravg, x=a] {12C_3-_I_T1_19fm.txt};
\addplot[mark=triangle*, color=PineGreen,  mark size=1.3, mark options={fill=PineGreen,rotate=180}] table [y=ravg, x=a] {12C_3-_I_T2_19fm.txt};
\addplot[color=OliveGreen, densely dotted] table [y=ravg, x=a] {12C_3-_I_MAVG_19fm.txt};
\end{axis}
\end{tikzpicture}}
\caption{Behaviour of the average $\alpha-\alpha$ distance of the $3_1^-$ eigenstates as a function of the lattice spacing for $Na \geq 19$~fm. Although slower than in the \ce{^8Be} case, convergence of the average values of $\mathcal{L}^2$ to its expected eigenvalues is attained in the zero-spacing limit.}\label{F-8.0-10}
\end{minipage}
\end{figure}

Concerning the $3_1^-$ multiplet, the three bound multiplets reach the asymptotic region after some oscillations at $a \lesssim 1.30$~fm, where they become degenerate within $0.12$~MeV, and eventually overlap with two digit accuracy at $a \approx 0.80$~fm, see Fig.~\ref{F-8.0-07}. All the PDF associated to the wavefunctions of the present multiplet are found to have well-developed principal maxima ($\approx 10^4$ times deeper than any any other PDF extremum), corresponding to $\alpha-\alpha$ separations $d^*$ of about $3.4$~fm. \\
\begin{figure}[ht!]
\hspace{1.6cm}
\begin{minipage}{0.48\columnwidth}
\includegraphics[height=0.83\columnwidth]{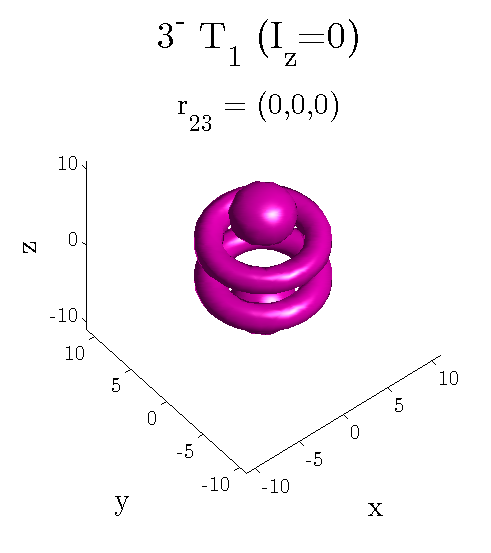}
\end{minipage}
\begin{minipage}{0.48\columnwidth}
\includegraphics[height=0.83\columnwidth]{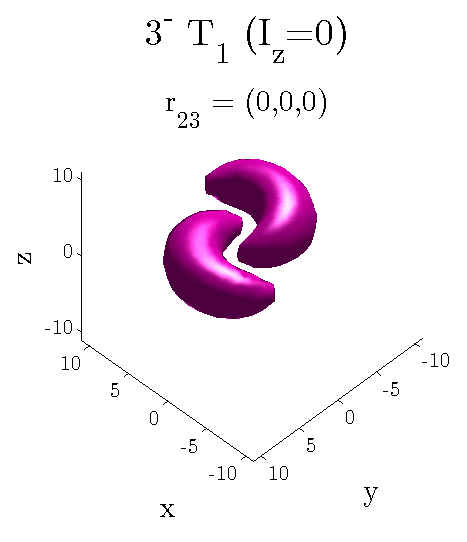}
\end{minipage}
\caption{Spatial distribution of the PDF of the $3_1^-$ $T_1$ state with $I_z=0$ in the configuration space slices with $\mathbf{r}_{23} = (0,0,0)$ (left) and $\mathbf{r}_{23} = (1,2,5)$ (right). The outer isohypsic surfaces of the two plots correspond to a probability density equal to the 10\% of the maximum value of the PDF ($\approx 4.4\cdot 10^{-9}$~fm$^{-6}$). Distances along the axes are measured in lattice spacing units ($a=0.65$~fm). In particular, the crescent-shaped regions in the right plot encompass one single local PDF extremum each, which correspond to principal maxima of the wavefunction. As for the $2_E^+$ $I_z = 0$ state, all the extrema in the $\mathbf{r}_{23} = (0,0,0)$ slice of the PDF are indeed saddle points. Consequently, the configurations with two overlapping $\alpha$-particles and the third one lying in the centre of one of the spheres or in the inner circle of one of the two tori are unstable. }
\label{F-8.0-06quater}
\end{figure}

Moreover, both the energy eigenvalue and the average value of the potential energy of the $3_{A_2}^-$ state is minimized for lattice spacings equal to $\approx 1.35$ and $2.35$~fm. In particular, for the latter value of the spacing $\mathcal{R} \approx 3.45$~fm (cf. Figs.~\ref{F-8.0-08} and \ref{F-8.0-09}), a reasonable agreement with $d^*$. On the other hand, for the former spacing the average value of the $\alpha$-$\alpha$ distance is strongly influenced by the tails of the wavefunction. Both the minima can be related to the exact inclusion of the principal maxima of the PDF associated to the aforementioned state into the lattice.\\
In the case of the $3_{T_1}^-$ states, the energy minima at $a \approx 1.45$, $2.40$ and $3.15$~fm are still found to be in good correspondence with the ones of $\mathcal{V}$. Again, not all the principal maxima detected in the PDFs can be exactly (or in good approximation) included in the cubic lattice, due to the non-trivial spatial orientation of the probability density surfaces encompassing the absolute maxima, cf. Fig.~\ref{F-8.0-06quater}. Nevertheless, the two minima of $\mathcal{E}_r$ at $2.40$ and $3.15$~fm correspond to values of the average interparticle distance $\mathcal{R}$ of about $3.45$~fm, again in good agreement with $d^*$.\\ Similarily to the previous case, not all the principal maxima of the probability density functions associated to the $3_{T_2}^-$ states can be exactly mapped in the cubic lattice. Although the shallow minimum of the energy eigenvalue of the multiplet between $a = 2.25$ and $2.3$~fm is shifted by about $0.2$~fm from the nearest minimum of $\mathcal{V}$, the remaining two energy minima at $a \approx 1.45$ and $3.15$~fm are in good correspondence with the ones of the average values of the potential energy. Concerning the avreage values of the interparticle distance, the agreement between $\mathcal{R}$ at $a \approx 2.3$ and $3.15$~fm and $d^*$ is worse than in the previous case (cf. Fig.~\ref{F-8.0-10}), due to the spatial distribution of the $3_{T_2}^-$ wavefunctions.\\
\begin{figure}[ht!]
\begin{minipage}[c]{0.47\columnwidth}
\scalebox{0.85}{
\begin{tikzpicture}
\begin{axis}[
	xmin = 0,
	xmax = 4.0,
	xlabel={\small $a$ [fm]},
	ylabel={\small $\hbar^{-2}\mathcal{L}^2$},
	xtick={0,0.5,1.0,1.5,2.0,2.5,3.0,3.5,4.0},
	ytick={0,2,4,6,8,10,12,14,16,18},
	legend columns=1,
	legend style ={at={(0.15,0.85)}, anchor=north west, draw=black,fill=white,align=left,/tikz/column 1/.style={
                column sep=4pt}},
	legend entries ={\textcolor{Violet}{$A_1$},\textcolor{Orange}{$E$},\textcolor{PineGreen}{$T_2$}},
]
\addlegendimage{Violet,mark=square*,mark size =1,mark options={fill=Violet}};
\addlegendimage{Orange,mark=pentagon*,mark size=1.3,mark options={fill=Orange}};
\addlegendimage{PineGreen,mark=triangle*,mark size=1.3,mark options={fill=PineGreen,rotate=180}};
\addplot[mark=square*, color=Violet,  mark size=1, mark options={fill=Violet}] table [y=J, x=a] {12C_0+_I_A1_19fm.txt};
\addplot[mark=pentagon*, color=Orange,  mark size=1.3, mark options={fill=Orange}] table [y=J, x=a] {12C_2+_I_E_19fm.txt};
\addplot[mark=triangle*, color=PineGreen,  mark size=1.3, mark options={fill=PineGreen,rotate=180}] table [y=J, x=a] {12C_2+_I_T2_19fm.txt};
\addplot[color=Red, solid] table [y=J, x=a] {12C_2+_I_MAVG_19fm.txt};
\end{axis}
\end{tikzpicture}}
\caption{Behaviour of the average vaules of the squared total angular momentum of the $0_1^+$ and $2_1^+$ eigenstates as a function of the lattice spacing for $Na \geq 19$~fm.}\label{F-8.0-09}
\end{minipage}\hfill
\begin{minipage}[c]{0.47\columnwidth}
\scalebox{0.85}{
\begin{tikzpicture}
\begin{axis}[
	xmin = 0,
	xmax = 4.0,
	xlabel={\small $a$ [fm]},
	ylabel={\small $\hbar^{-2}\mathcal{L}^2$},
	xtick={0,0.5,1.0,1.5,2.0,2.5,3.0,3.5,4.0},
	ytick={10,12,14,16,18,20,22,24,26,28,30,32,34,36,38},
	legend columns=1,
	legend style ={at={(0.10,0.82)}, anchor=north west, draw=black,fill=white,align=left,/tikz/column 1/.style={
                column sep=4pt}},
	legend entries ={\textcolor{Turquoise}{$A_2$},\textcolor{Magenta}{$T_1$},\textcolor{PineGreen}{$T_2$}},
]
\addlegendimage{Turquoise,mark=diamond*,mark size=1.3,mark options={fill=Turquoise}};
\addlegendimage{Magenta,mark=triangle*,mark size=1.3,mark options={fill=Magenta}};
\addlegendimage{PineGreen,mark=triangle*,mark size=1.3,mark options={fill=PineGreen,rotate=180}};
\addplot[mark=diamond*, color=Turquoise,  mark size=1.3, mark options={fill=Turquoise}] table [y=J, x=a] {12C_3-_I_A2_19fm.txt};
\addplot[mark=triangle*, color=Magenta,  mark size=1.3, mark options={fill=Magenta}] table [y=J, x=a] {12C_3-_I_T1_19fm.txt};
\addplot[mark=triangle*, color=PineGreen,  mark size=1.3, mark options={fill=PineGreen,rotate=180}] table [y=J, x=a] {12C_3-_I_T2_19fm.txt};
\addplot[color=OliveGreen, densely dotted] table [y=J, x=a] {12C_3-_I_MAVG_19fm.txt};
\end{axis}
\end{tikzpicture}}
\caption{Behaviour of the squared total angular momentum of the $3_1^-$ eigenstates as a function of the lattice spacing for $Na \geq 19$~fm. Even if slowly, convergence of the average values of $\mathcal{L}^2$ to its expected eigenvalues is attained in the zero-spacing limit.}\label{F-8.0-11}
\end{minipage}
\end{figure}

\begin{figure}[hb!]
\begin{minipage}[c]{0.61\columnwidth}
\scalebox{0.93}{
\begin{tikzpicture}
\begin{axis}[
	xmin = 0,
	xmax = 2.0,
	width = 5.3 cm,
	height = 7 cm,
	xlabel={\small $a$ [fm]},
	ylabel={\small $\hbar^{-2}|\Delta\mathcal{L}^2|$},
	xtick={0,0.5,1,1.5,2},
	ymode=log,
	log basis y = {2.718281828459},
	yticklabels={$e^{-3}$,$e^{-2}$,$e^{-1}$, $e^{0}$}
]
\addplot[mark=diamond*, color=Violet,  mark size=1.3, mark options={fill=Violet}] table [y=dJ, x=a] {12C_0+_I_A1_19fm.txt};
\addplot[mark=pentagon*, color=Orange,  mark size=1.3, mark options={fill=Orange}] table [y=dJ, x=a] {12C_2+_I_E_19fm.txt};
\addplot[mark=triangle*, color=PineGreen,  mark size=1.3, mark options={fill=PineGreen,rotate=180}] table [y=dJ, x=a] {12C_2+_I_T2_19fm.txt};
\addplot[color=Red, solid] table [y=dJ, x=a] {12C_2+_I_MAVG_19fm.txt};
\end{axis}
\end{tikzpicture}
\begin{tikzpicture}
\begin{axis}[
	xmin = 0,
	xmax = 2.0,
	width = 5.3 cm,
	height = 7 cm,
	xlabel={\small $a$ [fm]},
	xtick={0,0.5,1,1.5,2},
	ymode=log,
	log basis y = {2.718281828459},
	yticklabels={$e^{-3}$,$e^{-2}$,$e^{-1}$, $e^{0}$}
]
\addplot[mark=diamond*, color=Turquoise, mark size=1.3, mark options={fill=Turquoise} ] table [y=dJ, x=a] {12C_3-_I_A2_19fm.txt};
\addplot[mark=triangle*, color=Magenta, mark size=1.3, mark options={fill=Magenta} ] table [y=dJ, x=a] {12C_3-_I_T1_19fm.txt};
\addplot[mark=triangle*, color=PineGreen, mark size=1.3, mark options={fill=PineGreen,rotate=180} ] table [y=dJ, x=a] {12C_3-_I_T2_19fm.txt};
\addplot[color=OliveGreen, densely dotted] table [y=dJ, x=a] {12C_3-_I_MAVG_19fm.txt};
\end{axis}
\end{tikzpicture}}
\end{minipage}\hfill
\begin{minipage}[c]{0.36\columnwidth}
\caption{Difference between the average value and the expected eigenvalue of the squared angular momentum for the $2_1^+$ (left) and the $3_1^-$ states (right) as a function of the lattice spacing. The same convention on the markers for the cubic group irreps of Figs.~\ref{F-8.0-07}-\ref{F-8.0-08} is understood. It is worth remarking that the deviations from linearity for small values of the spacing in the $2_E^+$ curve (cf. the left panel of the figure) are sensibly larger than the ones of the other multiplets, an effect perhaps due to residual finite-volume effects.}\label{F-8.0-12}
\end{minipage}
\end{figure}
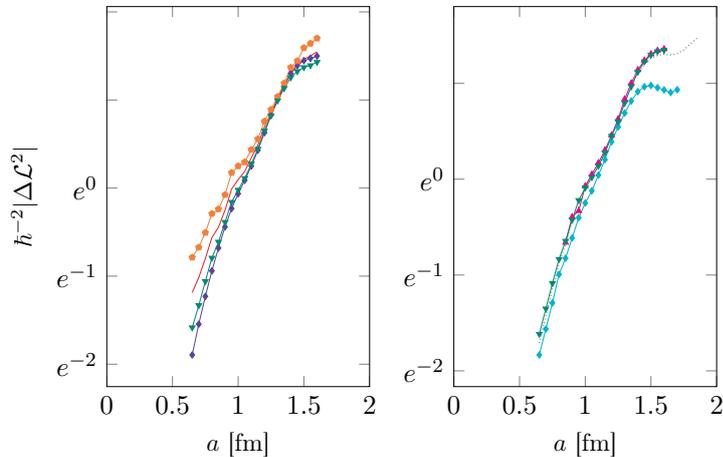

Switching now to the average values of the squared total angular momentum, the convergence rate of the $0_1^+$ and $2_1^+$ states to the expected $\mathcal{L}^2$ eigenvalues is sensibly slower than the one of the homologous states of beryllium, cf. Fig.~\ref{F-7.0-11}. In particular, one decimal digit agreement between the $\mathcal{L}^2$ average value on the ground state and the expected eigenvalue is reached for $a \approx 1.0$~fm, whereas two decimal digit precision is attained only at  $a \approx 0.65$~fm. Besides, for the $2_1^+$ multiplet one decimal digit precision in the angular momentum estimation is reached only at $a \approx 0.70$~fm, even if, for the $T_2$ multiplet convergence is slightly faster, as observed in the $-3.3$~MeV multiplet of \ce{^8Be} (cf. Fig.~\ref{F-7.0-11}). \\
For the $3_1^-$ state the situation is similar, since one-digit precision in the estimation of the eigenvalue of the squared total angular momentum is reached only at $a = 0.85$, $0.80$ and $0.75$~fm for the $3_{A_2}^-$, $3_{T_1}^-$ and $3_{T_2}^-$ multiplets respectively. Contrary to the case of the $0_1^+$ and $2_1^+$ states of \ce{^8Be}, it turns out that the computation of the average values of $\mathcal{L}^2$ does not provide more precise information on the transformation properties of the group of states under SO(3) rotations, since the energies themselves become degenerate with greater accuracy at larger lattice spacings. \\
Nevertheless, by subtracting the expected squared angular momentum eigenvalues from the $\mathcal{L}^2$ average values and then taking the absolute value the observations on the asymptotic corrections to the latter in Sec.~\ref{S-7.0} find another confirmation. If the spacing is small enough, \textit{i.e.} $a \lesssim 1.4$~fm for the $0_1^+$ and $2_1^+$ states or $a \lesssim 1.3$~fm for the $3^-$ multiplet, the $\log |\Delta\mathcal{L}^2|$ behave almost linearly with the lattice spacing, with a positive slope, see Fig.~\ref{F-8.0-12}. \\

\section{\textsf{Conclusion}}\label{S-9.0}

The transposition of any physical system on a cubic lattice may yield to shifts in the eigenvalues and in the average values of operators, due to finite volume and discretization effects. In particular, the breaking of rotational symmetry into cubic group summetry affects the average values of all the operators transforming as spherical tensors under the elements of SO(3) \cite{BNL15}. Nevertheless, the construction of the lattice counterpart of the squared total angular momentum operator allows for an unambiguous identification of the lattice Hamiltonian eigenstates in terms of SO(3) irreps, provided the spatial distribution of the eigenfunctions is localized and smooth enough to fit the size and the spacing of the lattice. This is exactly the case of the $2_E^+$ and $2_{T_2}^+$ multiplets of \ce{^8Be}, where the average value of the squared angular momentum operator reaches its expectation value with deviations of $0.01\%$ already at $ a \approx 1.8$~fm, see Fig.~\ref{F-7.0-11}, a spacing for which the energy eigenvalues of the two multiplets are still separated by more than $2$~MeV, Fig.~\ref{F-7.0-05}. Furthermore, the asymptotic finite volume corrections to the average values of the squared angular momentum operator approximately fit a negative exponential of the lattice size (cf. Figs.~\ref{F-7.0-04} and \ref{F-7.0-16}), like the leading-order ones for the energy \cite{KLH12}. Discretization corrections for the average values of the same operator turned out also to depend exponentially on $a$ in the zero lattice-spacing limit, although with a positive decay constant (cf. Figs.~\ref{F-7.0-16} and \ref{F-8.0-12}). \\
Besides exploring the role of $\mathcal{L}^2$ in the classification of the lattice Hamiltonian eigenstates in terms of the angular momentum quantum number, the model offered us also the possibility to test the interpretation of the local minima of energy eigenvalues in terms of the spatial distribution of the relevant eigenfunctions (cf. the $4_2^+$ and the $6_1^+$ multiplets of \ce{^8Be} and the $0_1^+$, $2_1^+$ and $3_1^-$ multiplets of \ce{^{12}C}) as well as the results presented in Ref.~\cite{BNL14} (cf. the $0_1^+$ and $2_1^+$ states of \ce{^8Be}). In case a local maximum of the squared modulus of a lattice eigenfunction is included within the mesh points, in fact, the corresponding energy eigenvalue as a function of the lattice spacing displays a minimum.\\
Moreover, we have shown that the use of multiplet-averaging (cf. Sec.~\ref{S-5.3}) for the energies and the average values of the squared angular momentum for states with $\ell = 0, 2, 3, 4$ and $6$ (cf. Secs.~\ref{S-7.0} and \ref{S-8.0}) reduces both discretization and finite-volume effects by evening the fluctuations about the continuum and infinite-volume counterparts, as predicted in Ref.~\cite{BNL14}. \\
Likewise interesting are the computational implications of this work. In the attempt of suppressing both discretization and finite-volume effects for the three-body system, considerable efforts have been devoted in developing memory-saving and fast codes for the diagonalization of the lattice Hamiltonian. The final choice of the \textit{Lanczos algorithm} and of the GPU as a support for the state vectors processing permitted us to monitor the evolution of the eigenergies and the average values of other physical observables concerning six bound state multiplets of the \ce{^{12}C} nucleus for a significant range of box-sizes and spacings. In addition, the extensive usage of projectors in the iterative diagonalization process allowed us to extend the analysis of Ref.~\cite{BNL14} to higher angular momentum multiplets, both for the \ce{^{8}Be} and the \ce{^{12}C}, discarding all the possible intermediate states devoid of the desired transformation properties under the elements of the permutation group and the cubic group. Eventually, the diagonalization techniques outlined here are expected to pave the way for the investigation of lattice artifacts on the spectrum of a four-body system, the \ce{^{16}O}, subject of a forthcoming paper.\\

\section*{\textsf{Acknowledgments}}

First of all, we express our gratitude to Bing Nan Lu and Dean Lee for the helpful and stimulating discussions and Timo A. L\"ahde, Andreas Nogga, Tom C. Luu and Alexander Strube for the technical assistance. Besides, we acknowledge financial support from the Deutsche Forschungsgemeinschaft (Sino-German collaboration CRC 110, grant No. TRR~110) and the VolkswagenStiftung (grant No. 93562). The work of UGM was also supported by the Chinese Academy of Sciences (CAS) President’s International Fellowship Initiative (PIFI) (Grant No. 2018DM0034). Finally, we acknowlegde computational resources provided by Forschungs\-zentrum J\"ulich (PAJ 1830 test project) and RWTH Aachen (JARA 0015 project).

\begin{appendix}

\section{\textsf{Technicalities}}\label{S-10.0}

\subsection{\textsf{Discretization of derivatives}}\label{S-10.1}

In the lattice environment, spatial derivatives have to be naturally expressed in terms of finite differences. As a consequence, all the differential operators are represented by non-commuting matrices, whose non diagonal elements are collectively referred as hopping terms. For the discretization of all the differential operators of interest the improvement scheme presented in sect.~9.1.1 \cite{GaL10} is implemented.\\ Any given $\mathcal{C}^{2K}$ function $f(x \pm ka)$ on the lattice with $k \in \mathbb{K}$ admits a Taylor expansion about any point $x$ of its domain,
\begin{equation}
f(x \pm ka) = f(x) \pm ka f^{(1)}(x) + \frac{k^2a^2}{2!} f^{(2)}(x) \pm \frac{k^3x^3}{3!}f^{(3)}(x) + ... \pm  \frac{(ka)^{2K-1}}{2K-1!}f^{(2K-1)}(x) + \mathcal{O}(a^{2K})~.
\label{10.1-01}
\end{equation}
From the subtraction of $f(x - ka)$ from $f(x + ka)$, it is possbile to construct an aprroximation scheme for the first derivative, 
\begin{equation}
\begin{split}
f_{ka}^{-} \equiv f(x + ka)-f(x-ka) & = 2ka f^{(1)}(x) + 2\frac{k^3a^3}{3!} f^{(3)}(x) \\ & +  2\frac{k^5a^5}{5!} f^{(5)}(x)  + \ldots + 2\frac{(ka)^{2K-1}}{2K-1!}f^{(2K-1)}(x) + \mathcal{O}(a^{2K+1})
\end{split}
\label{10.1-02}
\end{equation}
whose truncation error is given by $\mathcal{O}(a^{2K+1})$. Summing up a linear combination of $f_{ka}^{-}$ with k ranging from 1 to K, in fact, all the contributions from the odd derivatives up to order $2K-1$ in the discretized expression of the first derivative can be ruled out,  
\begin{equation}
\begin{split}
\sum_{k=1}^K C_k^{(1,K)} f_{ka}^{-} = 2af^{(1)}(x)\sum_{k=1}^K C_k^{(1,K)} k  & +  2\frac{a^3}{3!} f^{(3)}(x) \sum_{k=1}^K C_k^{(1,K)} k^3  \\  &+ \ldots + 2\frac{a^{2K-1}}{2K-1!} f^{(2K-1)}(x)\sum_{k=1}^K C_k^{(1,K)} k^{2K-1} + \mathcal{O}(a^{2K+1})~.
\end{split}
\label{10.1-03}
\end{equation}
At this stage, it is sufficient to impose to the unknown coefficients $C_k^{(1P,K)}$ the following constraints, 
\begin{equation}
\sum_{k=1}^K C_k^{(1,K)} k^{2l-1} =
\begin{dcases}
1/2a & \mathrm{if} \hspace{0.2cm} l = 1\\
0   & \mathrm{if} \hspace{0.2cm} 2 \leq l \leq K\\
\end{dcases}\label{10.1-04}
\end{equation}
in order to recover the desired approximated expression for $f^{(1)}(x)$,
\begin{equation}
f^{(1)}(x) \approx \sum_{k=1}^K C_k^{(1,K)} f_{ka}^{-}~.\label{10.1-05}
\end{equation}
Analytically, the coefficients take the form 
\begin{equation}
C_k^{(1,K)} = (-1)^{k+1} \frac{1}{2a} \frac{2}{k} \frac{(K!)^2}{ K+k! K-k!}\label{10.1-06}
\end{equation}
as it can be proven by solving the associated linear system in Eq.~\eqref{10.1-04} with the Cramer's rule and recalling the determinant formulas for Vandermonde-like matrices. \\
On the other hand, the sum between $f(x - ka)$ and $f(x + ka)$, permits to derive the aprroximation scheme for the second (pure) derivative, 
\begin{equation}
\begin{split}
f_{ka}^{+} \equiv f(x + ka)+f(x-ka) & = 2f(x) + k^2a^2 f^{(2)}(x) \\ & + 2\frac{k^4a^4}{3!} f^{(4)}(x) +  2\frac{k^6a^6}{6!} f^{(6)}(x)  + \ldots + 2\frac{(ka)^{2K}}{2K!}f^{(2K)}(x) + \mathcal{O}(a^{2K+2})
\end{split}
\label{10.1-07}
\end{equation}
whose truncation error is given by $\mathcal{O}(a^{2K+2})$. Again, summing a linear combination of $f_{ka}^{+}$ with k ranging from $1$ to K, in fact, all the contributions from the even derivatives up to order $2K$ to the discretized expression of the second derivative can be cancelled in the same fashion,  
\begin{equation}
\begin{split}
\sum_{k=1}^K C_k^{(2P,K)} f_{ka}^{+}  & = 2f(x)\sum_{k=1}^K C_k^{(2P,K)}  + a^2f^{(2)}(x)\sum_{k=1}^K C_k^{(2P,K)} k^2  \\ & +  2\frac{a^4}{4!} f^{(4)}(x)\sum_{k=1}^K C_k^{(2P,K)} k^4  + \ldots + 2\frac{a^{2K}}{2K!} f^{(2K)}(x)\sum_{k=1}^K C_k^{(2P,K)} k^{2K} + \mathcal{O}(a^{2K+2})~.
\end{split}
\label{10.1-08}
\end{equation}
The constraints on the $C_k^{(2P,K)}$ are, now, 
\begin{equation}
\sum_{k=1}^K C_k^{(2P,K)} k^{2l} =
\begin{dcases}
1/a^2 & \mathrm{if} \hspace{0.2cm} l = 1\\
0   & \mathrm{if} \hspace{0.2cm} 2 \leq l \leq K,\\
\end{dcases}\label{10.1-09}
\end{equation}
and enable us rewriting the second (pure) derivative on the lattice as 
\begin{equation}
f^{(2)}(x) \approx C_0^{(2P,K)}f(x) + \sum_{k=1}^K C_k^{(2P,K)} f_{ka}^{+}~, \label{10.1-10}
\end{equation}
where a coefficient for the diagonal term of the discretized operator has been introduced as in \cite{BNL14},
\begin{equation}
C_0^{(2P,K)} = -2\sum_{k=1}^K C_k^{(2P,K)}~.\label{10.1-11}
\end{equation}
Solving the linear system associated to the coefficients with nonzero subscript in Eq.~\eqref{10.1-09}, the analytic expression of the $C_k^{(2P,K)}$'s can be obtained,
\begin{equation}
C_k^{(2P,K)} = (-1)^{k+1} \frac{1}{a^2} \frac{2}{k^2} \frac{(K!)^2}{ K+k! K-k!}~.\label{10.1-12}
\end{equation}
Equipped with the approximation schemes for both the first and the second derivatives of a function of one variable, we conclude the section with the treatment of second mixed derivatives. Denoting henceforth the mixed derivatives of an analytic function in two variables (x,y) as 
\begin{equation}
\frac{\partial^{m+n}}{\partial^m x \partial^n y} f(x,y) = f^{(m,n)}(x,y)~, \label{10.1-13} 
\end{equation}
the Taylor expansion of the two-variables functions $f(x \pm ka ,y \pm ka)$ and $f(x \pm ka ,y \mp ka)$ about $(x,y)$ can be written as
\begin{equation}
\begin{gathered}
f(x\pm ka,y \pm ka) = f(x,y) \pm ak[f^{(1,0)}(x,y)+f^{(0,1)}(x,y)] \\ + \frac{a^2 k^2}{2}[f^{(2,0)}(x,y)+2f^{(1,1)}(x,y)+f^{(0,2)}(x,y)]  \\ \pm \frac{a^3 k^3}{2}[f^{(3,0)}(x,y)+3f^{(2,1)}(x,y) +3f^{(1,2)}(x,y)+f^{(0,3)}(x,y)] \\ + \dots + \frac{a^{2K}k^{2K}}{2K!}\sum_{i=0}^{2K}{{2K}\choose{i}}f^{(2K-i,i)}(x,y) + \mathcal{O}(a^{2K+1})~,
\end{gathered}
\label{10.1-14}
\end{equation}
and 
\begin{equation}
\begin{gathered}
f(x\pm ka,y \mp ka) = f(x,y) \pm ak[f^{(1,0)}(x,y)-f^{(0,1)}(x,y)] \\ + \frac{a^2 k^2}{2}[f^{(2,0)}(x,y)-2f^{(1,1)}(x,y)+f^{(0,2)}(x,y)] \\ \pm \frac{a^3 k^3}{2}[f^{(3,0)}(x,y)-3f^{(2,1)}(x,y) +3f^{(1,2)}(x,y)-f^{(0,3)}(x,y)] \\ + \dots + \frac{a^{2K}k^{2K}}{2K!}\sum_{i=0}^{2K}{{2K}\choose{i}}(-1)^if^{(2K-i,i)}(x,y) + \mathcal{O}(a^{2K+1})~,
\end{gathered}
\label{10.1-15}
\end{equation}
respectively. Now, by defining the following fourfold combination of displaced functions, 
\begin{equation}
f_{ka}^{M} \equiv f(x+ka,y+ka) -f(x-ka,y+ka) -f(x+ka,y-ka)+f(x-ka,y-ka)
\label{10.1-16}
\end{equation}
an expression for the second mixed derivative $f^{(1,1)}(x,y)$ in terms of mixed derivatives of higher order can be recovered, 
\begin{equation}
\begin{split}
\sum_{k=1}^K C_k^{(2M,K)} f_{ka}^{M} & = 4 a^2f^{(1,1)}(x)\sum_{k=1}^K C_k^{(2M,K)} k^2  +  4\frac{a^4}{3!} [f^{(1,3)}(x)+f^{(3,1)}(x)]\sum_{k=1}^K C_k^{(2M,K)} k^4  \\ & + \ldots + 4\frac{a^{2K}}{2K!} \sum_{i=1}^{K}{{2K}\choose{2i-1}}f^{\substack{(2K-2i+1,\\ 2i-1)}}(x)\sum_{k=1}^K C_k^{(2M,K)} k^{2K}  + \mathcal{O}(a^{2K+2})~.
\end{split}
\label{10.1-17}
\end{equation}
Thus, aiming at rewriting the latter as a superposition of $f_{ka}^{M}$'s truncated to order $2K$, 
\begin{equation}
f^{(1,1)}(x) \approx \sum_{k=1}^K C_k^{(2M,K)} f_{ka}^{M}~, \label{10.1-18}
\end{equation}
we get the following contraints on the coefficients of the expansion
\begin{equation}
\sum_{k=1}^K C_k^{(2M,K)} k^{2l} =
\begin{dcases}
1/4a^2 & \mathrm{if} \hspace{0.2cm} l = 1\\
0   & \mathrm{if} \hspace{0.2cm} 2 \leq l \leq K.\\
\end{dcases}\label{10.1-19}
\end{equation}
The solution of the linear system associated to the latter equation coincides with the one of the preceeding case except for a factor $1/4$,
\begin{equation}
C_k^{(2M,K)} = (-1)^{k+1} \frac{1}{4a^2} \frac{2}{k^2} \frac{(K!)^2}{ K+k! K-k!}~.\label{10.1-20}
\end{equation}
From a direct comparison between the expansion coefficients of the three differential operators, the following relationship, 
\begin{equation}
C_k^{(1,K)} = \frac{ak}{2}C_k^{(2P,K)} = 2ak\hspace{1mm}C_k^{(2M,K)}~,\label{10.1-21}
\end{equation}
can be inferred, thus allowing for a quicker evaluation of the former (cf. Tab.~\ref{T-10.1-01}).\\ Moreover, the discretization scheme for the first derivatives can be likewise exploited for the definition of second mixed derivatives on the lattice, thus expressing $f^{(1,1)}(x,y)$ in terms of $K(K-1)$ hopping terms of the kind $f(x+ma,y+na)$. Although straightforward, this alternative implementation is slower than the one presented here, due to repeated loops over non-diagonal terms. \\
\begin{table}[h!]
\begin{center}
{\renewcommand\arraystretch{1.2}
\begin{tabular}{c|ccccc}
\toprule
K & 1 & 2 & 3 & 4 & 5\\ 
\midrule
$C_1^{(1,K)}$ & $\frac{1}{2}$ & $\frac{2}{3}$ & $\frac{3}{4}$ & $\frac{4}{5}$ & $\frac{5}{6}$\\
$C_2^{(1,K)}$ & & -$\frac{1}{12}$ & -$\frac{3}{20}$ & -$\frac{1}{5}$ & -$\frac{5}{21}$\\
$C_3^{(1,K)}$ &  &  & $\frac{1}{60}$ & $\frac{4}{105}$ & $\frac{5}{84}$\\
$C_4^{(1,K)}$ &  &  &  & -$\frac{1}{280}$ & -$\frac{5}{504}$ \\
$C_5^{(1,K)}$ & & & & & $\frac{1}{1260}$ \\
\bottomrule
\end{tabular}}
\end{center}
\caption{Coefficients for the discretization of first derivatives with $K \leq 5$ and unitary lattice spacing.}\label{T-10.1-01}
\end{table}
Even if in most of the calculations the derivative improvement index $K$ has been kept equal to $4$, a source of concern can be the convergence of the Taylor expansions of the functions (cf. Eqs.~\eqref{10.1-01}, \eqref{10.1-14} and \eqref{10.1-15}). However the second derivative improvement scheme in the limit $K \rightarrow \infty$ converges uniformly to the exactly quadratic operator in the momentum space over the Briullouin zone \cite{BNL14}. Furthermore, both the exact kinetic energy in momentum space and the respective discretized operator in the configuration space in the latter limit gave no evidence of convergence or stability issues.\\ 

\subsection{\textsf{The cubic group}}\label{S-10.2}

In this section a short review on the cubic group is given, together with the transformation table for basis states of SO(3) irreps with $\ell \leq 8$ into the $\mathcal{O}$ ones.\\
\begin{table}[h!]
\begin{center}
\scalebox{0.85}{ 
{\renewcommand\arraystretch{1.2} 
\begin{tabular}{ccccc}
\toprule
 E & $6C_2''$ & $3C_4^2(\pi)$ & $8C_3'$ & $6C_4(\frac{\pi}{2})$ \\
\midrule
$(0,0,0)$ & $(0,\pi,\frac{\pi}{2})$ & $(\pi,\pi,0)$ & $(\frac{\pi}{2},\frac{\pi}{2},\pi)$ & $(\frac{\pi}{2},\frac{\pi}{2},\frac{3\pi}{2})$\\
& $(0,\pi,\frac{\pi}{2})$ & $(0,\pi,0)$ & $(\pi,\frac{3\pi}{2},\frac{3\pi}{2})$ & $(\frac{3\pi}{2},\frac{\pi}{2},\frac{\pi}{2})$\\
& $(0,\pi,\frac{3\pi}{2})$ & $(\pi,0,0)$ & $(\pi,\frac{3\pi}{2},\frac{\pi}{2})$ & $(\pi,\frac{\pi}{2},\pi)$\\
& $(\frac{3\pi}{2},\frac{\pi}{2},\frac{3\pi}{2})$ & & $(\frac{3\pi}{2},\frac{\pi}{2},\pi)$ & $(\pi,\frac{3\pi}{2},\pi)$\\
& $(0,\frac{\pi}{2},\pi)$ & & $(\pi,\frac{\pi}{2},\frac{3\pi}{2})$ & $(\frac{\pi}{2},0,0)$\\
& $(\pi,\frac{\pi}{2},0)$ & & $(\frac{\pi}{2},\frac{3\pi}{2},\pi)$ & $(\frac{3\pi}{2},0,0)$\\
& & & $(\pi,\frac{\pi}{2},\frac{\pi}{2})$ & \\
& & & $(\frac{3\pi}{2},\frac{3\pi}{2},\pi)$ & \\
\bottomrule
\end{tabular}}}
\end{center}
\caption{\textsl{Rappresentation of the group.} The elements belonging to each of the conjugacy classes are listed as terns of Euler angles. Accordingly the symmetry operation $(\alpha, \beta,\gamma)$ consists of a rotation of angle $\gamma$ about the $z$ lattice axis, followed by one of angle $\beta$ about the $y$ axis and by another of angle $\alpha$ about the $z$ axis.}\label{T-10.2-01}
\end{table}

The group in analysis consists of 24 rotations about the symmetry axes of the cube (or the octahedron), subdivided into five equivalence classes. Adopting Sch\"onflies notation \cite{Car97}, $E$ represents the identity, $3C_4^2(\pi)$ the rotations of $180^{\circ}$ about the three fourfold axes orthogonal to the faces of the cube (\textit{i.e.} the lattice axes), $6C_4(\pi/4)$ the $45^{\circ}$ and $135^{\circ}$ rotations about the latter axes (6 elements), $6C_4''$ the $180^{\circ}$ rotations about the six diagonal axes parallel to two faces of the cube and $8C_3'(2\pi/3)$ are rotations of $120^{\circ}$ and $240^{\circ}$ about the four diagonal axes passing to opposite vertexes of the lattice (8 elements).\\
Moreover, the characters of the 5 irreducible representations of $\mathcal{O}$ are presented in Tab.~\ref{T-10.2-02}. In the same table are also presented the characters of $2\ell+1$-dimensional irreps of SO(3), that, as known, induce reducible representations of the cubic group.  \\
\begin{table}[h!]
\begin{center}
\scalebox{0.85}{
{\renewcommand\arraystretch{1}
\begin{tabular}{c|ccccc}
\toprule
$\Gamma$ & E & $6C_2''$ & $3C_4^2(\pi)$ & $8C_3'$ & $6C_4(\frac{\pi}{2})$ \\
\midrule
$A_1$ & 1 & 1 & 1 & 1 & 1\\
$A_2$ & 1 & -1 & 1 & 1 & -1 \\
$E$ & 2 & 0 & 2 & -1 & 0 \\
$T_1$ & 3 & -1 & -1 & 0 & 1 \\
$T_2$ & 3 & 1 & -1 & 0 & -1 \\
$D^{\ell}$ & $2\ell+1$ & $(-1)^{\ell}$ & $(-1)^{\ell}$ & $1-\mathrm{mod}(\ell,3)$ & $(-1)^{[\frac{\ell}{2}]}$ \\
\bottomrule
\end{tabular}}}
\end{center}
\caption{Character table of the cubic group. The characters of the $2\ell+1$-dimensional irrep of SO(3) with respect to cubic group operations. With the exception of the $\ell=0,1$ cases, this representation is fully reducible with respect to the $\mathcal{O}$ operations.}\label{T-10.2-02}
\end{table}

The full decomposition of the $2\ell+1$-dimensional irreps of the rotation group, whose result for $\ell \leq 8$ are presented in Tab.~ \ref{T-4.0-01}, can be carried out by means of the Great Orthogonality Theorem for characters: if 
\begin{equation}
D^{\ell} = \sum_{\oplus} q_{\nu} D^{\nu}\label{10.2-01}
\end{equation}
is the decomposition of the irrep $\ell$ of SO(3) into the $\#\mathcal{C}l$ cubic group irreps, the multiplicity of the latter is given by
\begin{equation}
q_{\nu} = \frac{1}{|\mathcal{O}|} \sum_{i = 1}^{\#\mathcal{C}l}|\mathcal{C}l_i|[\chi_i^{\nu}]^*\chi_i^{\ell}\label{10.2-02}
\end{equation}
where the order of $\mathcal{O}$ is at the denominator, while $\chi_i^{\nu}$  and $\chi_i^{\ell}$ are respectively the characters of the irreps of the cubic and the rotation group related to the conjugacy class $\mathcal{C}l_i$ with $|\mathcal{C}l_i|$ elements. In particular, the map between the basis states of the latter and the SO(3) ones can be reconstructed via the projectors in Eq.~\eqref{4.0-10}. Denoting with $T_q^{(k)}$ the $q$ component of a spherical tensor of rank $2k+1$, the generic component of the irreducible cubic tensor obtained from it is 
\begin{equation}
T_q^{(\Gamma,k)} = \sum_{q' = -k}^k \sum_{g \in\mathcal{O}}\chi_{\Gamma}(g)D_{qq'}^{k}(g) T_{q'}^{(k)}\label{10.2-03}
\end{equation}
where the index $q$ ranges from $-k$ to $k$. Conversely, the \textit{transpose} transformation rule holds for the basis states of the two groups, 
\begin{equation}
|\ell, \Gamma, m\rangle = \sum_{m' = -\ell}^{\ell} \sum_{g \in\mathcal{O}}\chi_{\Gamma}(g)D_{m'm}^{\ell}(g) |\ell,m'\rangle.\label{10.2-04}
\end{equation}
Due to rank deficiency of the projector, the label $k$ in the cubic tensor does not represent any more its effective rank, but only the original irrep of SO(3) from which it has been obtained: the descent in symmetry, in fact, constrains the maximum rank of any irreducible tensor operator to run from one to three. As noticed in Sec.~\ref{S-4.0} for the energy eigenstates, the non-null components $q$ of  $T^{(\Gamma,k)}$ and $|\Gamma \ell \rangle$, admixture of the $q \mod 4$ components of their SO(3) counterparts, can be univocally labeled with the $I_z$ quantum number. The ensuing distribution of $m$ components of a spin-$l$ irrep into the $(\ell, \Gamma)$ irreps of the cubic group is known under the name of \emph{subduction} \cite{DEP09}. Furthermore, when the occurrence coefficient $q_{\Gamma}$ the irrep $\Gamma$ of $\mathcal{O}$ is greater than one, further linear combinations on the outcoming states (cf. Eq.~\eqref{10.2-03}) or cubic tensor components (cf. Eq.~\eqref{10.2-04}) should be considered, in order to block-diagonalize the relevant projector and disentangle the repeated multiplets of states.\\

\end{appendix}

\addto\captionsbritish{\renewcommand{\appendixname}{}}

\section*{\textsf{References}}

\begingroup
\renewcommand{\section}[2]{}

\end{document}